\documentclass[aps,preprint,byrevtex,nobibnotes,11pt]{revtex4}%
\usepackage{amsfonts}
\usepackage{amsmath}
\usepackage{amssymb}
\usepackage{graphicx}%
\begin{document}
\preprint{UATP/06-01}
\title[Solubility in Compressible Polymers]{Solubility in Compressible Polymers:\ Beyond the Regular Solution Theory}
\author{Albert A. Smith and P.D. Gujrati}
\email{pdg@arjun.physics.uakron.edu}
\affiliation{The Department of Physics, The Department of Polymer Science, The University
of Akron, Akron, OH 44325}

\pacs{PACS number}

\begin{abstract}
The age-old idea of "like dissolves like" requires a notion of "likeness" that
is hard to quantify for polymers. We revisit the concepts of pure component
cohesive energy density $c^{\text{P}}$ and mutual cohesive energy density
$c_{12}$\ so that they can be extended to polymers. We recognize the inherent
limitations of $c_{12}$ due to its very definition, which is based on the
assumption of no volume of mixing (true for incompressible systems), one of
the assumptions in the random mixing approximation (RMA); no such limitations
are present in the identification of $c^{\text{P}}.$ We point out that the
other severe restriction on $c_{12}$\ is the use of pure components in its
definition because of which $c_{12}$\ is not merely controlled by mutual
interactions. Another quantity $c_{12}^{\text{SRS}}$\ as a measure of mutual
cohesive energy density that does not suffer from the above limitations of
$c_{12}$ is introduced. It reduces to $c_{12}$ in the RMA limit$.$ We are able
to express $c_{12}^{\text{SRS}}$ in terms of $c_{12}$ and pure component
$c^{\text{P}}$'s.\ We also revisit the concept of the internal pressure and
its relationship with the conventional and the newly defined cohesive energy
densities. In order to investigate volume of mixing effects, we introduce two
different mixing processes in which volume of mixing remains zero. We then
carry out a comprehensive reanalysis of various quantities using a recently
developed recursive lattice theory that has been tested earlier and has been
found to be more accurate than the conventional regular solution theory such
as the Flory-Huggins theory for polymers. In the RMA limit, our recursive
theory reduces to the Flory-Huggins theory or its extension for a compressible
blend. Thus, it supersedes the Flory-Huggins theory. Consequently, the
conclusions based on our theory are more reliable and should prove useful.

\end{abstract}
\date{\today}
\maketitle

\section{Introduction}

\subsection{Pure Component and Athermal Reference State}

Understanding the solubility of a polymer in a solvent is a technologically
important problem. It is well documented that "like dissolves like" but it is
almost impossible to quantify the notion of "likeness" of materials. The
understanding of solubility requires a basic understanding of "likeness" that
is lacking at present. Solubility parameters in all their incarnations are
attempts to quantify this simple notion. There are several ways one can
proceed to understand solubility by considering various thermodynamic
quantities, not all of which are equivalent. However, almost all these current
approaches are based on our thermodynamic understanding of mixtures at the
level of the \emph{regular solution theory }(RST)
\cite{vanLaar,Hildebrand1916,Hildebrand}.

In practice, one usually introduces the Hildebrand solubility parameter
\[
\delta=\sqrt{c^{\text{P}}}%
\]
for a pure component in terms of $c^{\text{P}},$ the latter being the cohesive
energy density of the pure (P) component and is normally reported at its
boiling temperature. In this work, we will take this definition to be
operational at any temperature, thus treating $\delta$\ as a thermodynamic
quantity, and not just a parameter. Then its value at the boiling point will
be the customarily quoted solubility parameter. The cohesive energy is related
to the interaction energy $\mathcal{E}_{\text{int}}$\ obtained by subtracting
the energy $\mathcal{E}_{\text{ath}}$\ of the hypothetical \emph{athermal}
state of the pure component, the state in which the self-interactions (both
interparticle and intraparticle) are absent, from the total energy
$\mathcal{E}$:%
\begin{equation}
\mathcal{E}_{\text{int}}\ \equiv\mathcal{E-E}_{\text{ath}}.
\label{InteractionEnergy}%
\end{equation}

The athermal state is usually taken to be at the same temperature $T$ and the
volume $V$ as the system itself. In almost all cases of interest,
$\mathcal{E}_{\text{ath}}$ is nothing but the kinetic energy and depends only
on $T$ but not on $V.$ Thus, $\mathcal{E}_{\text{int}}$ depends directly on
the strength of the self-interaction, the only interaction present in a pure
component and vanishes as self-interactions disappear, since $\mathcal{E}%
\rightarrow\mathcal{E}_{\text{ath}}$ as this happens. Thus, $\mathcal{E}%
_{\text{int}}$\ can be used to estimate the strength of the self-interaction.
One can also take the hypothetical state to be at the same $T$ and the
pressure $P$. In this case, there would in principle be a difference in the
volume $V$ of the pure component and $V_{\text{ath}}$ of the hypothetical
state, but this difference will not change $\mathcal{E}_{\text{int}}$ in
(\ref{InteractionEnergy}). The hypothetical state is \emph{approximated} in
practice by the vapor phase at the boiling point in which the particles are
assumed to be so far apart that their mutual interactions can be neglected.
However, to be precise, it should be the vapor phase at zero pressure so that
the volume is infinitely large to make the particle separation also infinitely
large to ensure that they are non-interacting. This causes problems for
polymers \cite{Choi}. Our choice of the athermal state in
(\ref{InteractionEnergy}) to define $\mathcal{E}_{\text{int}}$ overcomes these
problems altogether. By definition,%
\begin{equation}
c^{\text{P}}\equiv-\mathcal{E}_{\text{int}}/V \label{cohesive_def}%
\end{equation}
is the negative of the interaction energy density (per unit volume $V)$ of the
system at a given temperature $T.$ (At the boiling point, $V$ is taken to be
the volume of the liquid)$.$ The negative sign is to ensure $c^{\text{P}}>0$
since usually $\mathcal{E}_{\text{int}}$\ is negative to give cohesion.
Because of its dimensions, $c^{\text{P}}$\ is also known as the \emph{cohesive
pressure}. In this form, $c^{\text{P}}$ is a thermodynamic quantity and
represents the thermodynamically averaged (potential) energy per unit volume
of the pure component. Thus, $c^{\text{P}}$ can be calculated even for
macromolecules like polymers, which is of central interest in this work or for
polar solvents. It is a macroscopic, i.e. a thermodynamic quantity
characterizing\ microscopic interparticle interactions in a pure component.
This is important as it is well known that $\delta$\ cannot be measured
directly as most polymers cannot be vaporized without decomposing \cite{Du}.
Thus, theoretical means are needed to evaluate $\delta,$ which is our goal in
this work.

\ It should be noted from the above definition that $c^{\text{P}}$ contains
the contributions from all kinds of forces (van der Waals, dipolar, and
hydrogen bonding forces) in the system \cite{Hansen}. In this work, we are
only interested in the weak van der Waals interactions for simplicity, even
though the investigation can be easily extended to include other interactions.
It should also be noted that the definition (\ref{cohesive_def}) does not
suffer from any inherent approximation, and can be used to calculate
$c^{\text{P}}$ in any theory or by experimental measurements. As we will see,
this is not true of the mutual cohesive density definition, which is
introduced below.

\subsection{Mixture and Self-interacting Reference State}

As it stands, the pure component quantity $c^{\text{P}}$ is oblivious to what
may happen in a mixture formed with another component. Despite this,
$c^{\text{P}}$ or $\delta$\ is customarily interpreted as a rough measure of a
solvent's ability to dissolve a solute ("like dissolves like"). This
interpretation of the solubility parameter is supposed to be reliable only for
non-polar solvents formed of small molecules, and one usually refrains from
using it for polar solvents such as esters, ketones, alcohol, etc. Our
interest is to investigate this quantity for macromolecules here, and its
significance for the solubility in a mixture. According to the famous
Scatchard-Hildebrand relation \cite{Hildebrand}, the energy of mixing $\Delta
E_{\text{M}}$ per unit volume for a binary mixture of two components $1$ and
$2$ must always be non-negative since it is given by
\begin{equation}
\Delta E_{\text{M}}=(\delta_{1}-\delta_{2})^{2}\varphi_{1}\varphi
_{2},\ \ \ \ (\varphi_{1}+\varphi_{2}\equiv1), \label{mixEnergy0}%
\end{equation}
where $\varphi_{i}$ are the volume fractions of the two components $i$, and
$\delta_{1}$ and $\delta_{2}$ are their solubility parameters. It is
implicitly assumed here that the volume of mixing $\Delta V_{\text{M}}=0$.
Thus, this expression does not contain the contribution from a non-zero volume
of mixing. We will be interested in investigating this additional contribution
in this work. Later, we will discover that (\ref{mixEnergy0}) can only be
justified [see (\ref{VolumeFraction}) below], if we take
\begin{equation}
\varphi_{i}\equiv\phi_{\text{m}i}/\phi_{\text{m}},\ \ \ \ \ (\phi_{\text{m}%
}\equiv\phi_{\text{m}1}+\phi_{\text{m}2}) \label{monomer_fraction}%
\end{equation}
as representing the monomer density ratios, or monomer fractions; see below
for precise definition. Only in the RST can we identify $\varphi_{i}$ with the
volume fraction of the $i$th component. The significance of (\ref{mixEnergy0})
is that the behavior of the mixture is completely determined by the pure
component properties and provides a justification for "like dissolves like".
This must be a gross approximation even for non-polar systems and cannot be
true in general since the energy of mixing can be negative in many cases, as
shown elsewhere \cite{Gujrati2003}, and as we will also see here later; see,
for example, Fig. \ref{F18}. What we discover is that $\Delta E_{\text{M}}$
can be negative even if $\Delta V_{\text{M}}=0$. Thus, zero volume of mixing
is not sufficient for the validity of (\ref{mixEnergy0}).

For the mixture, we need to introduce a thermodynamic or macroscopic quantity
$c_{12}$ characterizing the mutual interaction between the two components;
this quantity should be determined by the microscopic interaction between the
two components. Therefore, the introduction of $c_{12}$ will, in principle,
require comparing the mixture with a hypothetical mixture in which the two
components have no mutual interactions similar to the way the pure component
(having self-interaction) was compared with the athermal state above (in which
the self-interaction was absent) for the introduction of $c^{\text{P}}.$ The
hypothetical state of the mixture should not be confused with the athermal
state of the mixture; the latter will require even the self interaction of the
two components to be absent. The new hypothetical state will play an important
role in our investigation and will be called \emph{self-interacting reference
state} (SRS). The mutual interaction energy in the binary mixture is obtained
by subtracting the energy of SRS from that of the mixture:
\begin{equation}
\mathcal{E}_{\text{int}}^{(\text{M})}\ \equiv\mathcal{E-E}_{\text{SRS}};
\label{Mutual_InteractionEnergy}%
\end{equation}
compare with (\ref{InteractionEnergy}). Just as before, SRS can be taken at
the same $T$ and $V,$ or $T$ and $P$ as the mixture. This allows us to
separate out the two contributions, one due to the presence of mutual
interactions at the same volume $V_{\text{SRS}}=V$ of SRS, and the second
contribution due to merely a change in the volume from $V_{\text{SRS}}$ to
$V;$ this was not the case with $\mathcal{E}_{\text{int}}$ above.\ Each
contribution of $\mathcal{E}_{\text{int}}^{(\text{M})}$ vanishes as mutual
interactions disappear, since $\mathcal{E}\rightarrow\mathcal{E}_{\text{SRS}%
},$ and $V_{\text{SRS}}\rightarrow V$ as this happens. If $\mathcal{E}%
_{\text{int}}^{(\text{M})}$ is used to introduce a mutual cohesive energy
density (to be denoted later by $c_{12}^{\text{SRS}}$), then such a density
would most certainly vanish with vanishing mutual interaction strength.
However, this is not the conventional approach adopted in the literature when
introducing $c_{12}$. Rather, one considers the energy of mixing. We will
compare the two approaches in this work. Whether the conventionally defined
$c_{12}$ vanishes with the mutual interactions remains to be verified. In
addition, whether it is related to the pure component self interaction
cohesive energy densities $c_{11}^{\text{P}}$ and $c_{22}^{\text{P}}$ in a
trivial fashion such as (\ref{london_berthelot_Conj}), see below, needs to be
investigated$.$ As we will see, this will require further assumptions even
within RST to which we now turn.

\subsection{Regular Solution Theory (RST)}

The customary approach to introduce $c_{12}$\ is to follow the classical
approach developed by van Laar, and Hildebrand
\cite{vanLaar,Hildebrand1916,Hildebrand}, which is based on RST, a theory that
can be developed consistently only on a lattice. The theory describes an
incompressible lattice model or a compressible model in which the volume of
mixing $\Delta V_{\text{M}}$ is zero. The lattice model is introduced as
follows. One usually takes a homogeneous lattice of coordination number $q$
and containing $N$ sites. The volume of the lattice is $Nv_{0}$ where $v_{0}$
is the lattice cell volume. We place on this lattice the polymer (component
$1$) and the solvent (component $2$) molecules in such a way that the
connectivity of the molecules are kept intact. It should be stressed that the
solvent molecules themselves may be polymeric in nature. In addition, the
excluded volume interactions are enforced by requiring that no site can be
occupied by more than one monomer at a time. The monomer densities of the two
components $\phi_{\text{m}i}$ ($i=1,2)$ are the densities of sites occupied by
the $i$th component. Two monomers belonging to components $i$ and $j,$
respectively, interact with an energy $e_{ij}=$ $e_{ji}$\ only when they are
nearest-neighbor. (This nearest-neighbor restriction can be easily relaxed,
but we will not do that here for simplicity.) The interaction between the
polymer and the solvent is then characterized by a single excess energy
parameter $\varepsilon\equiv\varepsilon_{12}$ defined in general by the
combination%
\begin{equation}
\varepsilon_{ij}\equiv e_{ij}-(1/2)(e_{ii}+e_{jj}). \label{excess_E}%
\end{equation}
The origin of this combination is the fixed lattice connectivity as shown
elsewhere \cite{Gujrati2000}$.$ For an incompressible binary mixture for
which
\[
\phi_{\text{m}1}+\phi_{\text{m}2}=1,
\]
the excess energy $\varepsilon$ is sufficient for a complete thermodynamic
description on a lattice \cite{Gujrati2000}.

On the other hand, a compressible lattice model of the mixture, which requires
introducing voids as a separate component ($i=0$), will usually require two
additional excess energy parameters $\varepsilon_{01},$ and $\varepsilon_{02}$
\cite{Gujrati1998}. In the following, we will implicitly assume that a void
occupies a site of the lattice and has it occupies a volume $v_{0}.$ In our
picture, a pure component is a pseudo-binary mixture ($i=0,1$ or $i=0,2),$
while a compressible binary mixture is a pseudo-tertiary mixture ($i=0,1,2).$

Since voids do not interact with themselves or with any other component, we
must set%
\[
e_{00}=e_{0i}=0,
\]
so that the corresponding excess energy
\begin{equation}
\varepsilon_{0i}=-(1/2)e_{ii}, \label{pureExcess}%
\end{equation}
see (\ref{excess_E}), and is normally positive since $e_{ii}$ is usually negative.

Two of the three conditions for RST\ to be operative are

(i) \emph{Isometric Mixing: }$\Delta V_{\text{M}}$ $\equiv0$, and

(ii) \emph{Symmetric Mixture: }The two components must be of the \emph{same}
size and have the same interaction ($e_{11}\equiv e_{22}$). This is a
restatement of "like dissolves like."

The condition $\Delta V_{\text{M}}=0$ for \emph{isometric mixing} is always
satisfied in an incompressible system. For a compressible system, $\Delta
V_{\text{M}}$ need not be zero, and one can have either an isometric or a
non-isometric mixing depending on the conditions.

\subsubsection{Random Mixing Approximation (RMA)}

The third condition for RST to be valid is the fulfillment of the

iii) \emph{RMA\ Limit: }The interaction energy $\varepsilon_{ij}$ be extremely
weak ($\varepsilon_{ij}\rightarrow0$), and the coordination number $q$ of the
lattice extremely large ($q\rightarrow\infty$) simultaneously so that the
product%
\[
q\varepsilon_{ij}\text{ remains fixed and finite, as }\varepsilon
_{ij}\rightarrow0,\text{ and }q\rightarrow\infty.
\]
\ \ \ \ \ \ \ \ \ \ \ \ \ \ \ \ \ \ \ \ \ \ \ 

Indeed, if one introduces the dimensionless Flory-Huggins chi parameter
$\chi_{ij}\equiv\beta q\varepsilon_{ij},$ where $\beta\equiv1/k_{\text{B}}T,$
$T$\ being the temperature in the Kelvin scale, then one can also think of
keeping $\chi_{ij}$ fixed and finite, instead of $q\varepsilon_{ij},$ under
the simultaneous limit
\begin{equation}
\chi_{ij}\equiv\beta q\varepsilon_{ij}\text{ fixed and finite, as }%
\beta\varepsilon_{ij}\rightarrow0,\text{ and }q\rightarrow\infty.
\label{RMALimit}%
\end{equation}
It is quite useful to think of RST in terms of these limits as both
$\beta\varepsilon_{ij}$ and $q$ are dimensionless. The above simultaneous
limit gives rise to what is usually known as the \emph{random mixing
approximation} (RMA), and has been discussed in detail in a recent review
article \cite{Gujrati2003} in the context of polymer mixtures.\ For an
incompressible system, we need to keep only a single chi parameter $\chi
\equiv\beta q\varepsilon$ fixed and finite in the limit. For a compressible
system, one must also simultaneously keep the two additional chi parameters
related to $\varepsilon_{01}$ and $\varepsilon_{02},$ and $\beta Pv_{0}$ fixed
and finite, where $P$ is the pressure \cite{Gujrati1998,Gujrati2003}.

The RMA limit can be applied even to a pure component (for which the first two
conditions are meaningless). It can also be applied when mixing is not
isometric \cite{Note1} or when the mixture is not symmetric. Therefore, RST is
equivalent to the \emph{isometric RMA}. In the unusual case $\chi=0,$
($q\rightarrow\infty),$ the resulting isometric theory is known as the
\emph{ideal solution theory}, which will not be considered here as we are
interested in the case $\chi\neq0.$

\subsubsection{London-Berthelot and Scatchard-Hildebrand Conjectures}

The energy of mixing $\Delta E_{\text{M}}$ per unit volume is the central
quantity for solubility considerations, and can be used to introduce an
effective "energetic" chi \cite{Gujrati1998,Gujrati2003} as follows:%
\begin{equation}
\chi_{\text{eff}}^{\text{E}}\equiv\beta\Delta E_{\text{M}}v_{0}/\phi
_{\text{m}1}\phi_{\text{m}2}, \label{effective_EChi}%
\end{equation}
which is a measure of the Flory-Huggins $\chi(\equiv\chi_{12})$ parameter or
the excess energy $\varepsilon(\equiv\varepsilon_{12})$; the latter is
directly related to the mutual interaction energy $e_{12},$ see
(\ref{excess_E}), which explains the usefulness of $\Delta E_{\text{M}}$ for
solubility considerations. One of the important consequences of the
application of RST is the Scatchard-Hildebrand conjecture \cite{Hildebrand},
according to which the energy of mixing is given by a non-negative form
(\ref{mixEnergy0}). This is known to be violated; see \cite{Gujrati2003}, and
below. One of the reasons for its failure could be the much abused
London-Berthelot conjecture%
\begin{equation}
c_{12}=\sqrt{c_{11}^{\text{P}}c_{22}^{\text{P}}}\equiv\delta_{1}\delta_{2},
\label{london_berthelot_Conj}%
\end{equation}
relating the mutual cohesive energy density $c_{12}$\ with the pure component
cohesive energy densities $c_{11}^{\text{P}},$ $c_{22}^{\text{P}},$\ and used
in the derivation of (\ref{mixEnergy0}). In contrast, the London conjecture
\begin{equation}
e_{12}=-\sqrt{(-e_{11})(-e_{22})} \label{london_Conj}%
\end{equation}
deals directly with the microscopic interaction energies, and is expected to
hold for non-polar systems; see also \cite{Israelachvili} for an illuminating
derivation and discussion. In the isometric RMA limit, the two conjectures
become the same in that (\ref{london_Conj}) implies
(\ref{london_berthelot_Conj}). In general, they are two \emph{independent}
conjectures. As we will demonstrate here, (\ref{london_Conj}) does not imply
(\ref{london_berthelot_Conj}) once we go beyond the RMA limit. We will also
see that the non-negativity of the form (\ref{mixEnergy0}) is violated even
for isometric mixing.

Association of one component, such as through hydrogen bonding, usually makes
$e_{12}^{2}<\left\vert e_{11}\right\vert \left\vert e_{22}\right\vert .$ On
the other hand, complexing results in the opposite inequality $e_{12}%
^{2}>\left\vert e_{11}\right\vert \left\vert e_{22}\right\vert .$ It is most
likely that the binary interaction between monomers of two distinct species
will deviate from the London conjecture (\ref{london_Conj}) to some degree.
Thus, some restrictions have to be put on the possible relationship between
these energy parameters for our numerical calculations. We have decided to
consider only those parameters that satisfy the London conjecture
(\ref{london_Conj}) in this work for physical systems.

\subsubsection{ Deviation from London-Berthelot Conjecture}

The deviation from the London-Berthelot conjecture
(\ref{london_berthelot_Conj}) is usually expressed in terms of a binary
quantity $l_{12}$ defined via%
\begin{equation}
c_{12}=\sqrt{c_{11}^{\text{P}}c_{22}^{\text{P}}}(1-l_{12}), \label{Dev_l12}%
\end{equation}
and it is usually believed that the magnitude of $\ l_{12}\ $is very small:
\begin{equation}
\left\vert l_{12}\right\vert <<1. \label{Dev_Magnitude}%
\end{equation}
It is possible that $l_{12}$ vanishes at isolated points, so that the
London-Berthelot conjecture becomes satisfied. This does not mean that the
system obeys RST there.\ As we will demonstrate here, we usually obtain a
non-zero $l_{12}$, thus implying a failure of the London-Berthelot conjecture
(\ref{london_berthelot_Conj}), even if the London conjecture
(\ref{london_Conj}) is taken to be operative. Another root for the failure of
(\ref{london_berthelot_Conj}) under this assumption could be non-isometric
mixing. A third cause for the failure could be the corrections to the RMA
limit, since a real mixture is not going to truly follow RST. To separate the
effects of the three causes, we will pay particular attention to $l_{12}$ in
this work. We will assume the London conjecture (\ref{london_Conj}) to be
valid, and consider the case of isometric mixing. We will then
evaluate$\ l_{12}$ using a theory that goes beyond RST. A non-zero $l_{12}$ in
this case will then be due to the corrections to the RMA limiting behavior or RST.

\subsection{Internal Pressure}

Hildebrand \cite{Hildebrand1916} has also argued that the solubility of a
given solute in different solvents is determined by relative magnitudes of
internal pressures, at least for nonpolar fluids. Thus, we will also
investigate the internal pressure, which is defined properly by
\begin{equation}
P_{\text{IN}}=-P_{\text{int}}\equiv-(P-P_{\text{ath}}), \label{P_IN}%
\end{equation}
where $P_{\text{int}}$\ is the contribution to the pressure $P$ due to
interactions in the system, and is obtained by subtracting $P_{\text{ath}}$
from $P$, where $P_{\text{ath}}$ is the pressure of the hypothetical athermal
state. The volume $V$ of the system, which has the pressure $P$, may or may
not be equal to the volume of the hypothetical athermal state. This means that
$P_{\text{IN}}$\ will have different values depending on the choice of
athermal state volume. In this work, we will only consider the athermal state
whose volume is equal to the volume $V$ of the system; its pressure then has
the value $P_{\text{ath}},$ which is not equal to the pressure $P$ of the
system. As the interactions appear in the athermal system at constant volume,
its pressure $P_{\text{ath}}$ will reduces due to attractive interactions.
This will give rise to $P_{\text{int}},$\ which is negative so that
$P_{\text{IN}}$ will be a positive quantity for attractive interactions. For
repulsive interactions, which we will not consider here, $P_{\text{IN}}$ will
turn out to be a negative quantity. In either case, as we will show here,
$P_{\text{IN}}$ should be distinguished from ($\partial\mathcal{E}/\partial
V)_{T},$ with which it is usually equated. Their equality holds only in a very
special case as we will see here.

The availability of $PVT$-data makes it convenient to obtain ($\partial
\mathcal{E}/\partial V)_{T}.$ Therefore, it is not surprising to find it
common to equate it with $P_{\text{IN}}.$ In a careful analysis, Sauer and Dee
have shown a close relationship between ($\partial\mathcal{E}/\partial V)_{T}$
and $c^{\text{P}}$ \cite{Dee}. Here, we will investigate the relationship
between $P_{\text{IN}}$ and $c^{\text{P}}.$

\subsection{Internal Volume}

One can also think of an alternate scenario in which the pressure of the
athermal state is kept constant as interactions appear in the athermal system.
This pressure can be taken to be either $P$ or $P_{\text{ath}}$. Let us keep
its pressure to be $P$, and so that its volume is $V_{\text{ath}}.$ In this
case, the volume of the system $V$ will be smaller (greater) than
$V_{\text{ath}}$ of the athermal state because of the attractive (repulsive)
interactions, so that one can also use the negative of the change in the
volume $V-V_{\text{ath}}$ as a measure of the attractive (repulsive)
interactions. This allows us to introduce the following volume, to be called
\emph{internal volume }%
\begin{equation}
V_{\text{IN}}=-V_{\text{int}}\equiv-(V-V_{\text{ath}}) \label{V_IN}%
\end{equation}
as a measure of cohesiveness.

The two internal quantities are mutually related and one needs to study only
one of them.

\subsection{Beyond Regular Solution Theory: Solubility and Effective Chi}

In polymer solutions or blends of two components $1$ and $2$, non-isometry is
a rule rather than exception due to asymmetry in the degrees of polymerization
and in the pure component interactions. Thus, the regular solution theory is
most certainly inoperative. An extension of RST, to be described as the
non-isometric regular solution theory, allows for non-isometric mixing
($\Delta V_{\text{M}}\neq0$), and its successes and limitations have been
considered elsewhere \cite[where it is simply called the regular solution
theory]{RaneGuj2003}. It is the extended theory that will be relevant for polymers.

Let us suppress the pure component index $i$ in the following. Crudely
speaking, $c^{\text{P}}$ for a pure component is supposed to be related to the
pure component interaction parameter $\varepsilon(\equiv\varepsilon_{0i}),$ or
more correctly $\chi(\equiv\chi_{0i}).$ However, $\varepsilon$ is a
microscopic system parameter, independent of the thermodynamic state of the
system; thus, it must be independent of the composition, the degree of
polymerization, etc. On the other hand, $c^{\text{P}}$ is determined by the
thermodynamic state. Thus, its value will change with the thermodynamic state,
the degree of polymerization, etc. Only in the RMA limit, see
(\ref{RMA_Cohesive}) below, does one find a trivial proportionality
relationship between the two quantities $c^{\text{P}}$ and $\varepsilon,$ the
constant of proportionality being determined by the square of the pure
component monomer density $\phi_{\text{m}};$ otherwise, there is no extra
dependence on temperature and pressure of the system. This RMA behavior is
certainly not going to be observed in real systems, where we expect a complex
relationship between $c^{\text{P}}$ and $\varepsilon.$ A similar criticism
also applies to the behavior of $c_{12},$ for which one considers the energy
of mixing $\Delta E_{\text{M}};$ see below.

\subsubsection{Solubility}

In the RMA limit of an incompressible binary mixture, it is found that
$c_{12}$ is proportional to the mutual interaction energy $e_{12}$ between the
two components, see (\ref{RMA_CohesiveDensities}) below. Thus, the sign of
$c_{12}$ is determined by $e_{12}.$\ On the other hand, $\Delta E_{\text{M}}$
in the RMA limit is proportional to the excess energy $\varepsilon
(\equiv\varepsilon_{12})$ or $\chi(\equiv\chi_{12}),$ so that its sign is
determined by $\varepsilon$. However, the solubility of component $1$ in $2$
in a given state is determined not by their mutual interaction energy
$e_{12},$ which is usually attractive, but by the sign of the excess energy
$\varepsilon,$ and the entropy of mixing. For an incompressible binary
mixture, we have only one exchange energy $\varepsilon.$ Even away from the
RMA limit for the incompressible mixture, a positive $\varepsilon$ implies
that the two components will certainly phase separate at low enough
temperatures. Their high solubility (at constant volume) at very high
temperatures is mostly due to the entropy of mixing, but the energy of mixing
will play an important role at intermediate temperatures. The solubility
increases as $\varepsilon$\ decreases. It also increases as $T$ increases,
unless one encounters a lower critical solution temperature (LCST) in which
case the solubility will decrease with $T$. It is well known that LCST can
occur\ in a blend due to compressibilty; see, for example, \cite{Mukesh}.
Similarly, the solubility at constant pressure will usually decrease with
temperature. However, it is also possible for isobaric solubility to first
increase and then decrease with $T.$ Thus, a properly defined mutual cohesive
energy\ for a compressible blend should be able to capture such a features. A
negative $\varepsilon$ implies that the two components will never phase
separate.\ Thus, a complete thermodynamic consideration is needed for a
comprehensive study of solubility even when we are not in the RMA limit, and
requires investigating thermodynamic quantities such as $c_{12}$ or the
(energetic) effective chi (\ref{effective_EChi}) to which we now turn$.$ This
is even more so important when we need to account for compressibility.

\subsubsection{Energetic Effective Chi}

We have recently investigated a similar issue by considering the behavior of
the effective chi in polymers \cite{RaneGuj2005}, where an effective chi,
relevant in scattering experiments, was defined in terms of the excess second
derivative of the free energy with respect to some reference state. This
investigation is different in spirit from other investigations in the
literature, such as the one carried out by Wolf and coworkers \cite{Wolf}
where one only considers the free energy of mixing. As said above, $\Delta
E_{\text{M}}$ or $\chi_{\text{eff}}^{\text{E}}$ is determined by the excess
energy $\varepsilon_{12},$ and not by $e_{12}.$ However, since $c_{12}$ is
supposed to be a measure of $e_{12},$ it is defined indirectly by that part of
the energy of mixing $\Delta E_{\text{M}}$ or $\chi_{\text{eff}}^{\text{E}},$
see (\ref{effective_EChi}), that is supposedly determined by the mutual energy
of interaction $e_{12}$. For this, one must subtract the contributions of the
pure components from $\Delta E_{\text{M}}$ or $\chi_{\text{eff}}^{\text{E}};$
see below for clarity. Because of this, even though the previous investigation
of the effective chi \cite{RaneGuj2005} provides a clue to what might be
expected as far as the complex behavior of $\chi_{\text{eff}}^{\text{E}}$ is
involved, a separate investigation is required for the behavior of the
cohesive density $c_{12}$, which we undertake here. We will borrow some of the
ideas developed in \cite{RaneGuj2005}, especially the requirements that

(i) the cohesive energy density $c_{ij}$ for an $i$-$j$ mixture vanish with
$e_{ij},$ and

(ii) the formulas to determine $c_{ij}$ reduce to the standard RMA form under
the RMA limit.

The first condition replaces the requirement in \cite{RaneGuj2005} that the
effective chi vanish with $\varepsilon_{12}.$ The second requirement is the
same as in \cite{RaneGuj2005}. Its importance lies in the simple fact that any
thermodynamically consistent theory must reduce to the same unique theory in
the RMA limit; see \cite{Gujrati2000,RaneGuj2005} for details. We also borrows
the idea of \emph{reference states} that would be fully explained below.

\subsubsection{Symmetric and Asymmetric Blends}

It has been shown in \cite{RaneGuj2003} that the non-isometric regular
solution theory is more successful than RST but again only for a symmetric
blend. A symmetric blend is one in which not only the two polymers have the
same degree of polymerization ($M_{1}=M_{2}$), but they also have identical
interactions in their pure states ($e_{11}=e_{22}$). For asymmetric blends
(blends that are not symmetric), even the non-isometric regular solution
theory is qualitatively wrong. The significant conclusion of the previous work
is that the recursive lattice theory developed in our group \cite[and
references therein]{Gujrati1995b,RyuGuj1997}, and briefly discussed in the
next section to help the reader, is successful in explaining several
experimental results where the non-isometric regular solution theory fails. A
similar conclusion is also arrived at when we apply our recursive theory to
study the behavior of effective chi \cite{Gujrati2000,RaneGuj2005}.

It was shown a while back that the recursive lattice theory is more reliable
than the conventional mean field theory \cite{Gujrati1995a}; the latter is
formulated by exploiting RMA, which is what the regular solution theories are.
The recursive lattice theory goes beyond RMA and is successful in explaining
several experimental observations that could not be explained by the mean
field theory. Our aim here is to apply the recursive theory to study
solubility and to see the possible modifications due to

1. finite $q,$

2. non-weak interactions ($\varepsilon>0$),

3. non-isometric mixing, and

4. disparity in size (asymmetry in the degree of polymerization) and/or pure
component interactions.

\section{Recursive Lattice Theory}

\subsection{Lattice Model}

We will consider a lattice model for a multicomponent polymer mixture in the
following, in which only nearest-neighbor interactions are permitted. As
above, $q$ denotes the lattice coordination number. The number of lattice
sites will be denoted by $N$, and the lattice cell volume by $v_{0}$, so that
the lattice volume $V=Nv_{0}.$ The need for using the same coordination number
and cell volume for the mixture and for the pure components has been already
discussed elsewhere \cite{Gujrati2003} in order to have a consistent
thermodynamics. The monomers and voids are allowed to reside on lattice sites.
The excluded volume restrictions are accounted for as described above. As
shown elsewhere \cite{Gujrati2000}, one only needs to consider excess energies
of interaction $\varepsilon_{ij}$ between monomers of two distinct components
$i$, and $j;$ here, $e_{ij}$ represents the interaction energy between two
nearest-neighbor monomers of components $i$, and $j$ to investigate the
model$.$ To model free volume, one of the components will represent voids or
holes, always to be denoted by $i=0$ here$.$ Thus, for a pure component of,
for example, component $i=1$, the excess interaction energy $\varepsilon\ $is
$\varepsilon_{01}=-(1/2)e_{11}.$ Usually, $e_{11}$ is negative, which makes
$\varepsilon$ positive for a pure component. Let $N_{\text{m}i}$ denote the
number of monomers belonging to the $i$th species and $N_{\text{m}}$ the
number of monomers of all species, so that the number of voids $N_{0}$ is
given by $N_{0}\equiv N-N_{\text{m}}.$ Similarly, let $N_{ij}$ denote the
number of nearest-neighbor contacts between monomers of components $i$, and
$j.$ The densities in the following are defined with respect to the number of sites.

\subsection{Recursive Theory}

In the present work, we will use for calculation the results developed by our
group in which we solve the lattice model of a multicomponent polymer mixture
by replacing the original lattice by a Bethe lattice
\cite{Gujrati1995b,RyuGuj1997}, and solving it exactly using the recursive
technique, which is standard by now. The calculation is done in the grand
canonical ensemble. Thus, the volume $V$ is taken to be fixed. We will assume
that all material components ($i>0$) are linear polymers in nature. The degree
of polymerization of the $i$th component, i.e. the number of consecutive sites
occupied by it, is denoted by $M_{i}\geq1.$\ The linear polymers also include
monomers for which $M_{i}=1.$ Each void ($i=0)$ occupies a single site on the
lattice. Let $\phi_{0}\equiv N_{0}/N$ denote the density of voids,
$\phi_{\text{m}i}\equiv N_{\text{m}i}/N$ the monomer density of the $i$th
component, and $\phi_{ij}\equiv N_{ij}/N$ the density of nearest-neighbor
(chemically unbonded) contacts between the monomers of the two components $i$,
and $j$. It is obvious that
\begin{equation}
\phi_{\text{m}}\equiv\sum_{i>0}\phi_{\text{m}i}=1-\phi_{\text{0}}
\label{monomerD}%
\end{equation}
denotes the density of all material components ($i>0$) monomers. The density
of all chemical bonds is given by
\begin{equation}
\phi=\sum_{i>0}\phi_{\text{m}i}\nu_{i},\ \ \nu_{i}\equiv(M_{i}-1)/M_{i}.
\label{bondD}%
\end{equation}
The quantity $\phi_{\text{u}}\equiv q/2-\phi$ denotes the density of lattice
bonds not covered by the polymers. Let us also introduce $q_{i}\equiv
q-2\nu_{i},$ $\phi_{i\text{u}}\equiv q_{i}\phi_{\text{m}i}/2.$ As shown in
\cite{RyuGuj1997,Gujrati1998}, the pressure $P$ is given by%
\begin{equation}
P=(k_{\text{B}}T/v_{0})[-\ln\phi_{0}+(q/2)\ln(2\phi_{\text{u}}%
/q)]+(k_{\text{B}}Tq/2v_{0})\ln(\phi_{00}^{0}/\phi_{00})], \label{pressureEq}%
\end{equation}
where $k_{\text{B}}$ is the Boltzmann constant, $\phi_{00}$ is the density of
nearest-neighbor void-void contacts, and $\phi_{00}^{0}$ is its \emph{athermal
}value when all excess interactions $\epsilon_{ij}$ are identically zero. The
athermal values of $\phi_{ij}$ are given by
\[
\ \phi_{ij}^{0}=2\phi_{i\text{u}}\phi_{j\text{u}}/\phi_{\text{u}}%
(1+\delta_{ij}),
\]
where $\delta_{ij}$\ is the Kronecker delta, and give the values of the
contact densities in the atermal state when all $\varepsilon_{ij}=0$. The
athermal state has the same volume $V$ as of the original system. The first
term in (\ref{pressureEq}) give the athermal value $P_{\text{ath}}$ of \ $P,$
and the second term is the correction
\begin{equation}
P_{\text{int}}=(k_{\text{B}}Tq/2v_{0})\ln(\phi_{00}^{0}/\phi_{00})
\label{P_int}%
\end{equation}
to the athermal pressure due to interactions. For attractive interactions
responsible for cohesion, $P_{\text{int}}$ is going to be negative, as
discussed above. This is the correction to $P_{\text{ath}}$ due to
interactions and determines the internal pressure \ $P_{\text{IN}}$
$=-P_{\text{int}}.$ The identification also holds for pure components,\ so
that $P_{\text{IN}}$\ remains positive for attractive interactions.

Since there is no kinetic energy on a lattice, the internal energy in the
lattice model is purely due to interactions, and this energy per unit volume
$E_{\text{int}}$ is given by
\begin{equation}
E_{\text{int}}\equiv\sum_{i\geq j\geq0}e_{ij}\phi_{ij}/v_{0},
\label{energy_Def}%
\end{equation}
which will be used here to calculate the cohesive energy density, also known
as the cohesive pressure. Note that the form of the pressure in
(\ref{pressureEq}) does not explicitly depend on the number of components in
the mixture. It should also be clear from (\ref{pressureEq}) that the
incompressible state ($P\rightarrow\infty$) corresponds to $\phi
_{0}\rightarrow0$ at any finite temperature$.$ However, we will not be
interested in this limit in this work. In all our calculations, we will keep
$\phi_{0}>0.$

\subsection{RMA Limit}

The approximation has been discussed in details in
\cite{Gujrati2003,RaneGuj2005}, so we will only summarize the results. This
limit is very important since all thermodynamically consistent lattice
theories must reduce to the same unique theory in the RMA limit. Thus, the RMA
limit provides a unique theory, and can serve as a testing ground for the
consistency of any theory. We now show that our recursive theory reproduces
the known results in the RMA limit. To derive this unique theory, we note that
in this limit, we have $q_{i}\ ^{\underrightarrow{\text{RMA}}}\ q$, and that
the contact densities take the limiting form
\begin{equation}
\phi_{ij}^{0}\ ^{\underrightarrow{\text{RMA}}}\ q\phi_{\text{m}i}%
\phi_{\text{m}j}/(1+\delta_{ij}),\phi_{ij}\ ^{\underrightarrow{\text{RMA}}%
}\ q\phi_{\text{m}i}\phi_{\text{m}j}/(1+\delta_{ij}),\ q\ln(2\phi_{\text{u}%
}/q)\ ^{\underrightarrow{\text{RMA}}}\ -2\phi. \label{RMA1}%
\end{equation}
Finally, using $\phi_{\text{m}0}$ to also denote the void density $\phi_{0},$
we have in this limit%
\begin{subequations}
\begin{align}
&  E_{\text{int}}\ ^{\underrightarrow{\text{RMA}}}\ \sum_{i,j\geq1}qe_{ij}%
\phi_{\text{m}i}\phi_{\text{m}j}/2v_{0},\label{RMA3}\\
&  \beta Pv_{0}\ ^{\underrightarrow{\text{RMA}}}\ -\ln\phi_{0}-\phi-\sum
_{i>0}\chi_{0i}\phi_{\text{m}i}+\sum_{j>i\geq0}\chi_{ij}\phi_{\text{m}i}%
\phi_{\text{m}j}. \label{RMA2}%
\end{align}

For the pure component ($i$th component) quantities, we use an additional
superscript (P) for some quantities, such as $\phi_{ii}^{\text{P}}$
representing the contact density between unbonded monomers, or use the
superscript $i$ for other quantities, such as $E_{\text{int}}^{(i)}$
representing the pure component internal energy. In the RMA limit
\cite{Gujrati2003}, we obtain
\end{subequations}
\begin{equation}
\beta E_{\text{int}}^{(i)}\ ^{\underrightarrow{\text{RMA}}}\ -\chi_{0i}%
\phi_{\text{m}i}^{\text{P}2}/v_{0},\ \ \beta P_{\text{int}}^{(i)}%
\ ^{\underrightarrow{\text{RMA}}}\ -\chi_{0i}\phi_{\text{m}i}^{\text{P}%
2}/v_{0}, \label{RMAdensities1}%
\end{equation}
by restricting (\ref{RMA3},\ref{RMA2}) to a single material component $i>0$ in
addition to the species $0.\ $Here, we have used the fact that $\chi
_{0i}=-(1/2)\beta qe_{ii},$ and that $1-\phi_{\text{m}i}^{\text{P}}$
represents the pure component free volume density$.$ We notice that in the RMA
limit,
\[
E_{\text{int}}^{(i)}\ ^{\underrightarrow{\text{RMA}}}\ P_{\text{int}}^{(i)}%
\]
for a pure component. But this equality will not hold when we go beyond RMA.

\subsection{Infinite Temperature Behavior}

In the limit $\beta\rightarrow0$ at fixed $e_{ij},$ the limiting form of
$E_{\text{int}},$ and $P$ are
\begin{align*}
E_{\text{int}}  &  \rightarrow\sum_{i\geq j}e_{ij}\phi_{ij}^{0}/v_{0},\\
\beta Pv_{0}  &  \rightarrow\beta P_{\text{ath}}v_{0}\equiv-\ln\phi
_{0}+(q/2)\ln(2\phi_{\text{u}}/q).
\end{align*}
and
\[
\beta P_{\text{int}}v_{0}\rightarrow0.
\]
For any finite pressure $P\,,$ we immediately note that $\phi_{0}%
\rightarrow1,$ that is the entire lattice is covered by voids with probability
1.\ This shows that at a fixed and finite pressure $P,$ $\phi$ and $\phi
_{ij}^{0},$ and therefore, $E_{\text{int}}$ vanish as $T\rightarrow\infty$. On
the other hand, if the volume is kept fixed, which requires the free volume
density $\phi_{0}$ to be strictly less than 1, then $P\rightarrow\infty,$ as
$T\rightarrow\infty,$ and $\phi$ and $\phi_{ij}^{0},$ and therefore,
$E_{\text{int}}$ does not vanish as $T\rightarrow\infty.$ Thus, whether there
is cohesion or not at infinite temperatures depend on the process carried out.
This will emerge in our calculation below.

\subsection{Choice of parameters for Numerical Results}

In the following, we will take $v_{0}=1.6\times10^{-28}$ m$^{3}$ for all
calculations. We will take $e_{11}=-2.6\times10^{-21}$ J, and $e_{22}%
=-2.2\times10^{-21}$ J for the two components, unless specified otherwise. The
degree of polymerization $M$ will be allowed to take various values between
$10$, and $500,$ and $q$ is allowed to take various values between $6$, and
$14$ as specified case by case In some of the results, we will keep the
product $eq$ fixed, for example $eq=1.56\times10^{-20}$ J$,$ as we change $q$,
so that $e$ changes with $q,$ since the cohesive energy density is determined
by this product; see below.

\section{Internal Pressure and ($\partial\mathcal{E}/\partial V$)$_{T}$}

The internal pressure $P_{\text{IN}}=-P_{\text{int}}$ is obtained by
subtracting $P_{\text{ath}},$ see (\ref{P_IN}), from the isentropic volume
derivative of the energy $P=-(\partial\mathcal{E}/\partial V)_{S}$, which
follows from the first law of thermodynamics; the derivative is also supposed
be carried out at fixed number of particles $N_{\text{P}}$. We will assume
here that $P_{\text{ath}}$\ is the pressure of the athermal system at the same
volume as the real system\ $(V_{\text{ath}}=V)$.

The internal pressure is not the same as the isothermal derivative
$-(\partial\mathcal{E}/\partial V)_{T}.$ To demonstrate this, we start with
the thermodynamic identity
\[
-(\partial\mathcal{E}/\partial V)_{T}=P-T(\partial P/\partial T)_{V}.
\]
The derivatives here and below are also at fix $N_{\text{P}}$. We now note
that the internal energy is a sum of kinetic and potential (interaction)
energies $K$ and $U,$ respectively. If $U$ does not depend on particle
momenta, which is the case we consider here, then the canonical partition
function in classical statistical mechanics becomes a product of two terms,
one of which depends only on particle momenta, and the other one on $U$. Thus,
the entropy $S$ of the system in classical statistical mechanics is a sum of
two terms $S_{\text{KE}}$ and $S_{\text{conf}}$, where $S_{\text{KE}}$ depends
on $K$, and $S_{\text{conf}}$ on $U;$ see for example, \cite{Fedor}. The
entropy $S_{\text{KE}}$ due to the kinetic energy depends on $K$ and is
independent of the volume and the interaction energy. The configurational
entropy $S_{\text{conf}}$ of the system, on the other hand, is determined by
$U$ and is a function only of the volume $V$\ when there is \emph{no}
interaction ($U=0$), i.e. in the athermal state. Thus, for a system in the
athermal state,
\[
S_{\text{ath}}(K,V)=S_{\text{conf,ath}}(V)+S_{\text{KE}}(K),
\]
where $K(T)$ is the kinetic energy of the system at a given temperature $T$
\cite{Fedor}, so that%
\begin{equation}
(\partial S_{\text{ath}}/\partial V)_{K}=(\partial S_{\text{conf,ath}%
}/\partial V)=\beta P_{\text{ath}}. \label{ath_S_der}%
\end{equation}
Since the identity (\ref{ath_S_der}) is valid in general for the athermal
entropy, we conclude that $\beta P_{\text{ath}}$ is a function independent of
the temperature. Rather, it is a function of $N_{\text{P}},$ and $V$; indeed,
it must be simply a function of the number density $n_{\text{P}}\equiv
N_{\text{P}}/V$ of particles. Consequently, the athermal pressure
$P_{\text{ath}}$ depends linearly on the temperature, so that
\[
T(\partial P_{\text{ath}}/\partial T)_{V}=P_{\text{ath}}.
\]
Using this observation, we find that
\begin{equation}
(\partial\mathcal{E}/\partial V)_{T}=P_{\text{IN}}-T(\partial P_{\text{IN}%
}/\partial T)_{V}=[\frac{\partial}{\partial\beta}(\beta P_{\text{IN}})]_{V},
\label{EV_Derivative}%
\end{equation}
clearly establishing that $P_{\text{IN}}\neq(\partial\mathcal{E}/\partial
V)_{T},$ unless $(\partial P_{\text{IN}}/\partial T)_{V}$ vanishes, which will
happen if $P_{\text{int}}$ becomes independent of the temperature (at constant
volume). (As we will see below, this will happen in the RMA limit.) It is also
possible for $P_{\text{IN}}\ $and $(\partial\mathcal{E}/\partial V)_{T}$ to be
the same at isolated points, the extrema of $P_{\text{IN}}$ as a function of
$T,$ where $(\partial P_{\text{IN}}/\partial T)_{V}=0.$

In the RMA limit, we have%
\begin{equation}
P_{\text{IN}}\equiv-P_{\text{int}}\ ^{\underrightarrow{\text{RMA}}}%
\ -qe\phi_{\text{m}}^{2}/2v_{0}, \label{Pin_RMA}%
\end{equation}
so that $(\partial P_{\text{int}}/\partial T)_{V}=0.$ Thus, we conclude that
\[
P_{\text{IN}}\ ^{\underrightarrow{\text{RMA}}}\ (\partial\mathcal{E}/\partial
V)_{T}\text{, and }c^{\text{P}}\ ^{\underrightarrow{\text{RMA}}}%
\ P_{\text{IN}}%
\]
in the RMA limit, as claimed earlier.

Considering the RMA behavior of $P_{\text{IN}},$ and its equality with
$c^{\text{P}}$ in the same limit, we infer that we can also use the internal
pressure $P_{\text{IN}}$ as an alternative quantity to measure cohesiveness,
as was first noted by Hildebrand \cite{Hildebrand1916}. However, in general,
the two quantities $P_{\text{IN}}$ and $c^{\text{P}}$ are not going to be the same.

\begin{figure}
[ptb]
\begin{center}
\includegraphics[
trim=0.749561in 2.993454in 1.336173in 2.672081in,
height=3.0242in,
width=3.6806in
]%
{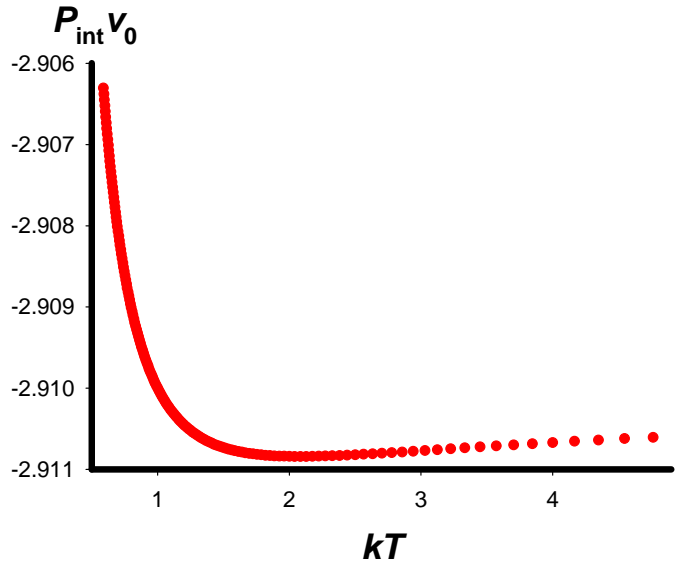}%
\caption{Non-monotonic behavior of $v_{0}P_{\text{int}}$ with $kT$ (arbitrary
energy scale). }%
\label{F1}%
\end{center}
\end{figure}

We can directly calculate the internal pressure and its temperature-dependence
either at constant volume or at constant pressure in our theory from
(\ref{pressureEq}). The internal pressure $P_{\text{IN}}$ is given by the
negative of $P_{\text{int}}$ in (\ref{P_int}), and can be used as an
alternative quantity that is just as good a measure of the cohesion as the
cohesive energy density \cite{Hildebrand1916}, as was discussed above. In Fig.
\ref{F1}, we show the interaction pressure $P_{\text{int}}v_{0}$ for a
symmetric blend ($M=100$) as a function of $kT$. Both axes are in arbitrary
energy unit. In the same energy unit, the excess energies are $e_{11}%
=-0.25=e_{22},$ and $e_{12}=-0.249,$ and the coordination number is $q=6$. The
free volume density is kept fixed at $\phi_{0}=0.1,$ as we change the
temperature$.$ This means that we also keep the total monomer density fixed.
Thus, if we consider the case of a fixed amount of both polymers, then this
analysis corresponds to keeping the volume fixed as the temperature is varied.
In other words, $P_{\text{int}}$ in Fig. \ref{F1} is for an isochoric process.
We immediately notice that $P_{\text{int}}$ not only changes with $T,$ so that
the second term in (\ref{EV_Derivative}) does not vanish, but it is also
non-monotonic. Moreover, the difference between $(\partial\mathcal{E}/\partial
V)_{T}$ and $P_{\text{IN}}$ could be substantial, especially at low
temperatures (see Fig. \ref{F1}) so that using $(\partial\mathcal{E}/\partial
V)_{T}$ for $P_{\text{IN}}$ could be quite misleading.

The minimum in $P_{\text{int}}$\ occurs at $kT\approx2.0$ in Fig. \ref{F1}. At
this point, $(\partial\mathcal{E}/\partial V)_{T}$ and $P_{\text{IN}}$ become
identical, but nowhere else. However, they become asymptotically close to each
other at very high temperatures where the slope $(\partial P_{\text{IN}%
}/\partial T)_{V}$ gradually vanishes. Let us fix the arbitrary energy unit to
be $1.38\times10^{-21}$J (equal in magnitude to $100k$), so that $kT=2.0$ in
this energy unit$,$ the approximate location of the minimum in $P_{\text{int}%
}$ in Fig. \ref{F1}. The minimum occurs around $T=200$ K, i.e., $t_{\text{C}%
}=-73.15$ $^{\circ}$C. Thus, $P_{\text{IN}}$ is an decreasing function of
temperature approximately above $t_{\text{C}}\simeq-73.15$ $^{\circ}$C, and an
increasing function below it. The location of the maximum in $P_{\text{IN}}$
will change with the applied pressure, the two degrees of polymerization, the
interaction energies, etc. The significant feature of Fig. \ref{F1} is the
very broad flat region near the minimum, which implies a very broad flat range
in temperature $t_{\text{C}}$ near the maximum of $P_{\text{IN}}$.

\begin{figure}
[ptb]
\begin{center}
\includegraphics[
trim=1.128415in 6.022026in 1.130859in 0.878988in,
height=3.768in,
width=5.9171in
]%
{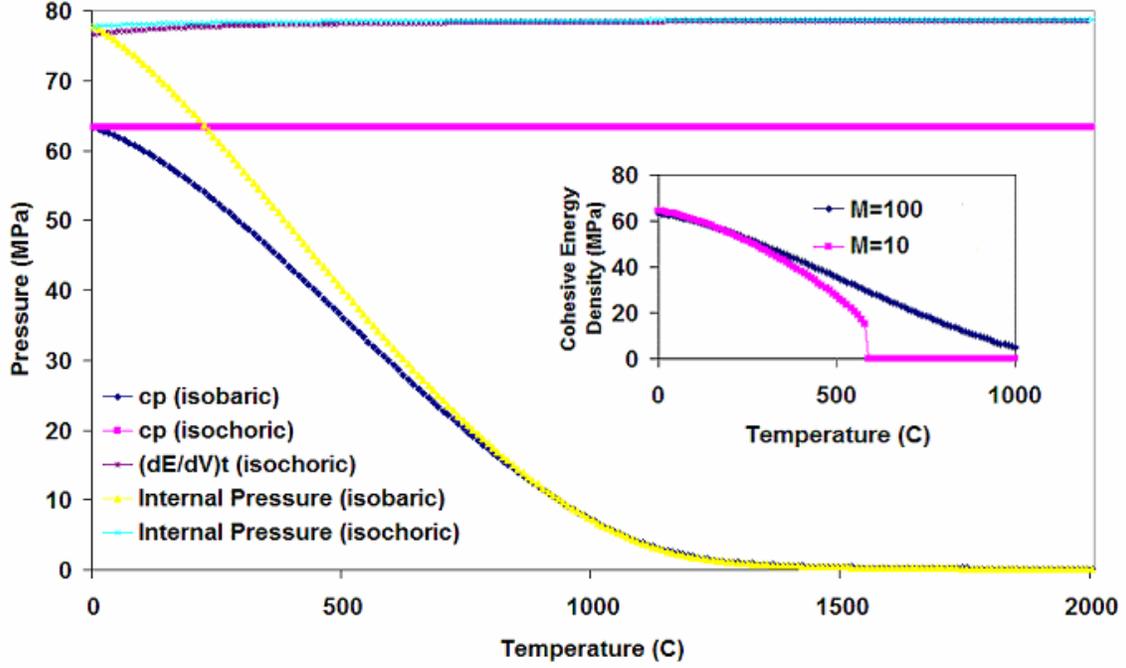}%
\caption{Various isochoric and isobaric pressures vs temperature for a pure
component $(M=100,q=10,v_{0}=1.6\times10^{-28}m^{\text{3}},$and $e=-2.6\times
10^{-21}~J)$. In the inset, we show cohesive pressure at $1.0$ atm for two
different $M$ ($M=100:$component 1 and $M=10:$component 2), with the smaller
$M$ showing a discontinuity at its boiling point around $600^{\circ}$C.}%
\label{F2}%
\end{center}
\end{figure}

In Fig. \ref{F2}, we show $P_{\text{IN}}$ (MPa) as a function of the
temperature $t_{\text{C}}$ ($^{\circ}$C) for a pure component ($M=100,$
$q=10,$ $v_{0}=1.6\times10^{-28}$ m$^{\text{3}},$ and $e=-2.6\times10^{-21}%
~$J) for a constant $V$ ($\phi_{0}=0.012246$)$,$ and constant $P$ ($1.0$ atm).
At 0$^{\circ}$C, the system has $\phi_{0}=0.012246,$ and $P=1.0$ atm in both
cases so that isobaric and isochoric $P_{\text{IN}}$ match there, as seen in
Fig. \ref{F2}. We notice that isochoric $P_{\text{IN}}$ has a very week
dependence on $t_{\text{C}}$ (suggesting that the temperature range in Fig.
\ref{F2} may be near the maximum of $P_{\text{IN}}),$ while isobaric
calculation provides a strongly dependent $P_{\text{IN}},$ which
asymptotically goes to zero. This difference in the isobaric-isochoric
behavior is consistent with our claim above. The difference between the two
internal pressures finally reaches a constant equal to about $80$ MPa at very
high temperatures$.$ We also show the derivative $(\partial\mathcal{E}%
/\partial V)_{T}$ calculated for the isochoric case for comparison. We use
isochoric $P_{\text{IN}}$ and (\ref{EV_Derivative}) to obtain isochoric
$(\partial\mathcal{E}/\partial V)_{T}.$ We notice that it differs from
isochoric $P_{\text{IN}}$ by a small amount over the entire temperature range
considered. Near 0$^{\circ}$C, they differ by about 1 MPa, and this difference
decreases as the temperature rises so that they approach each other at higher
temperatures. This is consistent with what we learned from Fig. \ref{F1}
above. We also notice that $P_{\text{IN}}>(\partial\mathcal{E}/\partial
V)_{T}$ over the temperature range in Fig. \ref{F2}, and the difference
gradually vanishes. From (\ref{EV_Derivative}), we conclude that $(\partial
P_{\text{IN}}/\partial T)>0$ for this to be true. This corresponds to
$(\partial P_{\text{int}}/\partial T)<0$ in Fig. \ref{F2}. Thus, the
temperature range is below the minimum of$\ P_{\text{int}}$; refer to Fig.
\ref{F1}.%

\section{Pure Component Cohesive Energy Density or Pressure}

\subsection{Athermal Reference State}

The following discussion in this section is general and not restricted to a
lattice except for the numerical results, which are on a lattice. Therefore,
we will include the kinetic energy of the system in this discussion. The
definition of $\delta$ requires considering a pure component consisting of one
particular component. To be specific, we consider this to be the component
$i=1,$ and we will not exhibit the index unless needed for clarity. At a given
temperature $T$, and pressure $P$ or volume $V$, one calculates the
interaction energy $E_{\text{int}}$ per unit volume by subtracting the kinetic
energy $K$\ due to motion from the total internal energy $\mathcal{E}$ of the
system. We will assume in the following that the interaction energy depends
only on the coordinates and not on the momenta of the particles. In classical
statistical mechanics, $K$ is a function only of the temperature $T$ but not
of the volume \cite{Fedor}, and can be obtained by considering a
\emph{fictitious} pure component at the same $T$, and $P$ (or $V$)$,$
containing the same number of particles, but with particles having \emph{no}
interaction ($U=0$)$.$ This fictitious reference system is, as said above,
known as the \emph{athermal }reference\emph{ }state. Then, $K=\mathcal{E}%
_{\text{ath}},$ where $\mathcal{E}_{\text{ath}}$ is the energy of the athermal
system. This reference state is equivalent to the conventional view according
to which one considers the gaseous state of the system at the same $T$, but at
zero pressure, so that $V\rightarrow\infty,$ which allows for infinite
separation between particles to ensure $\mathcal{E}_{\text{int}}=0.$ The
energy of the gaseous state is exactly $\mathcal{E}_{\text{ath}},$ which is
independent of the volume, even though the physical state requires
$V\rightarrow\infty$\ for the absence of interactions. We will continue to use
the athermal system instead of the gas phase since the latter requires
considering different pressure or volume than the pressure or volume of the
physical system under consideration. Thus, we find from
(\ref{InteractionEnergy})
\[
E_{\text{int}}\equiv E-(V_{\text{ath}}/V)E_{\text{ath}},
\]
where $E\equiv\mathcal{E}/V$ is the energy density of the pure component (with
interactions), and $E_{\text{ath}}\equiv\mathcal{E}_{\text{ath}}%
/V_{\text{ath}}$ is the energy density of the athermal reference state
($e\equiv e_{11}=0$), and $V$ and $V_{\text{ath}}$ are the corresponding
volumes. As said above, $E_{\text{int}}$ has different limits at infinite
temperatures depending on whether we consider an isochoric or isobaric process.

\subsection{Cohesive Energy Density}

The cohesive density $c^{\text{P}}$ is the interaction energy of the particles
in the pure component. We set $e\equiv e_{11},$ which must be negative for
cohesion. The cohesive energy density $c^{\text{P}}$ for the pure component
can be thought of as functions of $T$, and $P,$ or $T$, and $V,$ as the case
may be$.$ It is obvious that $E_{\text{int}}$ vanishes as the interaction
energy $e$ vanishes. This will also be a requirement for the cohesive energy
density: $c^{\text{P}}$ should vanish with the interaction energy. We have
calculated $c^{\text{P}}$ as a function of $T$ for isochoric (constant volume)
and for isobaric (constant pressure) processes. We denote the two quantities
by $c_{V}^{\text{P}}$\ and $c_{P}^{\text{P}},$ respectively, and show them in
Fig. \ref{F2}, where they can also be compared with $P_{\text{IN}}$ for
isochoric and isobaric processes, respectively. The conditions for the two
processes are such that they correspond to identical states at $0^{\circ}$C.
We find that while the two quantities $c^{\text{P}}$\ and $P_{\text{IN}}$ are
very different for the two processes, they are similar for each process. We
again note that $c_{V}^{\text{P}}$\ and $c_{P}^{\text{P}}$ behave very
differently, as discussed above, with $c_{V}^{\text{P}}$\ $\geq$
$c_{P}^{\text{P}}.$ Not surprisingly, the same inequality also holds for
$P_{\text{IN}}.$

The almost constancy of isochoric $c^{\text{P}}$\ and $P_{\text{IN}}$ provides
a strong argument in support of their usefulness as a suitable candidate for
the cohesive pressure. This should not be taken to imply that $c^{\text{P}}%
$\ and $P_{\text{IN}}$ remain almost constant in every process, as the
isobaric results above clearly establish. Unfortunately, most of the
experiments are done under isobaric conditions; hence the use of isochoric
cohesive quantities may not be useful, and even misleading and care has to be exercised.

In the lattice model with only nearest-neighbor interactions, $E_{\text{int}%
}=e\phi_{\text{c}}$ where $\phi_{\text{c}}\equiv\phi_{11}$ is the contact
density between monomers (of component $i=1);$ hence
\begin{equation}
\delta^{2}\equiv c^{\text{P}}\equiv-e\phi_{\text{c}}/v_{0}.
\label{cohesive_defL}%
\end{equation}

\paragraph{RMA\ Limit}

In the RMA limit, we find from (\ref{RMAdensities1}) that%
\begin{equation}
c^{\text{P}}\ ^{\underrightarrow{\text{RMA}}}\ -qe\phi_{\text{m}}^{2}/2v_{0},
\label{RMA_Cohesive}%
\end{equation}
where $\phi_{\text{m}}$\ represents the pure component monomer density in this
section. We thus find that the ratio
\[
c^{\text{P}}/qe\phi_{\text{m}}^{2}\overset{\text{RMA}}{=}\text{const}%
\]
in this limit. Since in this limit, the product $qe$ remains a constant, the
ratio
\[
\widetilde{c}\equiv c^{\text{P}}/\phi_{\text{m}}^{2}\overset{\text{RMA}}%
{=}\text{const}%
\]
in this limit. Both these properties will not hold true in general.

\paragraph{End Group Effects}

In the following, we will study the combinations
\begin{equation}
\widehat{c}\equiv2c^{\text{P}}v_{0}/(q-2\nu)e\phi_{\text{m}}^{2},c^{\prime
}\equiv c^{\text{P}}/(q-2\nu) \label{mod_cp}%
\end{equation}
which incorporate the end group effects along with the linear connectivity via
the correction $2\nu,$ where
\[
\nu\equiv(M-1)/M.
\]
In the RMA limit,%
\[
\widehat{c}\ ^{\underrightarrow{\text{RMA}}}\ 1,
\]
and%
\[
c^{\prime}/e\ ^{\underrightarrow{\text{RMA}}}\ \phi_{\text{m}}^{2}/2v_{0}.
\]
We will investigate how close $\widehat{c}$ is to $1$ in our calculation for
finite $q,$ and how close to a quadratic form in terms of $\phi_{\text{m}}$
does $c^{\prime}$ have.

\subsection{van der Waals Fluid}

To focus our attention, let us consider a fluid which is described by the van
der Waals equation. It is known that for this fluid, $E_{\text{int}%
}=-N_{\text{P}}^{2}a/V^{2},$ where $a>0$ is determined by the integral
\cite{Landau}
\begin{equation}
a\equiv-(1/2)\int_{2r_{0}}^{\infty}udV; \label{vdW_a}%
\end{equation}
here $u$ is a two-body potential function. It is clear that $a$ is
\emph{independent} of $T$ for the van der Waals fluid. We can also treat it to
be independent of $V\ ($any $V$-dependence must be very weak and can be
neglected). The lower limit of the integral is the zero of the two-body
potential energy. Thus,%
\begin{equation}
c^{\text{P}}=an_{\text{P}}^{2}, \label{vdWc}%
\end{equation}
where $n_{\text{P}}$ is the number density per unit volume; for polymers,
$n_{\text{P}}\equiv\phi_{\text{m}}/Mv_{0}$. It is interesting to compare
(\ref{vdWc}) with (\ref{RMA_Cohesive}) derived in the RMA limit. They both
show the \emph{same} quadratic dependence on the number density. We also note
that $c^{\text{P}}$ vanishes as $a$ vanishes, as we expect, but most
importantly the ratio $c^{\text{P}}/n_{\text{P}}^{2}$ is independent of the
temperature, just as $\widetilde{c}$ is a constant in the RMA limit$.$
Deviation from the quadratic dependence will be observed in most realistic
systems, since they cannot be described by this approximation.

An interesting observation about this fluid is that $P_{\text{int}}$ is
exactly equal to $E_{\text{int}};$ thus, $c^{\text{P}}=-P_{\text{int}}$ for
this fluid. In addition, it is well known that for this fluid
\begin{equation}
(\partial\mathcal{E}/\partial V)_{T}=(\partial\mathcal{E}_{\text{int}%
}/\partial V)_{T}=an_{\text{P}}^{2}, \label{vdWdEdV}%
\end{equation}
since $a$ is temperature-independent. Thus,
\begin{equation}
P_{\text{IN}}\equiv(\partial\mathcal{E}/\partial V)_{T} \label{PINdEdV}%
\end{equation}
for the the van der Waals fluid, in which $a$ must be taken as $T$%
-independent. The equality (\ref{PINdEdV}) is not always valid as discussed above.

\subsection{The Usual Approximation}

Usually, one approximates $E_{\text{int}}$ by the energy density of
vaporization at the boiling point at $T=T_{\text{B}}.$ This means that one
approximates $c^{\text{P}}(T,P)$ by its value $c^{\text{P}}(T_{\text{B}},P),$
which will be a function of $P$, but not of $T.$ On the contrary,
$c^{\text{P}}(T,P)$ will show variation with respect to both variables. In
addition, it will also change with the lattice coordination number $q,$ and
the interaction energy $e$. It is clear that isobaric $c^{\text{P}}(T,P)$ will
show the discontinuity at the boiling point, as shown in the inset in Fig.
\ref{F2}, where we report isobaric $c^{\text{P}}$ for two different molecular
weights $M=10,$ and $M=100$ at $1.0$ atm$.$ The shorter polymer system boils
at about $600^{\circ}$C (and $1.0$ atm$)$, but not the longer polymer over the
temperature range shown there. At $0^{\circ}$C,\ the smaller polymer has a
slightly higher value of isobaric $c^{\text{P}}.$ If we calculate isochoric
$c^{\text{P}}$ at a volume equal to that in the isobaric case at $0^{\circ}$C
and $1.0$ atm$,$ then the corresponding isochoric $c^{\text{P}},$ which is
almost a constant (as shown in the main figure in Fig. \ref{F2}), will be
about $65$ MPa, close to its value at $0^{\circ}$C. This value is much larger
than $c^{\text{P}}(T_{\text{B}},P)$ of about $15$ MPa at the boiling point.
Thus, the conventional approximation cannot be taken as a very good estimate
of the cohesiveness of the system, which clearly depends on the state of the
system$.$Cohesive energy density as a function of pressure for different $q$.
$\qquad\qquad\qquad\qquad\qquad\qquad\qquad$%
\begin{figure}
[ptb]
\begin{center}
\includegraphics[
trim=0.935322in 6.018832in 0.937766in 0.937516in,
height=3.7135in,
width=6.3027in
]%
{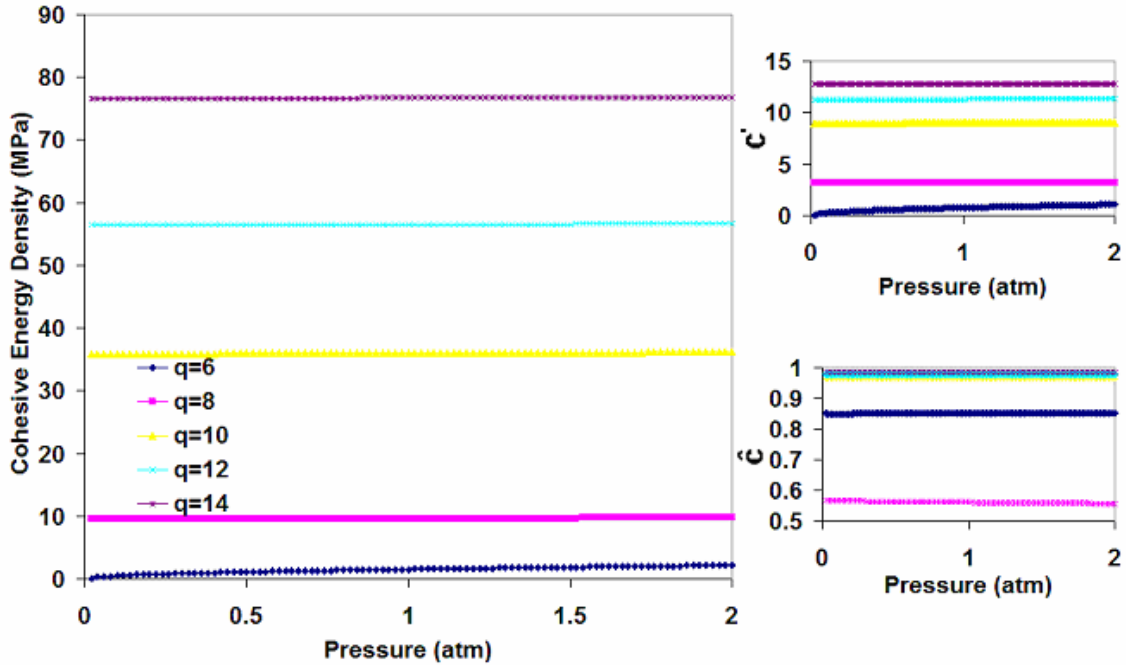}%
\caption{Cohesive energy density as a function of pressure for different $q$.
We also show $c^{\prime}$ and $\widehat{c}.$}%
\label{F3}%
\end{center}
\end{figure}
%

\begin{figure}
[ptb]
\begin{center}
\includegraphics[
trim=1.070568in 5.620840in 0.803333in 0.804498in,
height=4.2436in,
width=6.3019in
]%
{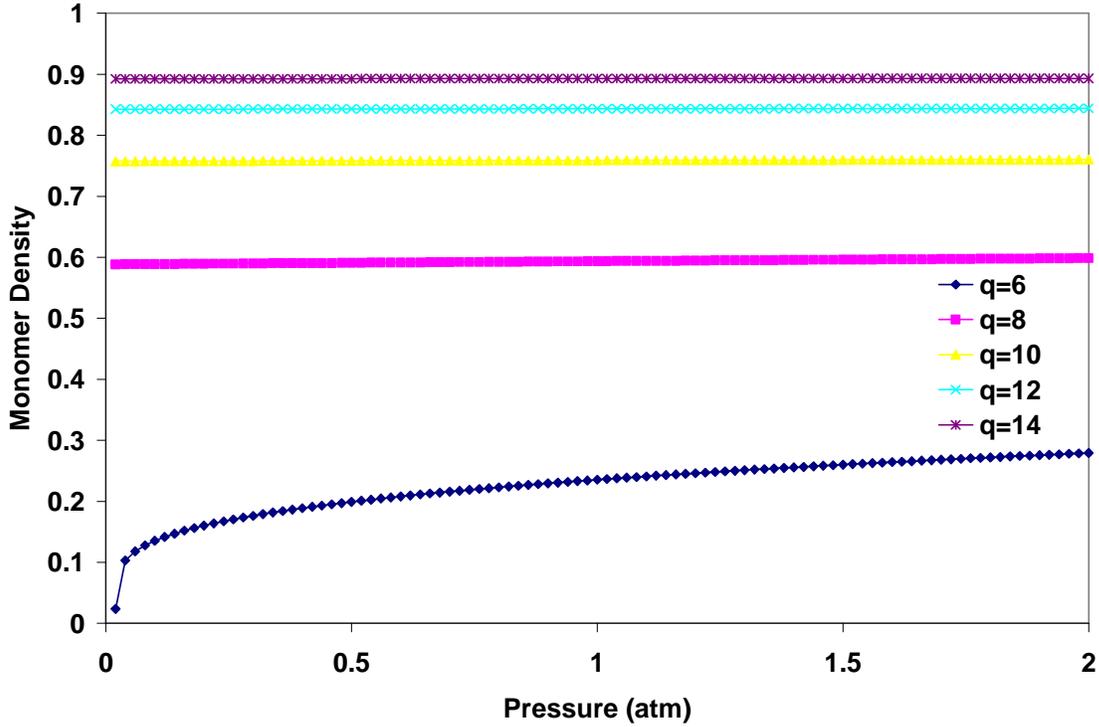}%
\caption{Monomer density as a function of presssure for different $q$. For
$q\geq8,$ we have a liquid which gets denser as $q$ increases. For $q=6,$ we
have a gas phase at low pressure which becomes a lighter liquid at higher
pressure, with a liquid gas transition around 0.05 atm. }%
\label{F4}%
\end{center}
\end{figure}
\begin{figure}
[ptbptb]
\begin{center}
\includegraphics[
trim=0.902732in 5.818773in 0.804963in 0.706596in,
height=4.1451in,
width=6.4679in
]%
{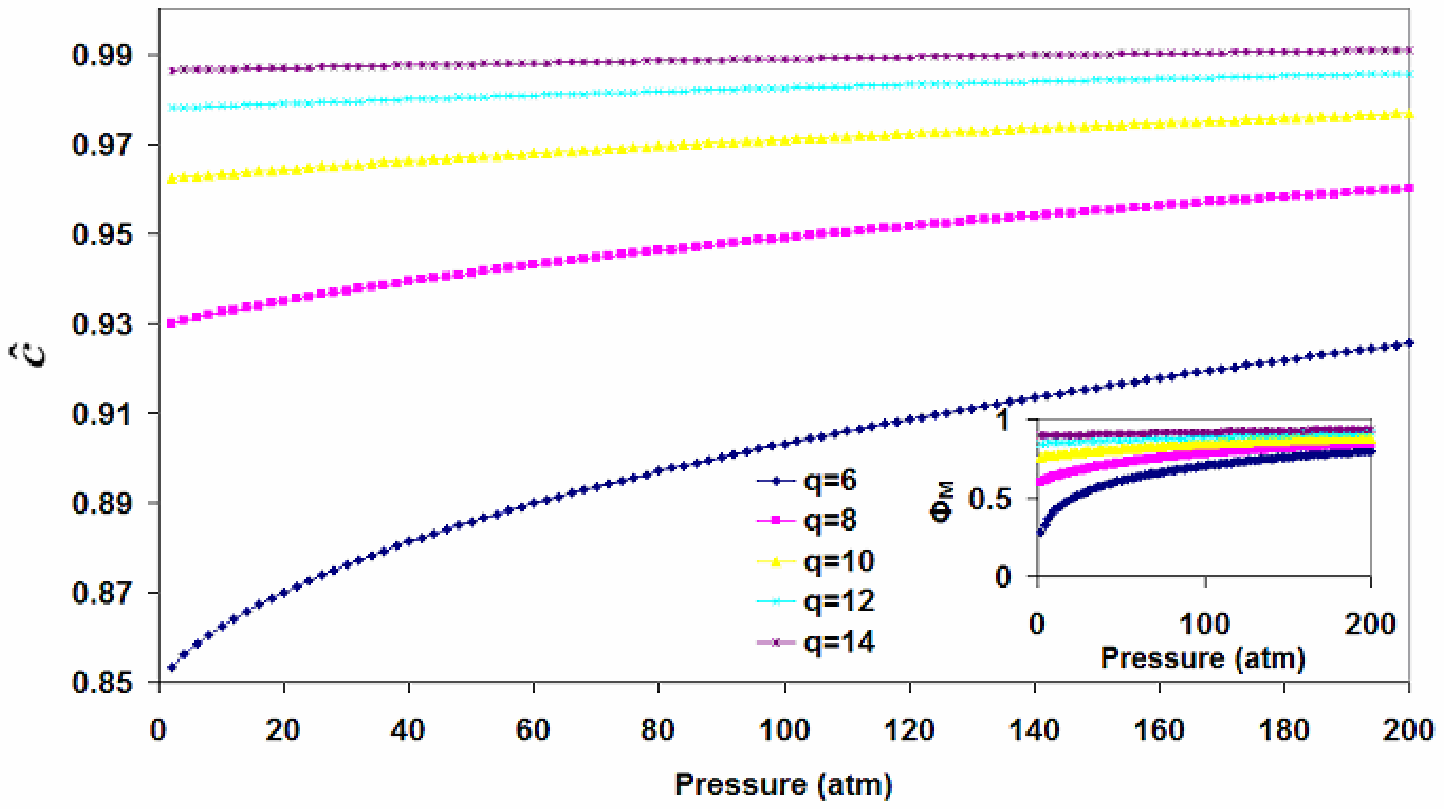}%
\caption{$\widehat{c}$ as a function of pressure at $500^{\circ}$C for
different $q$. As $q$ increases, $\widehat{c}$ approaches $1,$ its limiting
RMA value, which also forms its upper limit. For any finite $q$, $\widehat{c}$
is strictly lower. }%
\label{F5}%
\end{center}
\end{figure}

\subsection{Numerical results}

We will evaluate various quantities for isochoric and isobaric processes$.$
For isochoric calculations, we calculate $\phi_{\text{m}}$ at some reference
$T,P.$ Usually, we take it to be at $1^{\circ}$C, and $1.0$ atm. We then keep
$\phi_{\text{m}}$ fixed as we change the temperature. This is equivalent to
keeping the volume fixed for a given amount of the polymers. For isobaric
(constant $P$) calculations, we again start at $1^{\circ}$C, and take a
certain pressure such as $1.0$ atm, and keep the pressure fixed by adjusting
$\phi_{\text{m}}$\ as we change the temperature. For a given amount of
polymers, this amounts to adjusting the volume of the system.

We show $c^{\text{P}}$ in Fig. \ref{F3}, and the monomer density
$\phi_{\text{m}}$ in Fig. \ref{F4}, as a function of $P$ ($P\leq2$ atm) for
different values of $q$ from $q=6$ to $q=14.$ We have taken $M=500,$
$v_{0}=1.6\times10^{-28}$ m$^{\text{3}},$ and $e=-2.6\times10^{-21}~$J$,$ and
have set $t_{\text{C}}=500^{\circ}$C. We see that $c^{\text{P}}$ increases
with $q$, which is expected; see (\ref{RMA_Cohesive}). To further analyze this
dependence, we plot $c^{\prime}$ in Fig. 3. There is still a residual increase
with $q.$ The increase is also partially due to the fact that $\phi_{\text{m}%
}$ increases with $q,$ as shown in Fig. \ref{F4}. Therefore, we also plot
$\widehat{c}$ as a function of $P$ in Fig. \ref{F3}$.$ We notice that there is
still a residual dependence on $q$ with $\widehat{c}$ increasing with $q,$ and
reaching $1.0$ from below as $q$ increases, which is consistent the RMA limit.
In Fig. \ref{F5}, we plot $\widehat{c}$ and $\phi_{\text{m}}$\ (in the inset)
for the same system over a much wider range of pressure for different $q.$
From the behavior of $\ \widehat{c}$ and $\phi_{\text{m}},$\ we easily
conclude that $c^{\text{P}}$ changes strongly with $P$ for smaller $q,$ and
the dependence gets weaker as $q$ increases.

\begin{figure}
[ptb]
\begin{center}
\includegraphics[
trim=0.902732in 5.820900in 0.903547in 0.903464in,
height=3.9444in,
width=6.3693in
]%
{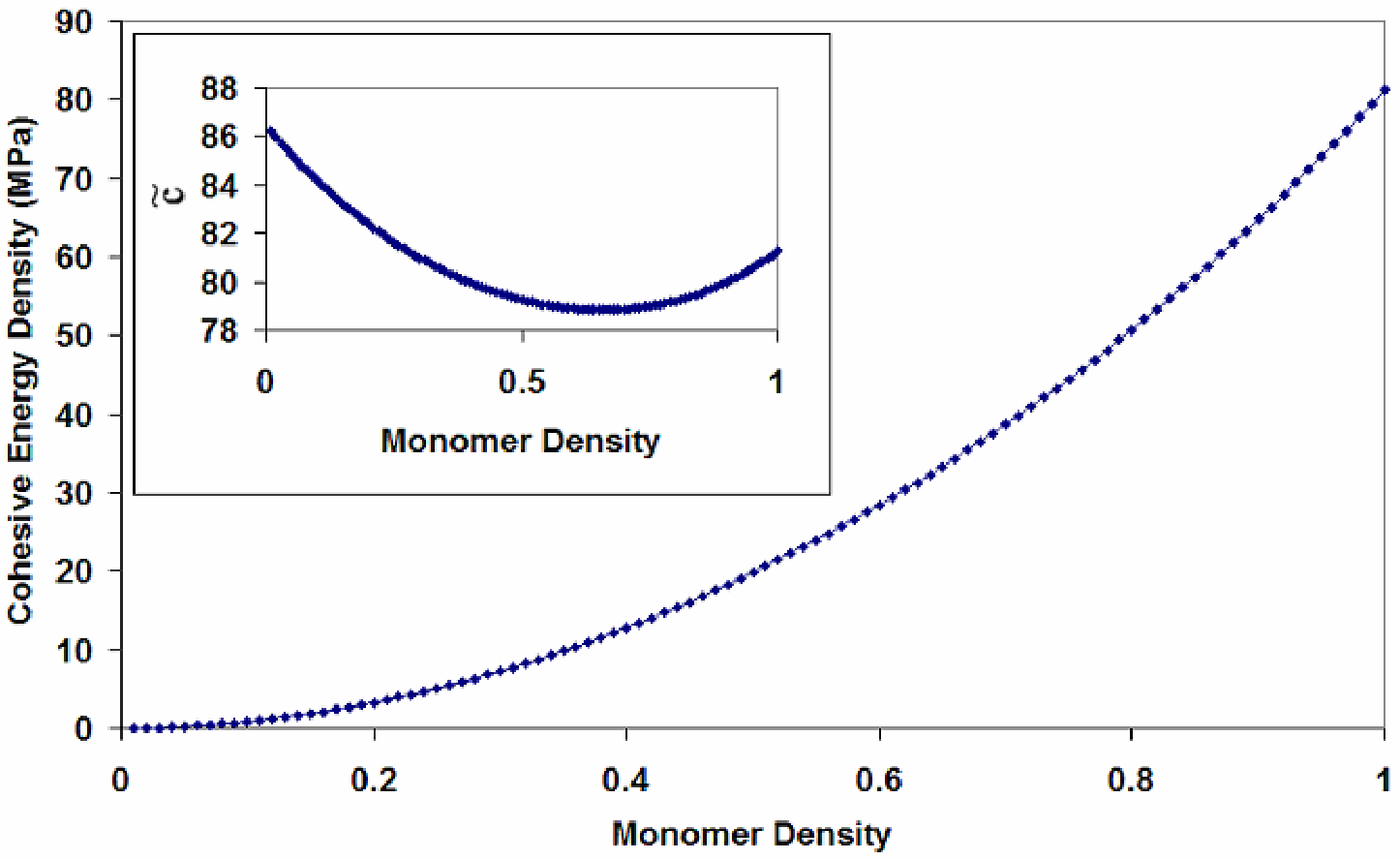}%
\caption{$c^{\text{P}}$ as a function of $\phi_{\text{m}}.$ In the inset, we
show $\widetilde{c},$ showing that $c^{\text{P}}$ is not quadratic in
$\phi_{\text{m}},$ as suggested by the regular solution theory.}%
\label{F6}%
\end{center}
\end{figure}

In Fig. \ref{F6}$,$ we plot $c^{\text{P}}$ for $q=12$ as a function of the
monomer density $\phi_{\text{m}}.$ We have taken $M=500,$ $v_{0}%
=1.6\times10^{-28}$ m$^{\text{3}},$ and $e=-2.6\times10^{-21}~$J$,$ and have
set $t_{\text{C}}=500^{\circ}$C. According to (\ref{RMA_Cohesive}), it should
be a quadratic function of the monomer density. To see if this is true for the
present case of finite $q,$ we plot the ratio $c^{\text{P}}/\phi_{\text{m}%
}^{2}$\ in the inset, which clearly shows that there is still a strong
residual dependence left in the ratio. Thus, $c^{\text{P}}$ in a realistic
system should not be quadratic in $\phi_{\text{m}}.$%

%

\begin{figure}
[ptb]
\begin{center}
\includegraphics[
trim=0.901917in 5.921995in 0.904361in 0.904527in,
height=3.8432in,
width=6.3693in
]%
{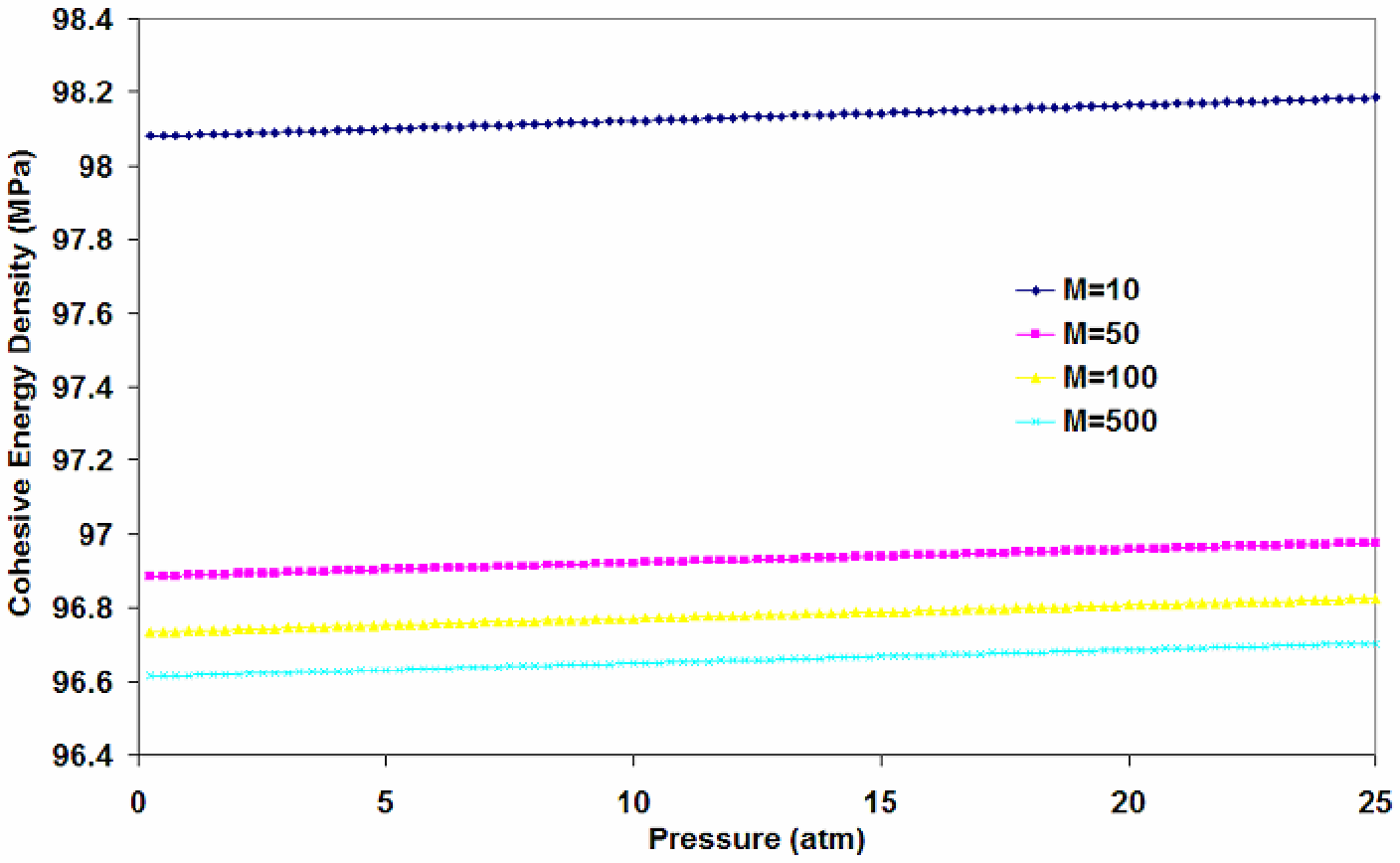}%
\caption{Cohesive energy density as a function of pressure for different
molecular weights. }%
\label{F7}%
\end{center}
\end{figure}

\begin{figure}
[ptb]
\begin{center}
\includegraphics[
trim=0.902732in 5.820900in 0.903547in 0.903464in,
height=3.9444in,
width=6.3693in
]%
{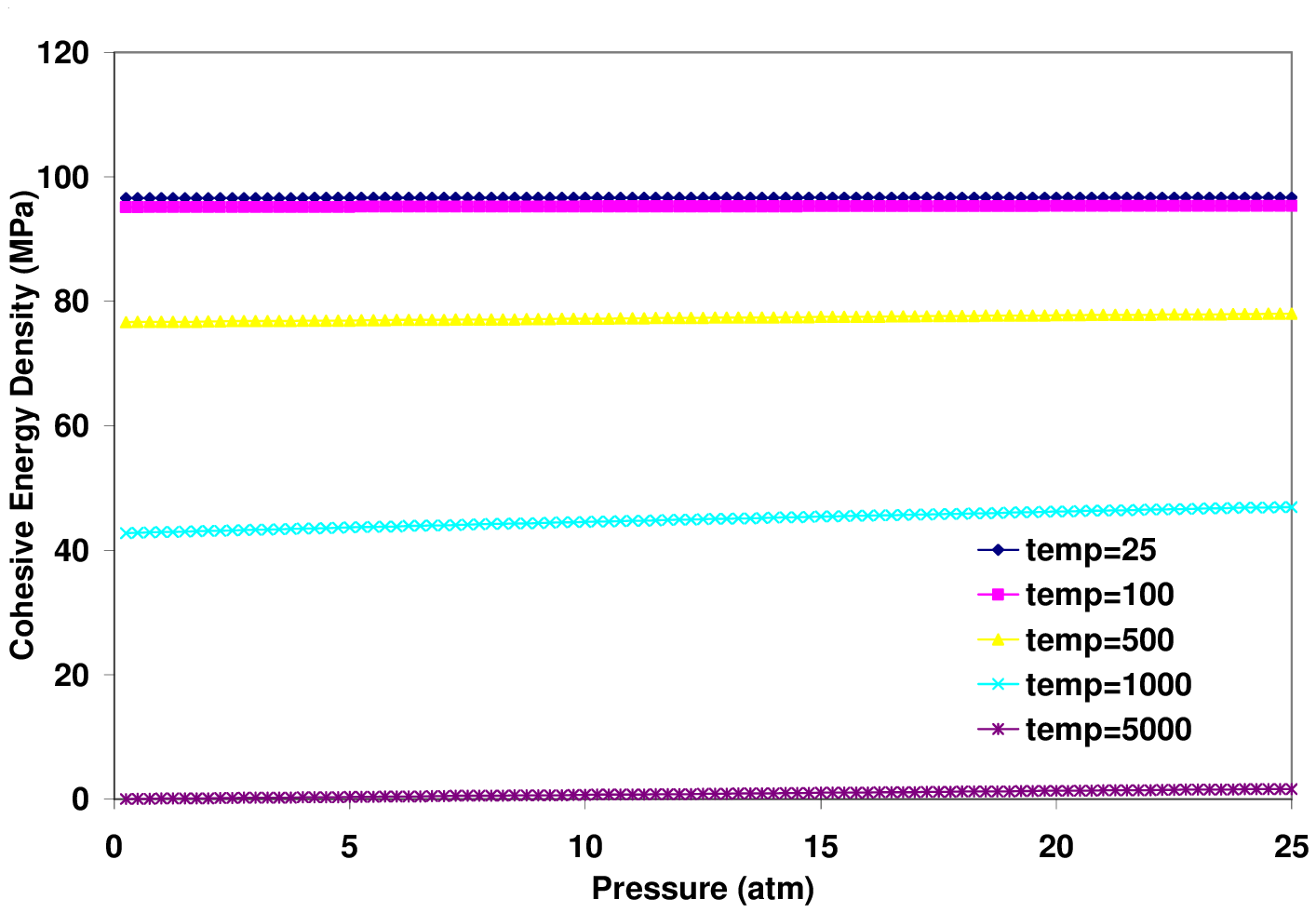}%
\caption{Cohesive energy density as a function of pressure at different
temperatures. }%
\label{F8}%
\end{center}
\end{figure}

In Fig. \ref{F7}, we show $c^{\text{P}}$ as a function $P$ for small pressures
for different $M$ ranging from $M=10$ to $M=500;$ we keep $q=14,$
$v_{0}=1.6\times10^{-28}$ m$^{\text{3}},$ and $e=-2.6\times10^{-21}~$J$,$ and
have set the temperature $t_{\text{C}}=25^{\circ}$C. We see that $c^{\text{P}%
}$ decreases as $M$ increases, but different curves are almost parallel,
indicating that it is not the slope, but the magnitude that is affected by
$M.$ In Fig. \ref{F8}, we show $c^{\text{P}}$ as a function $P$ for small
pressures for different temperatures ranging from $t_{\text{C}}=25^{\circ}$C
to $t_{\text{C}}=5000^{\circ}$C$;$ we keep $q=14,$ $M=100,~v_{0}%
=1.6\times10^{-28}$ m$^{\text{3}},$ and $e=-2.6\times10^{-21}~$J. We
immediately note that the pressure variation is quite minimal over such a
small range of pressure between 1 to 25 atm. However, the temperature
variation is quite pronounced, again affecting the magnitude and not the
slope. There are two important observations. (i) For the highest temperature
$5000^{\circ}$C, the state corresponds to a gas, since $c^{\text{P}}\simeq0.$
(ii) At low temperatures, $c^{\text{P}}$ reaches an asymptotic value since the
system has become almost incompressible.%

%

\begin{figure}
[ptb]
\begin{center}
\includegraphics[
trim=0.902732in 5.820900in 0.903547in 0.903464in,
height=3.9444in,
width=6.3693in
]%
{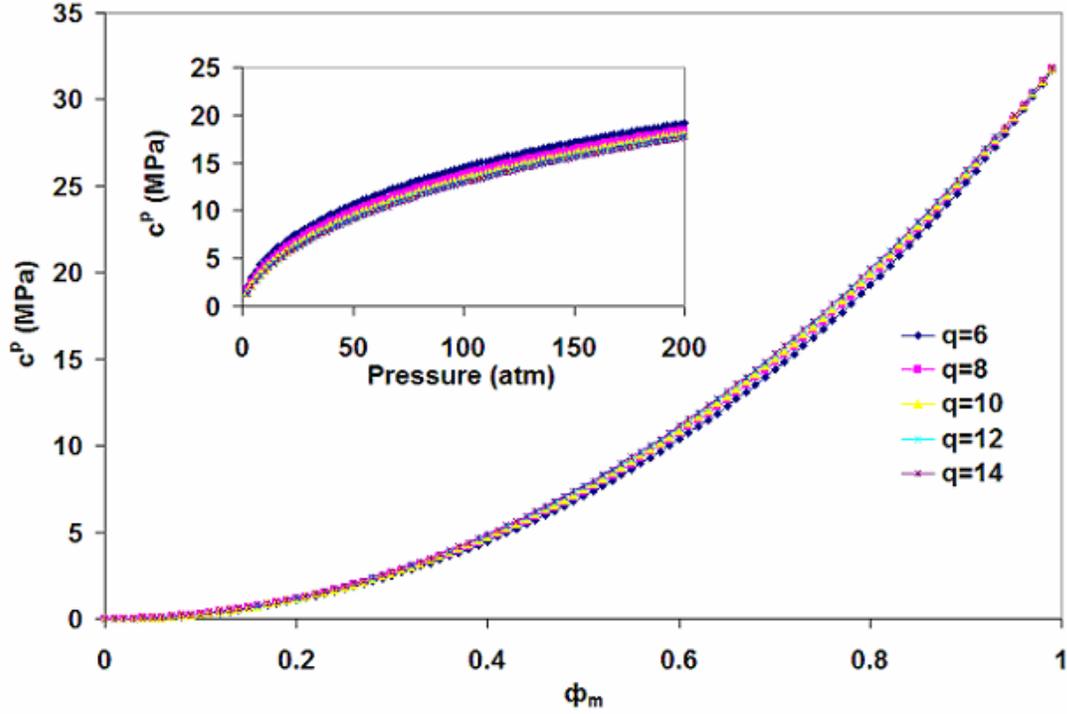}%
\caption{$c^{\text{P}}$ as a function of $\phi_{\text{m}}$ for different $q$
when $e(q/2-\nu)$ is kept fixed at ($-5.21\times10^{-21}$J).}%
\label{F9}%
\end{center}
\end{figure}
%

\begin{figure}
[ptb]
\begin{center}
\includegraphics[
trim=0.902732in 5.820900in 0.903547in 0.903464in,
height=3.9444in,
width=6.3693in
]%
{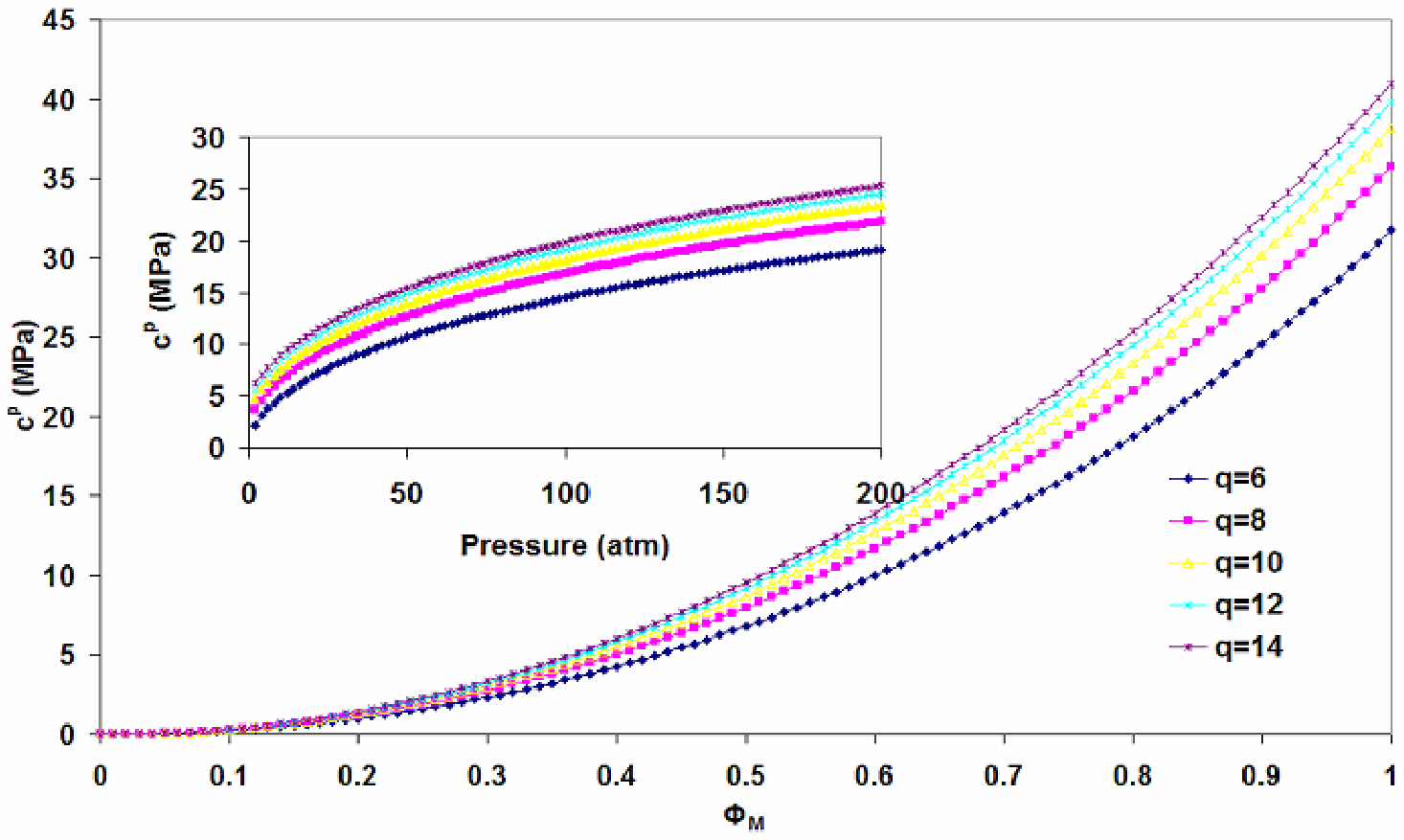}%
\caption{$c^{\text{P}}$ as a function of $\phi_{\text{m}}$ for different $q$
when $eq$ is kept fixed at ($-1.56\times10^{-20}$J).}%
\label{F10}%
\end{center}
\end{figure}

In Fig. \ref{F9}, which is for a pure component studied in Fig. \ref{F3}, we
keep the product $e(q/2-\nu)$ fixed at ($-5.21\times10^{-21}$J)$,$ as we
change $q$, so that $e$ changes with $q.$ We keep $M=500,$ $v_{0}%
=1.6\times10^{-28}$ m$^{\text{3}},$ and $t_{\text{C}}=500^{\circ}$C. The
effect of changing $q$ is now minimal. The cohesive energy density shows
minimal variation with $q$, except that it is higher for larger $q$, but the
difference first increases with $\phi_{\text{m}}$ and then decreases as
$\phi_{\text{m}}\rightarrow1.$\ In Fig. \ref{F10}, we keep instead the product
$eq$ fixed, so that there is no endpoint correction. In contrast with Fig.
\ref{F9}, we find that $c^{\text{P}},$ while still increasing with $q$, has
the property that the its difference for different $q$ continues to increases
with $\phi_{\text{m}}$ as $\phi_{\text{m}}\rightarrow1.$ It is clear from both
figures that $c^{\text{P}}$ increases with $q$ at a given $\phi_{\text{m}}$
and saturates as $q$ becomes large, which is the direction in which $q$ must
increase to obtain the RMA limit. Thus, $c^{\text{P}}$ achieves its maximum
value at a given density $\phi_{\text{m}}$ in the RMA approximation:\ for any
realistic system in which $q$ is some finite value, the value of the cohesive
energy at that density $\phi_{\text{m}}$\ will be strictly smaller. It is also
evident from the inset in both figures that at a given pressure $P$,
$c^{\text{P}}$ achieves its minimum value in the RMA approximation.

\begin{figure}
[ptb]
\begin{center}
\includegraphics[
trim=0.902732in 5.820900in 0.903547in 0.903464in,
height=3.9444in,
width=6.3693in
]%
{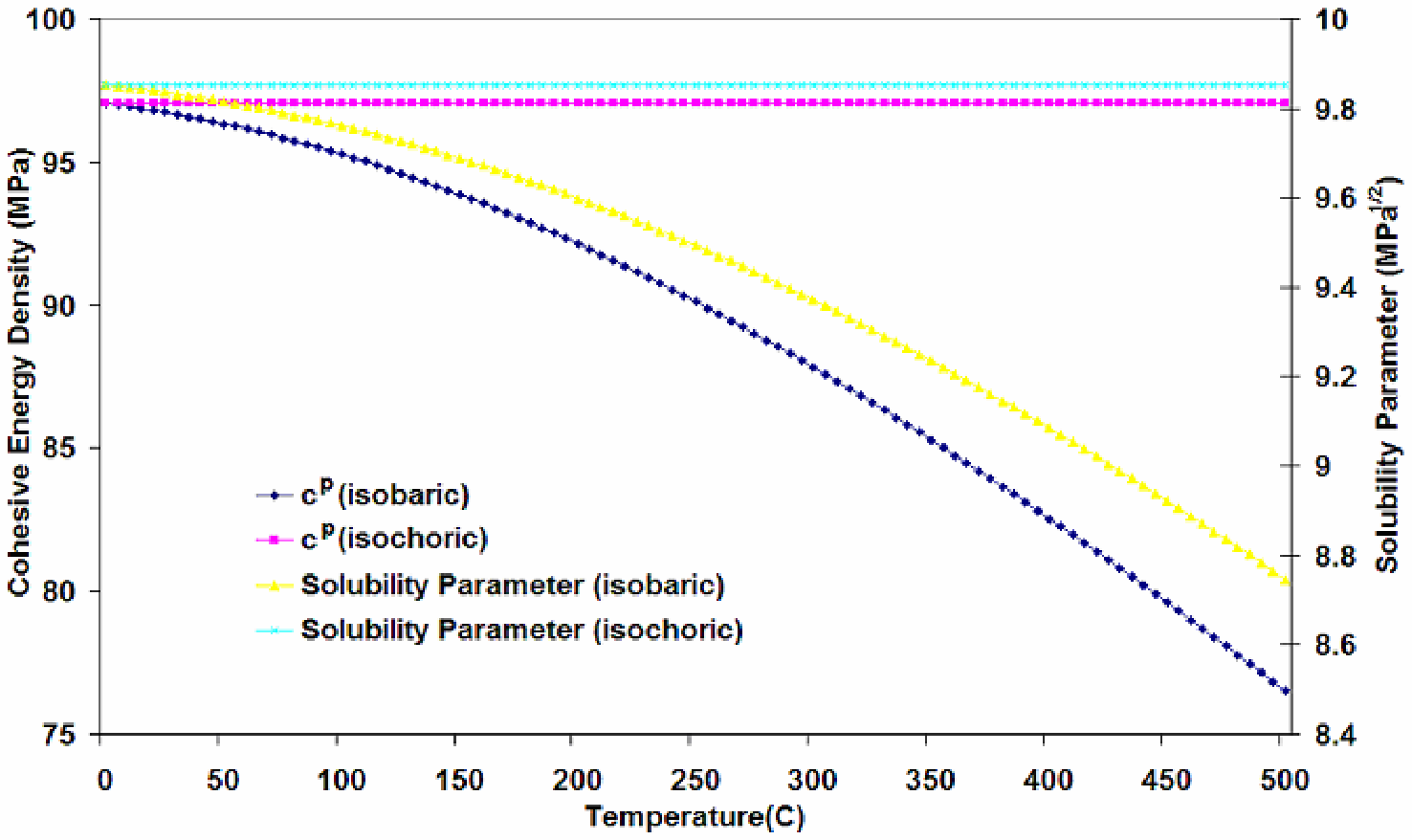}%
\caption{Various $c^{\text{P}}$ as a function of temperature. }%
\label{F11}%
\end{center}
\end{figure}

The results for isochoric and isobaric calculations are shown in Fig.
\ref{F11}. We consider $q=14,$ $M=100,~v_{0}=1.6\times10^{-28}$ m$^{\text{3}%
},$ and $e=-2.6\times10^{-21}~$J$,$ and have taken the pressure to be $P=1.0$
atm\ at the initial temperature $t_{\text{C}}=0^{\circ}$C. This corresponds to
the monomer density $\phi_{\text{m}}=0.997115.$ For the isochoric calculation,
we maintain the same monomer density ($\phi_{\text{m}}=0.997115$) at all
temperatures. For the isobaric calculation, we maintain the same pressure
($P=1$ atm) at all temperatures. This ensures that isochoric $c_{V}^{\text{P}%
}$ \ and isobaric $c_{P}^{\text{P}}$ are identical at $t_{\text{C}}=0^{\circ}%
$C. The figure clearly shows that $c_{V}^{\text{P}}$ is always higher than
$c_{P}^{\text{P}}$, as discussed above$.$ This behavior is not hard to
understand. As we raise $T$, the pressure increases in the constant volume
calculation above $P=1$ atm, making particles get closer together, thereby
increasing the cohesive energy.%

\section{Mutual Cohesive Energy Density or Pressure}

\subsection{Mutual Interaction}

The solubility of one of the components in a binary mixture of components
$i=1,2$ increases as the excess energy $\varepsilon_{12}$ decreases for the
simple reason that a larger $\varepsilon_{12}$ corresponds to a stronger
excess repulsion between the two components. In particular, the two components
will not phase separate at any temperature if $\varepsilon_{12}\leq0$. If
$\varepsilon_{12}>0,$ the two components will phase separate at low
temperatures. Thus, knowing whether $\varepsilon_{12}>0$\ immediately allows
us to conclude that solubility will not occur everywhere.

Usually, all energies $e_{ij}$\ are negative, but the sign of $\varepsilon
_{12}$\ depends on their relative magnitudes. If, however, we assume the
London conjecture (\ref{london_Conj}), then the corresponding excess energy
becomes
\begin{equation}
\varepsilon_{12}=(\sqrt{\left\vert e_{11}\right\vert }-\sqrt{\left\vert
e_{22}\right\vert })^{2}/2\geq0. \label{Mutual_Energy}%
\end{equation}
Thus, the London conjecture implies that the two components experience a
repulsive excess interaction so that the solubility decreases as
$\varepsilon_{12}$ increases, i.e., as $e_{11}$ and $e_{22}$ become more
disparate. The maximum solubility in this case occurs when $\varepsilon
_{12}=0,$ i.e., when the two components have identical interactions ("like
dissolves like"). In this case, the size or architectural disparity cannot
diminish the solubility because the entropy of mixing will always promote
miscibility. In general, $\varepsilon_{12}>0,$ and it becomes necessary to
study solubility under different thermodynamic conditions.

To simplify our investigation, we will assume the London conjecture
(\ref{london_Conj}) for the mixture in all the calculations. Thus, the two
components always experience a repulsive excess interaction in our
calculation. (This is also true when we intentionally violate the London
conjecture (\ref{london_Conj}) and set $e_{12}=0,$ as is the case for SRS.) In
the RMA limit, the conjecture (\ref{london_Conj}) immediately leads to
(\ref{london_berthelot_Conj}), as we have seen above. This need not remain
true when we go beyond RMA. Thus, we will inquire if
(\ref{london_berthelot_Conj}) is satisfied in general for cohesive energies
that are calculated from our theory under the assumption that the London
conjecture (\ref{london_Conj}) is valid. Any failure of
(\ref{london_berthelot_Conj}) under this condition will clearly have
significant implications for our basic understanding of solubility.

\subsection{van Laar-Hildebrand Approach using Energy of Mixing}

We follow van Laar \cite{vanLaar} and Hildebrand \cite{Hildebrand}, and
introduce $c_{12}$ by exploiting the energy of mixing $\Delta E_{\text{M}}$
per unit volume. According to the isometric regular solution theory
\cite{Hildebrand}, the two are related by%
\begin{equation}
\Delta E_{\text{M}}=(c_{11}^{\text{P}}+c_{22}^{\text{P}}-2c_{12})\varphi
_{1}\varphi_{2}, \label{mixEnergy1}%
\end{equation}
where $\varphi_{i}$ are supposed to denote the volume fractions. Using the
London-Berthelot conjecture (\ref{london_berthelot_Conj}), we immediately
retrieve (\ref{mixEnergy0}), once we recognize that $\delta_{1}^{2}\equiv
c_{11}^{\text{P}},$ and $\delta_{2}^{2}\equiv c_{22}^{\text{P}},$ see
(\ref{cohesive_def}) above. We now take (\ref{mixEnergy1}) as the general
definition of the mutual cohesive energy density $c_{12}$ in terms of the pure
component cohesive energy densities$.$ The extension then allows us to
evaluate $c_{12}$ by calculating $\Delta E_{\text{M}},$ provided we know
$c_{11}^{\text{P}}$ and $c_{22}^{\text{P}}$ for the pure components; the
latter are independent of the composition. It can be argued that since
(\ref{mixEnergy1}) is valid only for isometric mixing, it should not be
considered a general definition of $c_{12}$ for non-isometric mixing. However,
since one of our objectives is to investigate the effects due to isometric and
non-isometric mixing, we will adopt (\ref{mixEnergy1}) as the general
definition of $c_{12}.$

\subsubsection{Lattice model}

The kinetic energy of the mixture is the sum of the kinetic energies of the
pure components, all having the same temperature, and will not affect the
energy of mixing. Thus, we need not consider the kinetic energy in our
consideration anymore. In other words, we can safely use a lattice model in
calculating $\Delta E_{\text{M}}.$ This is what we intend to do below.%
\begin{figure}
[ptb]
\begin{center}
\includegraphics[
trim=0.902732in 5.820900in 0.903547in 0.903464in,
height=3.9444in,
width=6.3693in
]%
{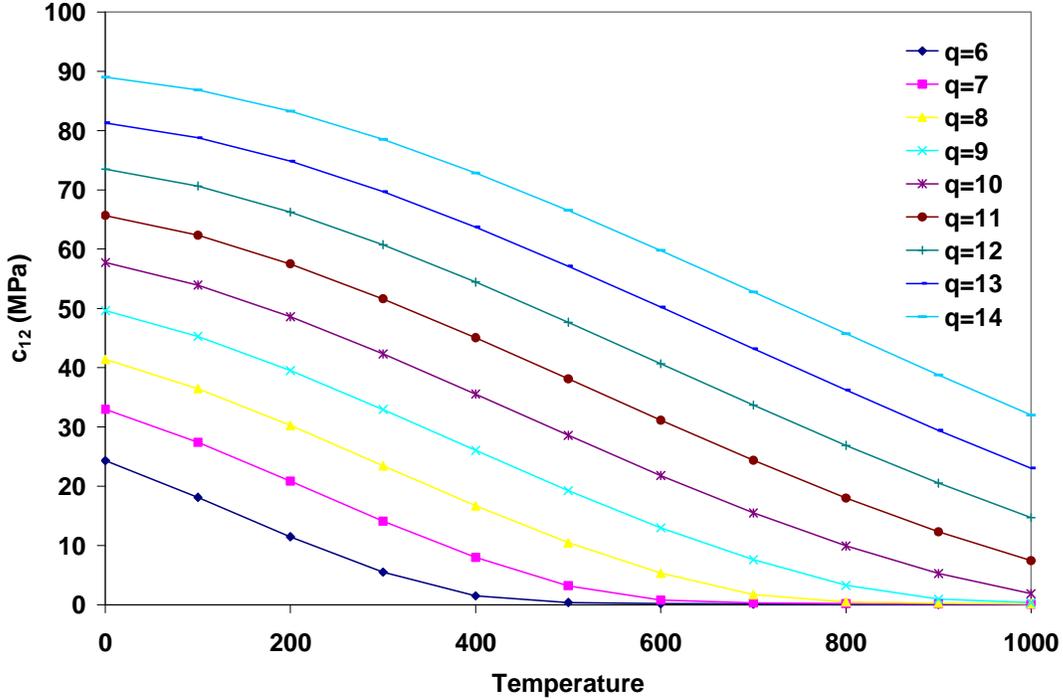}%
\caption{Mutual cohesive energy density as a function of temperature for
different $q.$ We have taken a blend at 50-50 composition ($M_{1}%
=M_{2}=100;e_{11}=-2.2\times10^{-21}$ J, $e_{22}=-2.6\times10^{-21}$ J,
$v_{0}=1.6\times10^{-28}$ m$^{3}$) at 1.0 atm. }%
\label{F12}%
\end{center}
\end{figure}

The definition of $c_{12}$ depends on a form (\ref{mixEnergy1}) whose validity
in general is questionable, as it is based on RST. To appreciate this more
clearly, let us find out the conditions under which the RMA limit of our
recursive theory will reproduce (\ref{mixEnergy1}). We first note that the
interaction energy density (per unit volume) $E_{\text{int}}$ of the mixture
from (\ref{energy_Def}) is given by
\begin{equation}
E_{\text{int}}v_{0}\equiv e_{11}\phi_{11}+e_{22}\phi_{22}+e_{12}\phi
_{12}=-c_{11}^{\text{P}}v_{0}\phi_{11}/\phi_{11}^{\text{P}}-c_{22}^{\text{P}%
}v_{0}\phi_{22}/\phi_{22}^{\text{P}}+e_{12}\phi_{12},
\label{energy_Def_Mixture}%
\end{equation}
where we introduced the pure component cohesive energy densities in the last
equation. The energy of mixing per unit volume is
\begin{equation}
\Delta E_{\text{M}}v_{0}=e_{11}\phi_{11}+e_{22}\phi_{22}+e_{12}\phi
_{12}-(V_{1}^{\text{P}}/V)e_{11}\phi_{11}^{\text{P}}-(V_{2}^{\text{P}%
}/V)e_{22}\phi_{22}^{\text{P}}, \label{mixEnergyL}%
\end{equation}
where $V_{1}^{\text{P}},$ and $V_{2}^{\text{P}}$ are the pure component
volumes. It is easy to see that in general%
\begin{equation}
V_{1}^{\text{P}}/V=x\phi_{\text{m}}/\phi_{\text{m}1}^{\text{P}}\equiv
\phi_{\text{m1}}/\phi_{\text{m}1}^{\text{P}},V_{2}^{\text{P}}/V=(1-x)\phi
_{\text{m}}/\phi_{\text{m}2}^{\text{P}}\equiv\phi_{\text{m2}}/\phi_{\text{m}%
2}^{\text{P}}, \label{volumeRatios}%
\end{equation}
where $x\equiv\phi_{\text{m1}}/\phi_{\text{m}}$ is the monomer fraction of
species $i=1$ introduced earlier in (\ref{monomer_fraction})$,$and
$\phi_{\text{m}i}^{\text{P}}$ the pure component monomer density of the $i$th
species. The monomer density of both species in the mixture is $\phi
_{\text{m}}\equiv\phi_{\text{m1}}+\phi_{\text{m2}}$.

\subsubsection{RMA Limit and Monomer Density Equality:}

In the RMA limit, it is easy to see by the use of (\ref{RMA2}) that the last
equation in (\ref{energy_Def_Mixture}) reduces to%
\[
E_{\text{int}}~\ ^{\underrightarrow{\text{RMA}}}~-c_{11}^{\text{P}}%
(V_{1}^{\text{P}}/V)^{2}-c_{22}^{\text{P}}(V_{2}^{\text{P}}/V)^{2}%
-2c_{12}(V_{1}^{\text{P}}/V)(V_{2}^{\text{P}}/V),
\]
where
\begin{equation}
c_{ii}^{\text{P}}~\ ^{\underrightarrow{\text{RMA}}}-(1/2)qe_{ii}\phi
_{\text{m}i}^{\text{P}2}/v_{0},\text{ }c_{12}\ \ ^{\underrightarrow
{\text{RMA}}}\ -qe_{12}\phi_{\text{m}1}^{\text{P}}\phi_{\text{m}2}^{\text{P}%
}/2v_{0} \label{RMA_CohesiveDensities}%
\end{equation}
are the cohesive energy densities in this limit. The form of $c_{ii}%
^{\text{P}}$ is exactly the same as the one derived above for the pure
component; see (\ref{RMAdensities1}), and (\ref{vdWc}). It is also a trivial
exercise to see that the RMA limit of the energy of mixing (\ref{mixEnergyL})
will exactly reproduce (\ref{mixEnergy1}) provided we identify%
\[
\varphi_{1}\equiv V_{1}^{\text{P}}/V,\varphi_{2}\equiv V_{2}^{\text{P}%
}/V\text{ ,}%
\]
and further assume%
\[
\text{ }V_{2}^{\text{P}}+V_{2}^{\text{P}}=V,
\]
so that $\varphi_{1}+$ $\varphi_{2}=1.$ The last condition is nothing but the
requirement that mixing be isometric and can be rewritten using
(\ref{volumeRatios}) as%
\[
\phi_{\text{m}}\phi_{\text{m}1}^{\text{P}}+x\phi_{\text{m}}(\phi_{\text{m}%
2}^{\text{P}}-\phi_{\text{m}1}^{\text{P}})\equiv\phi_{\text{m}1}^{\text{P}%
}\phi_{\text{m}2}^{\text{P}}.
\]
This should be valid for all $x$ including $x=0$ and $x=1.$ This can only be
true if we require the \emph{monomer density equality}:
\begin{equation}
\phi_{\text{m}}=\phi_{\text{m}1}^{\text{P}}=\phi_{\text{m}2}^{\text{P}}.
\label{monomer_DenEq}%
\end{equation}
This condition is nothing but the \emph{equality of the free volume densities}
in the mixture and the pure components, and is a consequence of isometric
mixing, as is easily seen from (\ref{MixingVolume}) obtained below. We finally
conclude that
\begin{equation}
\varphi_{1}\equiv x,\varphi_{2}\equiv1-x, \label{VolumeFraction}%
\end{equation}
as was also the case discussed earlier in the context of (\ref{mixEnergy0}).

\subsubsection{Isometric RMA\ Limit}

Let us now consider the isometric RMA limit for which we have the
simplification
\[
V_{1}^{\text{P}}/V=x,V_{2}^{\text{P}}/V=1-x;
\]
see (\ref{monomer_DenEq}). Thus,%
\[
E_{\text{int}}{}\ ^{\underrightarrow{\text{RMA}}}\ -[c_{11}^{\text{P}}%
\varphi_{1}^{2}+2c_{12}\varphi_{1}\varphi_{2}+c_{22}^{\text{P}}\varphi_{2}%
^{2}],
\]
as is well known \cite{Hildebrand}. This form can only be justified in the
isometric RMA limit, with the volume fractions given by (\ref{VolumeFraction}%
). Similarly,%
\[
\Delta E_{\text{M}}v_{0}{}\ ^{\underrightarrow{\text{RMA}}}\ q[e_{12}%
-(1/2)(e_{11}+e_{22})]\phi_{\text{m}1}\phi_{\text{m}2}=(\chi/\beta
)\phi_{\text{m}1}\phi_{\text{m}2},
\]
where we have introduced the Flory-Huggins chi parameter $\chi\equiv
q\beta\varepsilon.$ Using (\ref{monomer_DenEq}), we can rewrite the above
energy of mixing in the form (\ref{mixEnergy1}). We finally have for $\Delta
E_{\text{M}}$ in the isometric RMA limit%
\begin{equation}
\Delta E_{\text{M}}{}\ ^{\underrightarrow{\text{RMA}}}\ (\chi\phi_{\text{m}%
}^{2}/\beta v_{0})\varphi_{1}\varphi_{2}, \label{RMA_EnergyMix}%
\end{equation}
again with the volume fractions given by (\ref{VolumeFraction}).

\subsubsection{Volume Fractions}

In the following, we will always take $\varphi_{i}$ to be given by
(\ref{VolumeFraction}). This ensures that $\varphi_{1}+$ $\varphi_{2}=1.$
Another possibility is to define $\varphi_{i}$ in terms of partial monomer
volumes $\overline{v_{i}}$:
\begin{equation}
\varphi_{i}\equiv\phi_{\text{m}i}\overline{v_{i}}/v_{0}.
\label{PartialVolumeFraction}%
\end{equation}
However, as shown in \cite{RaneGuj2003}, the error is not significant except
near $x=0$ or $x=1$. Since the calculation of $\overline{v_{i}}$ is somewhat
tedious, we will continue to use (\ref{VolumeFraction}) for $\varphi_{i}$ in
our calculation, as we are mostly interested in $x=0.5$.

\subsubsection{Beyond Isometric RMA}

Beyond isometric RMA, the energy of mixing will not have the above form in
(\ref{RMA_EnergyMix}); rather, it will be related to the energetic effective
chi introduced in (\ref{effective_EChi}) \cite{Gujrati1998,Gujrati2003} in
exactly the same form as above:%
\[
\Delta E_{\text{M}}\equiv(\chi_{\text{eff}}^{\text{E}}\phi_{\text{m}}%
^{2}/\beta v_{0})\varphi_{1}\varphi_{2},
\]
which ties the concept of cohesive energy density intimately with that of the
effective chi, as noted earlier. However, it is also clear that $c_{12}$ and
$\chi_{\text{eff}}^{\text{E}}$ are not directly proportional to each other. \ 

\subsection{van Laar-Hildebrand $c_{12}$}

Using (\ref{cohesive_defL}) for pure component cohesive energies, we obtain
\begin{equation}
c_{12}v_{0}=(e_{12}/2)\phi_{12}-e_{11}[(\phi_{11}-V_{1}^{\text{P}}/V\phi
_{11}^{\text{P}})/2\varphi_{1}\varphi_{2}-\phi_{11}^{\text{P}}]-e_{22}%
[(\phi_{22}-V_{2}^{\text{P}}/V\phi_{22}^{\text{P}})/2\varphi_{1}\varphi
_{2}-\phi_{22}^{\text{P}}] \label{Mutual_Cohesive_Density}%
\end{equation}
as the general expression for the cohesive energy density. We will assume that
$\varphi_{i}$ are as given in (\ref{VolumeFraction}). It is clear that the
definition of $c_{12}$ given in (\ref{mixEnergy1}) is such that it not only
depends on the state of the mixture, but also depends on pure component
states. This is an unwanted feature. In particular, $c_{12}$ will show a
discontinuity if a pure component undergoes a phase change (see Fig. \ref{F22}
later), even though the mixture does not.

In Fig. \ref{F12}, we show $c_{12}$ for a 50-50 blend ($M=100$) at $1.0$ atm
as a function of $t_{\text{C}}$ calculated for different values of $q$ from
$q=6$ to $q=14.$ The curvature of $c_{12}$ gradually changes at low
temperatures from concave upwards to downwards. As not only the magnitude but
also the shape of $c_{12}$ changes with $q$, $c_{12}$ is not simply
proportional to $q$; it is a complicated function of it. This is easily seen
from the values of $c_{12}/q$\ at $t_{\text{C}}=0%
{{}^\circ}%
$C, which is found to increase with $q.$ It is about $4.0$ at $q$=6 and
increases to about $6.4$ at $q=14.$ The temperature where $c_{12}$
asymptotically becomes very small, such as $0.1$ MPa occurs at higher and
higher values of the temperature as $q$ increases. This is not surprising. We
expect the cohesion to increase with $q$ at a given temperature.

\subsection{Isometric Mixing:\ EDIM and DDIM}

The volume of mixing is defined as
\[
\Delta V_{\text{M}}\equiv V-V_{1}^{\text{P}}-V_{2}^{\text{P}}.
\]
Using (\ref{volumeRatios}), we find that the volume of mixing per unit volume
($\Delta v_{\text{M}}\equiv\Delta V_{\text{M}}/V$), and per monomer
($\Delta\widehat{v}_{\text{M}}\equiv\Delta V_{\text{M}}/N_{\text{m}}$) are%
\begin{subequations}
\begin{align}
\Delta v_{\text{M}}  &  \equiv\phi_{\text{m}}[1/\phi_{\text{m}}-x/\phi
_{\text{m}1}^{\text{P}}-(1-x)/\phi_{\text{m}2}^{\text{P}}%
],\label{MixingVolume}\\
\Delta\widehat{v}_{\text{M}}  &  \equiv(1/\phi_{\text{m}}-x/\phi_{\text{m}%
1}^{\text{P}}-(1-x)/\phi_{\text{m}2}^{\text{P}})v_{0}. \label{MixingVolume1}%
\end{align}

One of the conditions for RST to be valid is that this quantity be zero
(isometric mixing). The condition (\ref{monomer_DenEq}), which ensures
isometric mixing at all $x$, is much stronger than the isometric mixing
requirement at a given fixed value of $x$. There are situations in which
mixing is isometric at a given $x$, but (\ref{monomer_DenEq}) is not
satisfied. For a given pure component monomer densities $\phi_{\text{m}%
1}^{\text{P}},$ and $\phi_{\text{m}2}^{\text{P}},$ that are to be mixed at a
given composition $x,$ we must choose the mixture density $\phi_{\text{m}}$ to
satisfy
\end{subequations}
\begin{equation}
1/\phi_{\text{m}}=x/\phi_{\text{m}1}^{\text{P}}+(1-x)/\phi_{\text{m}%
2}^{\text{P}} \label{Mixture_Den_Def}%
\end{equation}
in order to make the mixing isometric; see (\ref{MixingVolume}). In this case,
(\ref{mixEnergy1}) will not hold true despite mixing being isometric. In our
calculations, we will consider both ways of ensuring isometric mixing. We will
call the mixing method satisfying (\ref{monomer_DenEq}) \emph{equal density
isometric mixing} (EDIM), and the mixing method satisfying
(\ref{Mixture_Den_Def}) \emph{different density isometric mixing} (DDIM).

For most of our computation, we fix the composition $x.$ Almost all of our
results are for a 50-50 mixture. We will consider both isometric mixing
processes noted above in our calculations. A variety of processes can be
considered for each mixing. In order to make calculations feasible, we need to
restrict the processes to a few selected ones. We have decided to investigate
the following processes with the hope that they are sufficient to illuminate
the complex behavior of cohesive energy densities and their usefulness.

\subsubsection{(Isometric) Isochoric Process}

The process should be properly called an isometric isochoric process, but we
will use the term isochoric process in short in this work. The volume of the
mixture is kept fixed as the temperature is varied. The energy of mixing can
be calculated for a variety of mixing processes. We have decided to restrict
this to isometric mixing. We calculate the mixture's monomer density
$\phi_{\text{m}}$ at each temperature. For each temperature, we use
(\ref{monomer_DenEq}) to determine the pure component monomer densities to
ensure isometric mixing for the selected $x$.

\subsubsection{(Isobaric) EDIM Process}

The process should be properly called an isobaric EDIM process, but we will
use the term EDIM process in short in this work. We keep the mixture at a
fixed pressure, which is usually $1.0$ atm, and calculate its monomer density
$\phi_{\text{m}}$ at each temperature. We then use EDIM to ensure isometric
mixing at each temperature and calculate the energy of mixing. In the process,
the mixture's volume keeps changing, and the pure component pressures need not
be at the mixture's fixed pressure. Thus, the mixing is not at constant
pressure, even though the mixture's pressure is constant.

\subsubsection{DDIM\ Process}

For DDIM, we keep the pressures of the pure components fixed, which is usually
$1.0$ atm$,$ and calculate $\phi_{\text{m}1}^{\text{P}},$ and $\phi
_{\text{m}2}^{\text{P}},$ from which we calculate $\phi_{\text{m}}$ using
(\ref{Mixture_Den_Def}) for the selected $xv$. This ensures isometric mixing,
but again the mixing process is not a constant pressure one since the mixture
need not have the same fixed pressure of the pure components. Even though the
energy of mixing is calculated for an isometric mixing, it is neither
calculated for an isobaric nor an isochoric process.

All the above three mixing processes correspond to isometric mixing at each
temperature. Thus, we can compare the calculated van Laar-Hildebrand $c_{12}$
for these three processes to assess the importance of isometric mixing on
$c_{12}$. It is not easy to consider EDIM for an isochoric (constant $V$)
process. Therefore, we have not investigated it in this work.

\subsubsection{Isobaric Process}

In this process, the pressure of the mixture is kept constant as the
temperature is varied. The calculation of the energy of mixing can be carried
out for a variety of mixing. We will restrict ourselves to mixing at constant
pressure so that the pure components also have the same pressure as the
mixture at all temperatures. Thus, the volume of mixing will not be zero in
this case. The process is properly described as a constant pressure mixing
isobaric process, but we will use the term isobaric process in short in this work.%

\begin{figure}
[ptb]
\begin{center}
\includegraphics[
trim=0.902732in 5.820900in 0.903547in 0.903464in,
height=3.9444in,
width=6.3693in
]%
{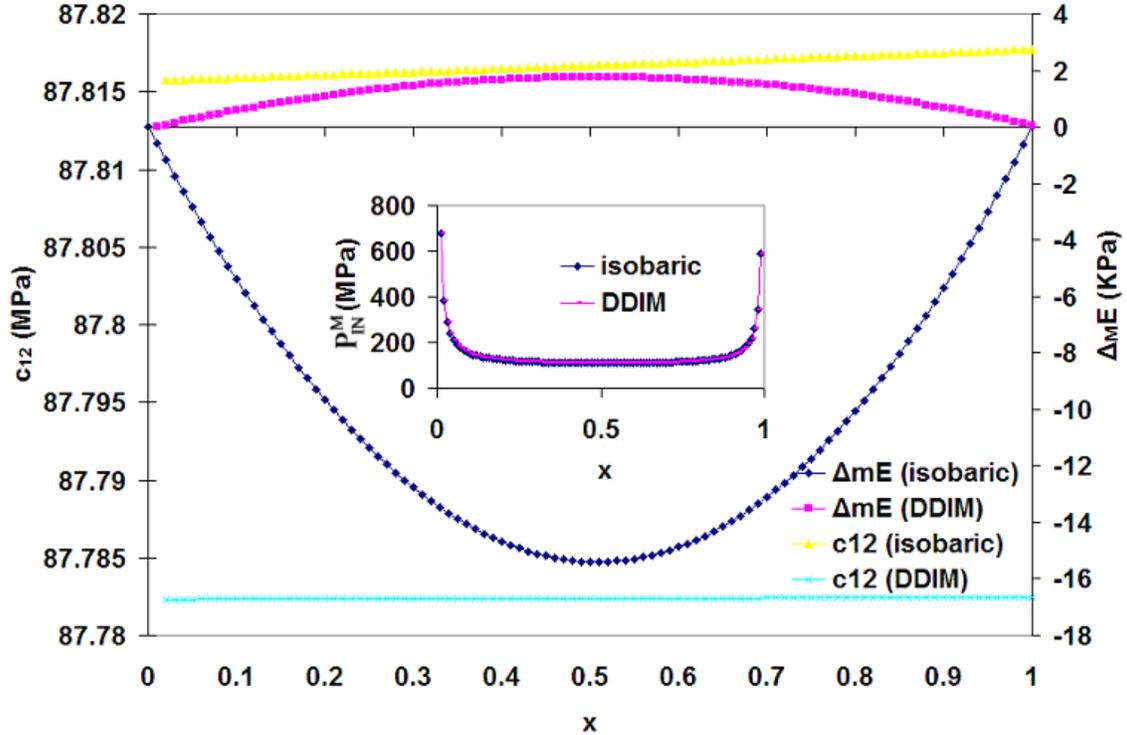}%
\caption{Energy of mixing and mutual cohesive energy density as a function of
composition $x$. We take $M_{1}=10,M_{2}=100$, $e_{11}=e_{22}=e_{12}%
=-2.6\times10^{-21}$ J, $v_{0}=1.6\times10^{-28}$ m$^{\text{3}}$ and $q=14.$
The pressure and temperature are fixed at $1.0$ atm and $300^{\circ}$C,
respectively.}%
\label{F15}%
\end{center}
\end{figure}

\subsection{Results}

\subsubsection{Size Disparity Effects}

The effects of size disparity alone are presented in Fig. \ref{F15}, where we
show the energy of mixing as a function of $x=\phi_{\text{m}1}/\phi_{\text{m}%
}$ for a blend ($M_{1}=10,M_{2}=100$) with $e_{11}=e_{22}=e_{12}%
=-2.6\times10^{-21}$ J, so that $\varepsilon_{12}=0.$ Thus, energetically,
there is no preference. We take $q=14,$ and the pressure and temperature are
fixed at $1.0$ atm and $300^{\circ}$C, respectively. We show isobaric and DDIM
results. While the energy of mixing is negative for the isobaric case, which
does not correspond to isometric mixing, it is positive everywhere for DDIM,
which does correspond to isometric mixing. This result should be contrasted
with the Scatchard-Hildebrand conjecture (\ref{mixEnergy0}) \cite{Hildebrand},
whose justification requires not only isometric mixing but also the
London-Berthelot conjecture (\ref{london_berthelot_Conj}). We also show
corresponding $c_{12}$, which weakly changes with $x.$ For $M_{1}=M_{2}=100,$
the energy and the volume of mixing vanish, which is not a surprising result
as we have a symmetric blend (both components identical in size and interaction).

It is important to understand the significance of the difference in the
behavior of $\Delta E_{\text{M}}$\ for the two processes in Fig. \ref{F15}.
From Fig. \ref{F7}, we find that the solubility parameter $\delta$\ at
$1.0$\ atm and $25^{\circ}$C is about 9.844 and 9.905 (MPa)$^{1/2}$ for
$M=100,$ and $10$ respectively. Thus, assuming the validity of the
Scatchard-Hildebrand conjecture (\ref{mixEnergy0}), we estimate $\Delta
E_{\text{M}}$\ $\cong0.9$ kPa at equal composition ($x=1/2$). However, a
correction for temperature difference needs to made, since the results in Fig.
\ref{F15} are for $t_{\text{C}}=300^{\circ}$C. From the inset in Fig.
\ref{F2}, we observe that $c^{\text{P}}$ almost decreases by a factor of 2/3,
while their difference has increased at $t_{\text{C}}=300^{\circ}$C relative
to $t_{\text{C}}=25^{\circ}$C. What one finds is that the corrected $\Delta
E_{\text{M}}$ is not far from the DDIM $\Delta E_{\text{M}}$ in Fig. \ref{F15}
at $x=1/2,$\ but has no relationship to the isobaric $\Delta E_{\text{M}},$
which not only is negative but also has a much larger magnitude$.$

It should be noted that the pressures of the pure components in both
calculations reported in Fig. \ref{F15} is the same: $1.0$ atm. Under this
condition, the Scatchard-Hildebrand conjecture (\ref{mixEnergy0}) cannot
differentiate between different processes as $\delta_{1\text{ }}$and
$\delta_{2}$ are unchanged. But this is most certainly not the case in Fig.
\ref{F15}. For a symmetric blend, $\Delta E_{\text{M}}=0$ even in an exact
theory. Since the symmetry requires $\delta_{1\text{ }}=\delta_{2},$ we find
that the London-Berthelot conjecture (\ref{london_berthelot_Conj}) is
satisfied. Thus, the violation of the Scatchard-Hildebrand conjecture we
observe in Fig. \ref{F15} is due to non-random mixing caused by size-disparity.

\subsubsection{Interaction Disparity Effects}

\begin{figure}
[ptb]
\begin{center}
\includegraphics[
trim=0.902732in 5.820900in 0.903547in 0.903464in,
height=3.9444in,
width=6.3693in
]%
{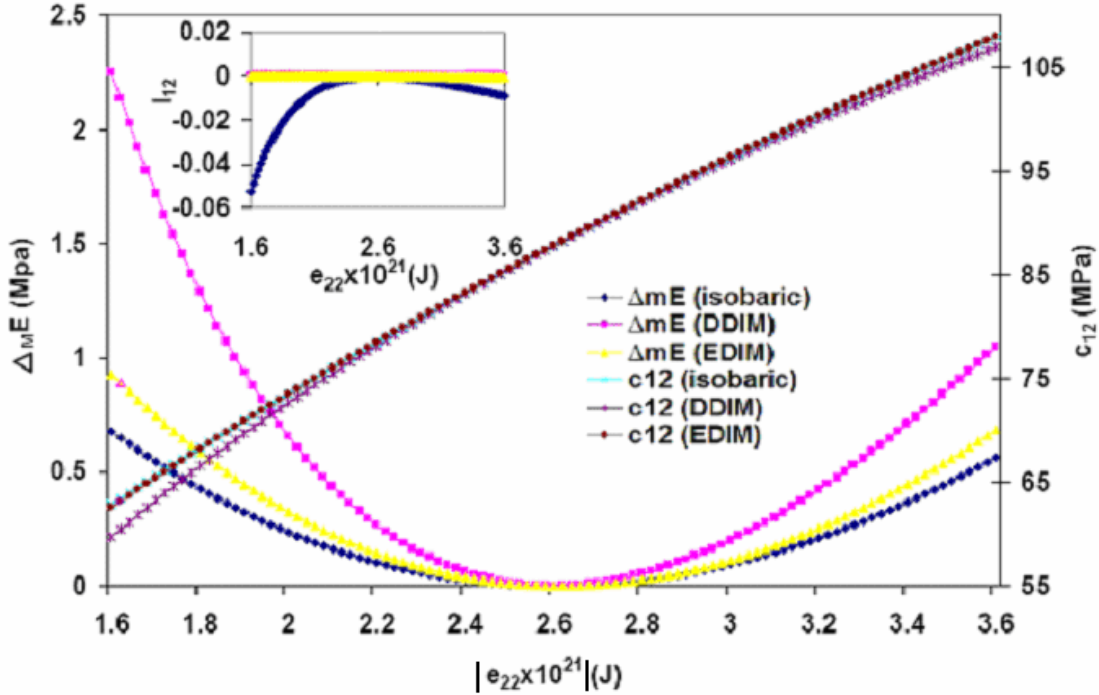}%
\caption{Energy of mixing, mutual cohesive energy density, and correction
$l_{12}$ as a function of \ $\left\vert e_{22}\right\vert $ for a 50-50 blend.
We take $M_{1}=M_{2}=100$, $e_{11}=-2.6\times10^{-21}$ J, $v_{0}%
=1.6\times10^{-28}$ m$^{\text{3}}$ and $q=14.$ The pressure and temperature
are fixed at $1.0$ atm and $300^{\circ}$C, respectively.}%
\label{F16}%
\end{center}
\end{figure}

In Fig. \ref{F16}, we fix $e_{11}=-2.6\times10^{-21}$ J, and plot the energy
of mixing as a function of $(-e_{22})$ for a blend with $M_{1}=M_{2}=100.$
Thus, there is no size disparity but the energy disparity is present except
when $e_{11}=e_{22}.$ We not only consider isobaric, and DDIM processes, but
also the EDIM process. At $e_{11}=e_{22}$, we have a symmetric blend; hence,
$\Delta E_{\text{M}}=0$ and\ $c_{11}^{\text{P}}=$\ $c_{22}^{\text{P}}$ for all
the processes. Correspondingly, we have $c_{12}=$ $c_{11}^{\text{P}}%
$or\ $c_{22}^{\text{P}},$ and the correction $l_{12}=0$ for all the
processes$.$ Away from this point, $\Delta E_{\text{M}}$ for the three
processes\ are different, but remain non-negative. The energy disparity has
produced much larger magnitudes of $\Delta E_{\text{M}}$\ than the size
disparity alone; compare with Fig. \ref{F15}. This difference in the
magnitudes of $\Delta E_{\text{M}}$ is reflected in the magnitudes of
$c_{12},$ as shown in the figure. We again find that isobaric and DDIM
processes are now quite different; compare the magnitudes of $\Delta
E_{\text{M}}$ in Figs. \ref{F15} and \ref{F16}. The difference between DDIM
and EDIM, though relatively small, is still present, again proving that
isometric mixing alone is not sufficient to validate the Scatchard-Hildebrand
conjecture. [We note that $\Delta E_{\text{M}}$ can become negative (results
not shown) if we add size disparity in addition to the interaction disparity.]
The non-negative $\Delta E_{\text{M}}$ is due to the absence of size
disparity, and is in accordance with the Scatchard-Hildebrand conjecture.
Since $\Delta E_{\text{M}}$\ is the highest for DDIM, the corresponding
$c_{12}$ is the lowest. Similarly, the isobaric energy of mixing is usually
the lowest, and the corresponding $c_{12}$ usually the highest. We note that
all $c_{12}$'s continue to increase with $\left\vert e_{22}\right\vert .$ We
observe that isobaric and EDIM $c_{12}$'s are closer to each other, and both
are higher than DDIM\ $c_{12}$. In the inset, we also show $l_{12},$ and we
learn that it is not a small correction to the London-Berthelot conjecture
(\ref{london_berthelot_Conj}) for the isobaric case (non-isometric mixing).
For isometric mixing, we have the usual behavior: a small correction $l_{12}.$%

\subsubsection{Variation with $P$}

\begin{figure}
[ptb]
\begin{center}
\includegraphics[
trim=0.902732in 5.820900in 0.903547in 0.903464in,
height=3.9444in,
width=6.3693in
]%
{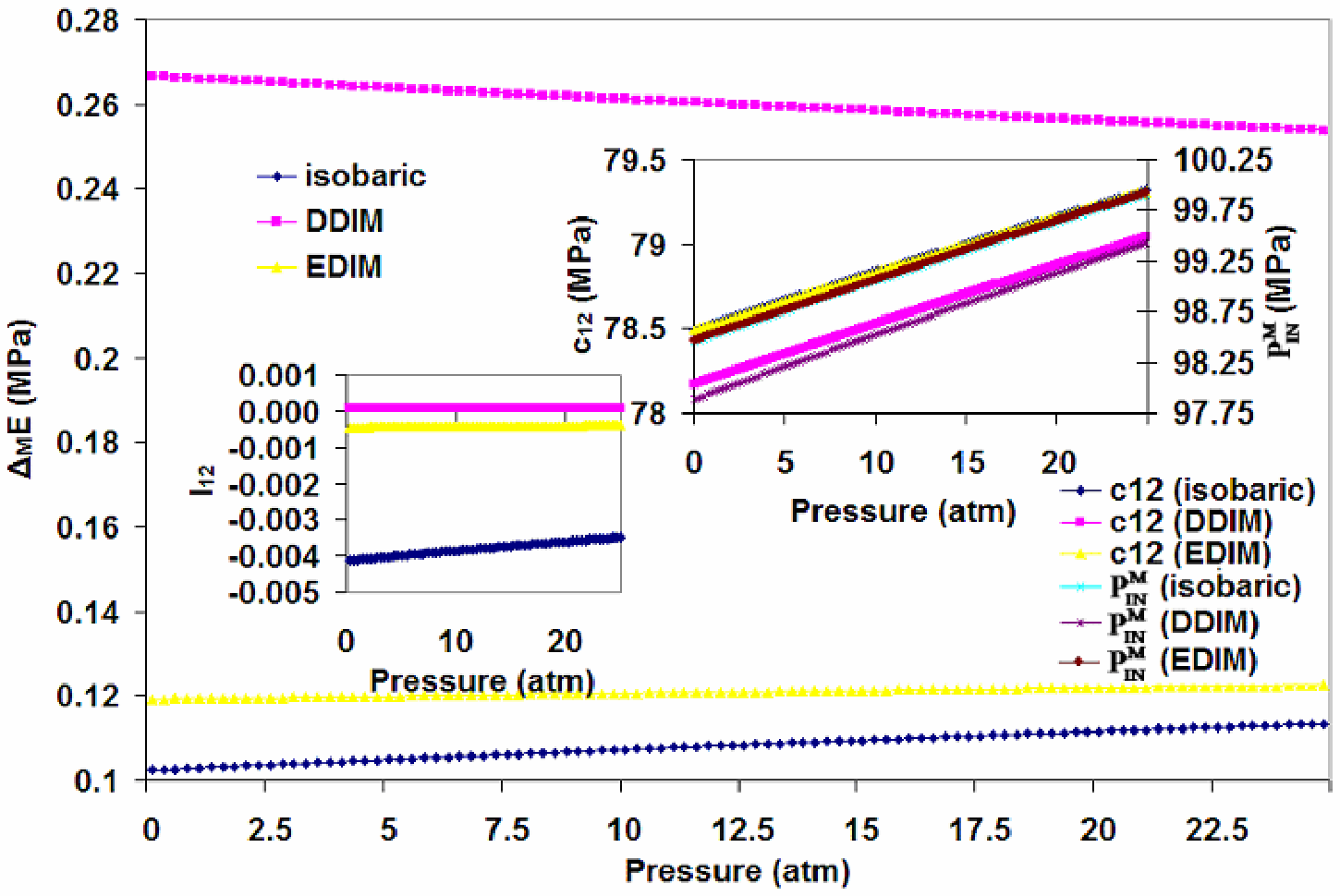}%
\caption{Energy of mixing, mutual cohesive energy density, and correction
$l_{12}$ as a function of \ pressure for a 50-50 blend. We take $M_{1}%
=M_{2}=100$, $e_{11}=-2.2\times10^{-21}$ J, $e_{22}=-2.6\times10^{-21}$ J,
$v_{0}=1.6\times10^{-28}$ m$^{\text{3}}$ and $q=14.$ The temperature is fixed
at $300^{\circ}$C. }%
\label{F17}%
\end{center}
\end{figure}

We plot $\Delta E_{\text{M}},$ $c_{12},$ and $l_{12}$\ as a function of $P$ in
Fig. \ref{F17} for isobaric, DDIM, and EDIM processes. We note that all these
quantities have a weak but monotonic dependence on $P$ over the range
considered$.$ Again, the isobaric $\Delta E_{\text{M}}$ remains the lowest,
and consequently isobaric $c_{12}$\ remains the highest over the range
considered. As noted above, isobaric and EDIM $c_{12}$'s are closer to each
other, but different from DDIM $c_{12}$ over the entire range in Fig.
\ref{F17}. The monotonic behavior in $P$ is correlated with a monotonic
behavior in $l_{12}.$ We observe that $l_{12}$ provides a small correction at
$300^{\circ}$C over the pressure range considered here. The interesting
observation is that isobaric $l_{12}$ provides the biggest correction, while
DDIM $l_{12}$\ the smallest correction. However, the EDIM $l_{12}$\ remains
intermediate, just as EDIM $c_{12}$ is.%

\subsubsection{Variation with $T$}

\begin{figure}
[ptb]
\begin{center}
\includegraphics[
trim=0.902732in 5.820900in 0.903547in 0.903464in,
height=3.9444in,
width=6.3693in
]%
{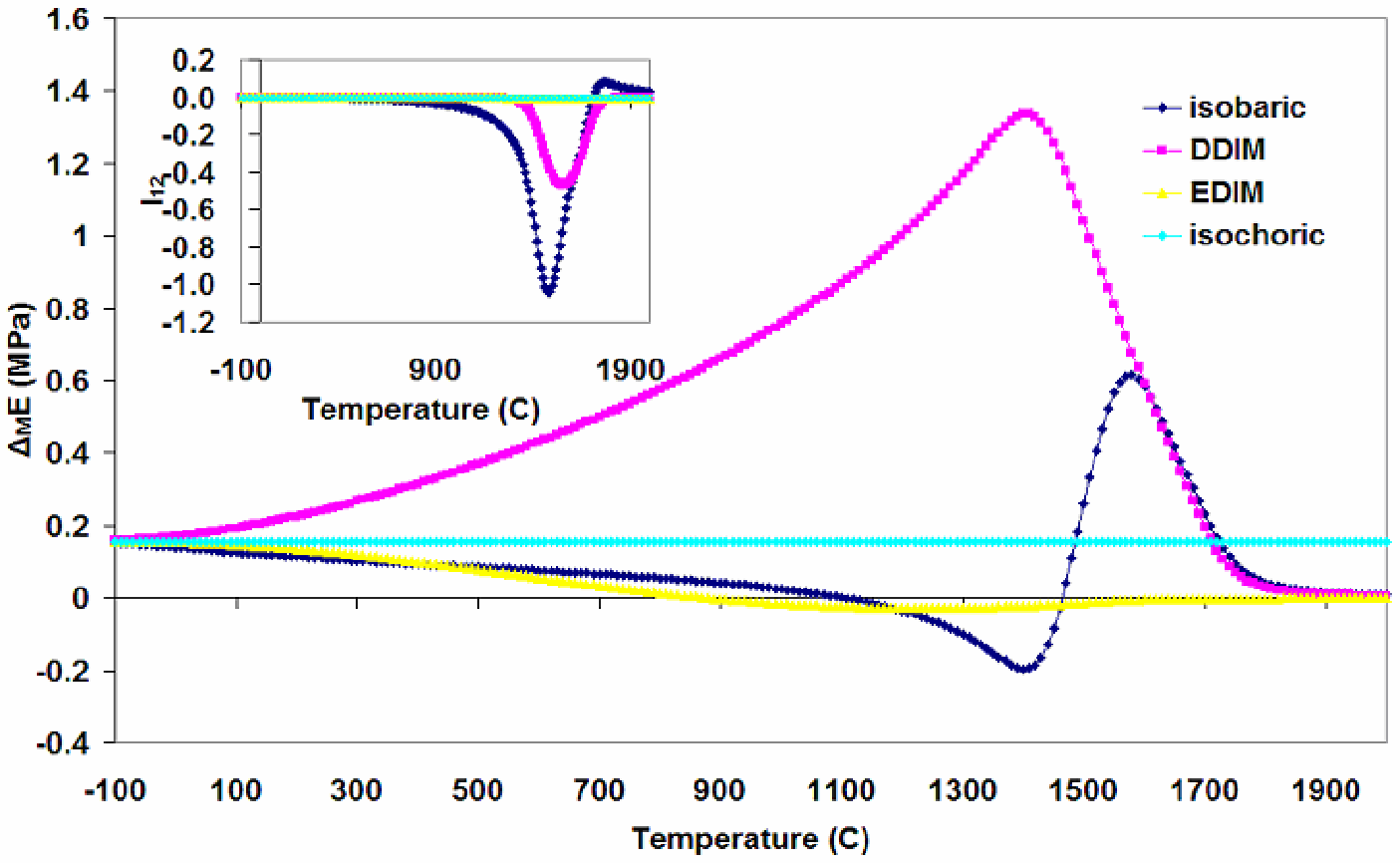}%
\caption{Energy of mixing, mutual cohesive energy density, and correction
$l_{12}$ as a function of \ temperature for a 50-50 blend. We take
$M_{1}=M_{2}=100$, $e_{11}=-2.2\times10^{-21}$ J, $e_{22}=-2.6\times10^{-21}$
J, $v_{0}=1.6\times10^{-28}$ m$^{\text{3}}$ and $q=14.$ The pressure is fixed
at $1.0$ atm. }%
\label{F18}%
\end{center}
\end{figure}

\begin{figure}
[ptb]
\begin{center}
\includegraphics[
trim=0.902732in 5.820900in 0.903547in 0.903464in,
height=3.9444in,
width=6.3693in
]%
{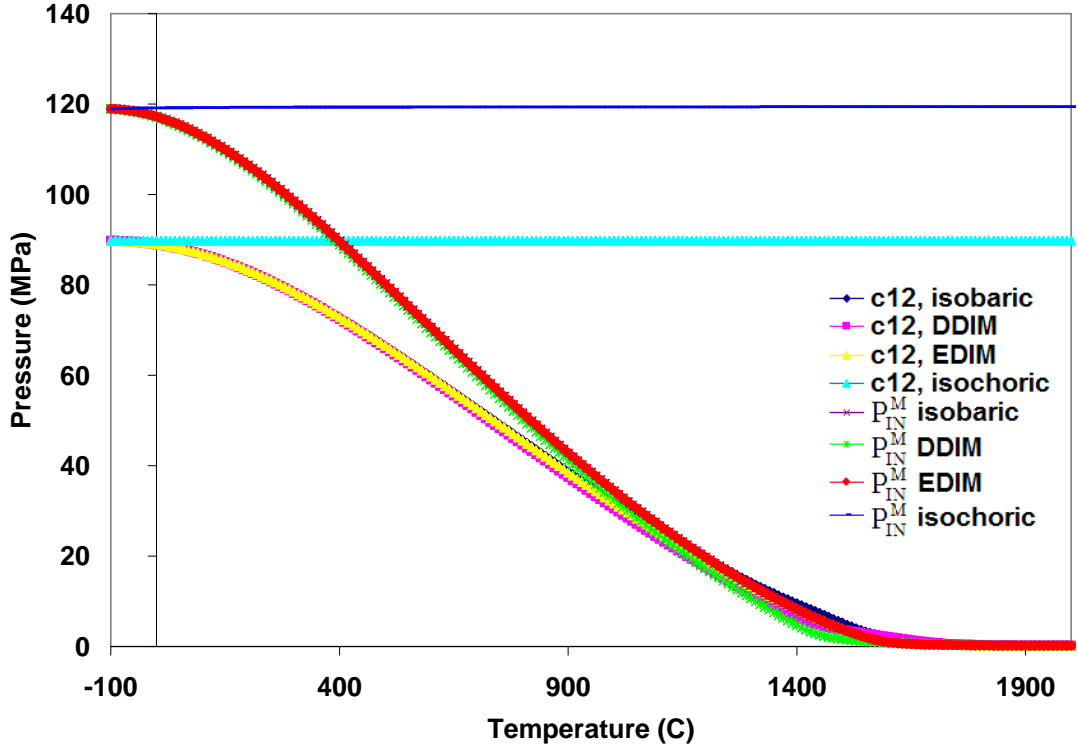}%
\caption{Different mutual pressures as a function of temperature; system as
described in Fig. \ref{F18}.}%
\label{F18-1}%
\end{center}
\end{figure}

We plot $\Delta E_{\text{M}}$ and $l_{12}$\ as a function of $T$ in Fig.
\ref{F18} for isobaric, DDIM, EDIM processes along with the isochoric process.
The corresponding $c_{12}$'s, all originating around 90 MPa at $-100^{\circ}$C
as a function of $T,$ are shown in Fig. \ref{F18-1}. All processes start from
the same state at the lowest temperature ($-100^{\circ}$C) in the figure.
There is a complicated dependence in $\Delta E_{\text{M}}$ on $T$ over the
range considered for some of the processes. Let us first consider the
isochoric process in which all quantities show almost no dependence on $T.$
The energy of mixing remains positive in accordance with the London-Berthelot
conjecture (\ref{london_berthelot_Conj}). This is further confirmed by an
almost constant $c_{12},$ and an almost vanishing correction $l_{12}$ in the
inset. Both isobaric and EDIM processes give rise to negative $\Delta
E_{\text{M}},$ thereby violating the London-Berthelot conjecture. The DDIM
$\Delta E_{\text{M}}$\ shows a peak, which is about eight times in magnitude
than its value at the lowest temperature, but remains positive throughout. The
corresponding $c_{12}$ shows a continuous decrease to zero with temperature
for isobaric, DDIM, and EDIM processes. However, it is almost a constant for
the isochoric process. The non-monotonic behavior in temperature of $\Delta
E_{\text{M}}$ is correlated with a similar behavior in $l_{12}.$ This behavior
is further studied in the next section. From Fig. \ref{F18}, we observe that
$l_{12}$ provides a small correction at $300^{\circ}$C. However, at much
higher temperatures, it is no longer a small quantity, and depends strongly on
the way the mixture is prepared, even if it remains isometric. In particular,
isobaric $l_{12}$ seems to provide the biggest correction. It is almost
constant and insignificant for the isochoric and EDIM\ processes.%

%

\subsection{Using Internal Pressure}

We now introduce a new quantity as another measure of the mutual cohesive
pressure by using the internal pressure. To this end, we consider the internal
pressure $P_{\text{IN}}$ in the RMA limit. We find from (\ref{RMA3}) that
\[
P_{\text{IN}}v_{0}\ ^{\underrightarrow{\text{RMA}}}\ q[-e_{11}\phi_{\text{m}%
1}^{2}/2-e_{22}\phi_{\text{m}1}^{2}/2-e_{12}\phi_{\text{m}1}\phi_{\text{m}%
2}],
\]
which can be expressed as%
\[
P_{\text{IN}}v_{0}{}\ ^{\underrightarrow{\text{RMA}}}\ c_{11}^{\text{P}}%
x^{2}\phi_{\text{m}}^{2}/\phi_{\text{m}1}^{\text{P}2}+c_{22}^{\text{P}%
}(1-x)^{2}\phi_{\text{m}}^{2}/\phi_{\text{m}2}^{\text{P}2}+2c_{12}x(1-x),
\]
where we have used (\ref{RMA_CohesiveDensities}) in the RMA limit. Now we take
this equation as a guide to define a new mutual cohesive energy density, the
\emph{mutual internal pressure} $P_{\text{IN}}^{\text{M}}$ in the general case
as%
\[
P_{\text{IN}}^{\text{M}}=[P_{\text{IN}}-c_{11}^{\text{P}}x^{2}\phi_{\text{m}%
}^{2}/\phi_{\text{m}1}^{\text{P}2}-c_{22}^{\text{P}}(1-x)^{2}\phi_{\text{m}%
}^{2}/\phi_{\text{m}2}^{\text{P}2}]/2x(1-x).
\]
In the RMA\ limit, $P_{\text{IN}}^{\text{M}}$ reduces to the RMA $c_{12}$
given in (\ref{RMA_CohesiveDensities}). We show $P_{\text{IN}}^{\text{M}}$ in
the inset in Figs. \ref{F15}, \ref{F17}, and \ref{F18-1}. What we discover is
that isochoric $P_{\text{IN}}^{\text{M}}$ is almost a constant as a function
of temperature around 120 MPa, and is larger than isochoric $c_{12},$ which is
around 90 MPa. Indeed, over most of the temperatures at the lower end, we find
that $P_{\text{IN}}^{\text{M}}$ $>$ $c_{12}$. However, the main conclusion is
that both $c_{12}$ and $P_{\text{IN}}^{\text{M}}$ are monotonic decreasing or
are almost constant, and behave identically except for their magnitudes.

\section{New Approach using SRS:\ Self-Interacting Reference State}

\begin{figure}
[ptb]
\begin{center}
\includegraphics[
trim=0.902732in 5.820900in 0.903547in 0.903464in,
height=3.9444in,
width=6.3693in
]%
{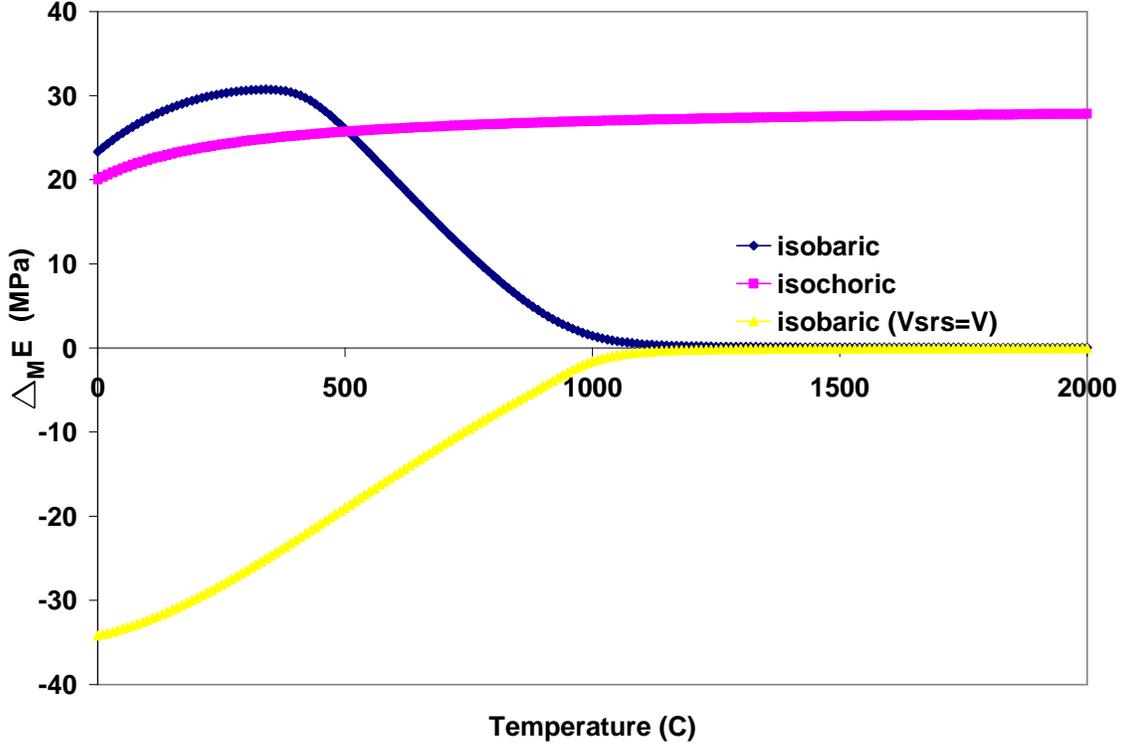}%
\caption{$\Delta E_{\text{M}}$ for IRS as a function of \ temperature for a
50-50 blend. We take $M_{1}=M_{2}=100$, $e_{11}=-2.2\times10^{-21}$ J,
$e_{22}=-2.6\times10^{-21}$ J $v_{0}=1.6\times10^{-28}$ m$^{\text{3}}$ and
$q=10.$ The pressure is fixed at $1.0$ atm. }%
\label{F19}%
\end{center}
\end{figure}

Unfortunately, the van Laar-Hildebrand cohesive energy $c_{12}$ does not have
the required property of vanishing with $e_{12},$ see
(\ref{RMA_CohesiveDensities}). The unwanted behavior of $c_{12}$\ is due its
definition in terms of the energy of mixing from which we need to subtract
pure component (for which $e_{12}$ may be thought to be zero) cohesive
energies $c_{ii}^{\text{P}}.$\ The subtracted quantity is used to define
$c_{12},$ and if this definition has to have any physical significance, it
should vanish in the hypothetical state, which we have earlier labelled SRS,
in which $e_{12}$ vanishes even though $e_{11}$ and $e_{22}$\ are non-zero.
The hypothetical state obviously violates the London condition
(\ref{london_Conj}), even though the real mixture does not. Let us demand the
subtracted quantity to vanish for SRS$.$ To appreciate this point, consider
(\ref{mixEnergyL}) for SRS. It is clear that $\Delta E_{\text{M}}$ continues
to depend on the thermodynamic state of the mixture via $\phi_{ii};$ this
quantity in general will not be equal to pure component quantity $\phi
_{ii}^{\text{P}}$. With the use of (\ref{volumeRatios}) in (\ref{mixEnergyL}),
we find that
\begin{equation}
\Delta E_{\text{M}}^{\text{SRS}}v_{0}=e_{11}[\phi_{11}-x\phi_{\text{m}}%
\phi_{11}^{\text{P}}/\phi_{\text{m}1}^{\text{P}}]+e_{22}[\phi_{22}%
-(1-x)\phi_{\text{m}}\phi_{22}^{\text{P}}/\phi_{\text{m}2}^{\text{P}}],
\label{SIRS_EnergyofMixing0}%
\end{equation}
which is usually going to depend on the process of mixing. On the other hand,
from (\ref{mixEnergy1}), we observe that
\begin{equation}
\Delta E_{\text{M}}^{\text{SRS}}\overset{c_{12}=0}{=}(c_{11}^{\text{P}}%
+c_{22}^{\text{P}})\varphi_{1}\varphi_{2}, \label{SIRS_EnergyofMixing1}%
\end{equation}
had $c_{12}=0.$\ In this case, $\Delta E_{\text{M}}$ would depend only on pure
component quantities $c_{ii}^{\text{P}},$ and its behavior in a given process
at fixed composition should be controlled by the behavior of $c_{ii}%
^{\text{P}}$\ in that process. This is not the case, as can be seen in Fig.
\ref{F19}, where we plot $\Delta E_{\text{M}}$ for the hypothetical state SRS
for a 50-50 mixture. We have ensured that the SRS state at the initial
temperature in Fig. \ref{F19} is the same in isobaric and isochoric processes.
But the pure components are slightly different. A new process is also
considered in Fig. \ref{F19}, in which we set the volume $V_{\text{SRS}}$ of
the hypothetical SRS at a given temperature to be equal to the volume $V$ of
the real mixture (nonzero $e_{12}$) at $1.0$ atm at that temperature. This
process is labelled isobaric ($V_{\text{SRS}}=$ $V$) in the figure. The pure
components are also at $1.0$ atm for this process. Since the pure components
for this process are the same as in the isobaric calculation,
(\ref{SIRS_EnergyofMixing1}) requires $\Delta E_{\text{M}}$ for the two
processes \ to be identical at all temperatures. This is evidently not the
case. Consider Fig. \ref{F11}. From this, we see that $c_{ii}^{\text{P}}$\ are
almost constant with $T$ for the isochoric case. Thus, according to
(\ref{SIRS_EnergyofMixing1}), $\Delta E_{\text{M}}$ should be almost a
constant, which is not the case. For the isobaric case, $c_{ii}^{\text{P}}%
$\ are monotonic decreasing with $T$, which will then make $\Delta
E_{\text{M}}$, according to (\ref{SIRS_EnergyofMixing1}), monotonic decreasing
with $T$, which is also not the case. Hence, we conclude that the van
Laar-Hildebrand $c_{12}$ does not vanish for SRS.

A similar unwanted feature is also present in the behavior of $P_{\text{IN}%
}^{\text{M}}$ introduced above for the same reason: it also does not vanish
for SRS.

It is disconcerting that the van Laar-Hildebrand $c_{12}(e_{12}=0)$ does not
vanish for SRS, even though the mutual interaction energy $e_{12}=0.$ This
behavior is not hard to understand. The mixture energy is controlled by the
process of mixing, since the mixture state varies with the process of mixing
even at $e_{12}$=0.%

\subsection{Mutual Energy of Interaction $c_{12}^{\text{SRS}}$}

To overcome this shortcoming, we introduce a new measure of the cohesive
energy that has the desired property of vanishing with $e_{12}.$ Let $E$
denote the energy per unit volume of the mixture. We will follow
\cite{RaneGuj2005}, and compare it with that of the hypothetical reference
state SRS$.$ Its energy per unit volume $E_{\text{SRS}}$ differs from $E$ due
to the absence of the mutual interaction between the two components. Let
$V_{\text{SRS}}$ denote the volume of the SRS. In general, we have%
\[
V_{\text{SRS}}/V=\phi_{\text{m}}/\phi_{\text{m,SRS}}.
\]
The difference
\[
E_{\text{int}}^{(\text{M})}\equiv E-(V_{\text{SRS}}/V)E_{\text{SRS}}%
\]
represents the mutual energy of interaction per unit volume due to $1$-$2$
contacts. From (\ref{energy_Def}), we obtain
\[
E_{\text{int}}^{(\text{M})}v_{0}\equiv e_{11}[\phi_{11}-(V_{\text{SRS}}%
/V)\phi_{11,\text{SRS}}]+e_{22}[\phi_{22}-(V_{\text{SRS}}/V)\phi
_{22,\text{SRS}}]+e_{12}\phi_{12},
\]
where the contact densities $\phi_{ij}$\ without SRS are for the mixture state
and with SRS are for the SRS state. These densities are evidently different in
the two states but approach each other as $e_{12}\rightarrow0$. The above
excess energy should determine the mutual cohesive energy density, which we
will denote by $c_{12}^{\text{SRS}}$ in the following to differentiate it with
the van Laar-Hildebrand cohesive energy density $c_{12}$. It should be evident
that $E_{\text{int}}^{\text{M}}\ $vanishes as $e_{12}$ vanishes due to its definition.%

\begin{figure}
[ptb]
\begin{center}
\includegraphics[
trim=0.902732in 5.820900in 0.903547in 0.903464in,
height=3.9444in,
width=6.3693in
]%
{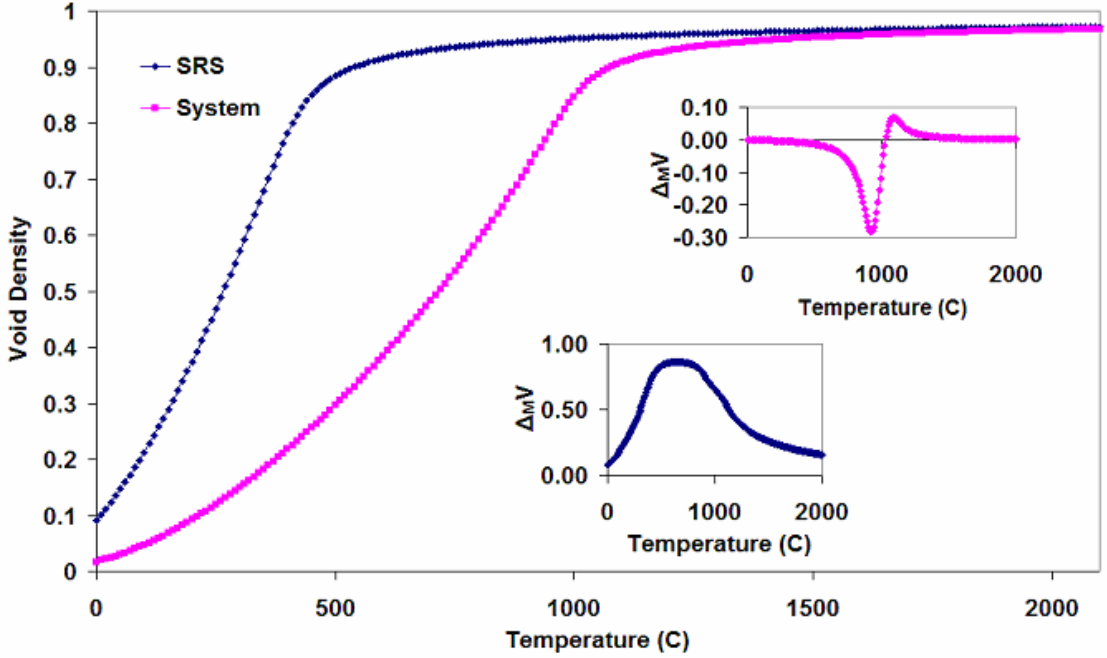}%
\caption{Void density and volume of mixing as a function of \ temperature for
a 50-50 blend. We take $M_{1}=M_{2}=100$, $e_{11}=-2.2\times10^{-21}$ J,
$e_{22}=-2.6\times10^{-21}$ J, $v_{0}=1.6\times10^{-28}$ m$^{\text{3}}$ and
$q=10.$ The pressure is fixed at $1.0$ atm. }%
\label{F14}%
\end{center}
\end{figure}

The absence of mutual interaction in SRS compared to the mixture causes SRS
volume to expand relative to the mixture. This is shown in Fig. \ref{F14} in
which the void density in SRS is larger than that in the mixture. The relative
volume of mixing at constant pressure ($1.0$ atm) for the mixture is shown in
the upper inset, and that for SRS is shown in the lower inset. It is clear
that the latter relative volume of mixing is always positive, indicating an
effective repulsion between the two components. This should come as no
surprise since the excess interaction $\varepsilon_{12}$ for SRS from
(\ref{excess_E})%
\[
\varepsilon_{12}^{\text{SRS}}=-(e_{11}+e_{22})/2>0,
\]
and has the value $2.4\times10^{-21}$ J in this case. Thus, it represents a
much stronger mutual repulsion than the mutual repulsion due to $\varepsilon
_{12}\cong0.1\times10^{-21}$ J in the mixture. The effect of adding the mutual
interaction $e_{12}$ to SRS is to add mutual "attraction" that results in
cohesion, and in the reduction of volume. Thus, the change in the volume can
also be taken as a measure of cohesion, as we will discuss below.%

\begin{figure}
[ptb]
\begin{center}
\includegraphics[
trim=0.902732in 5.820900in 0.903547in 0.903464in,
height=3.9444in,
width=6.3693in
]%
{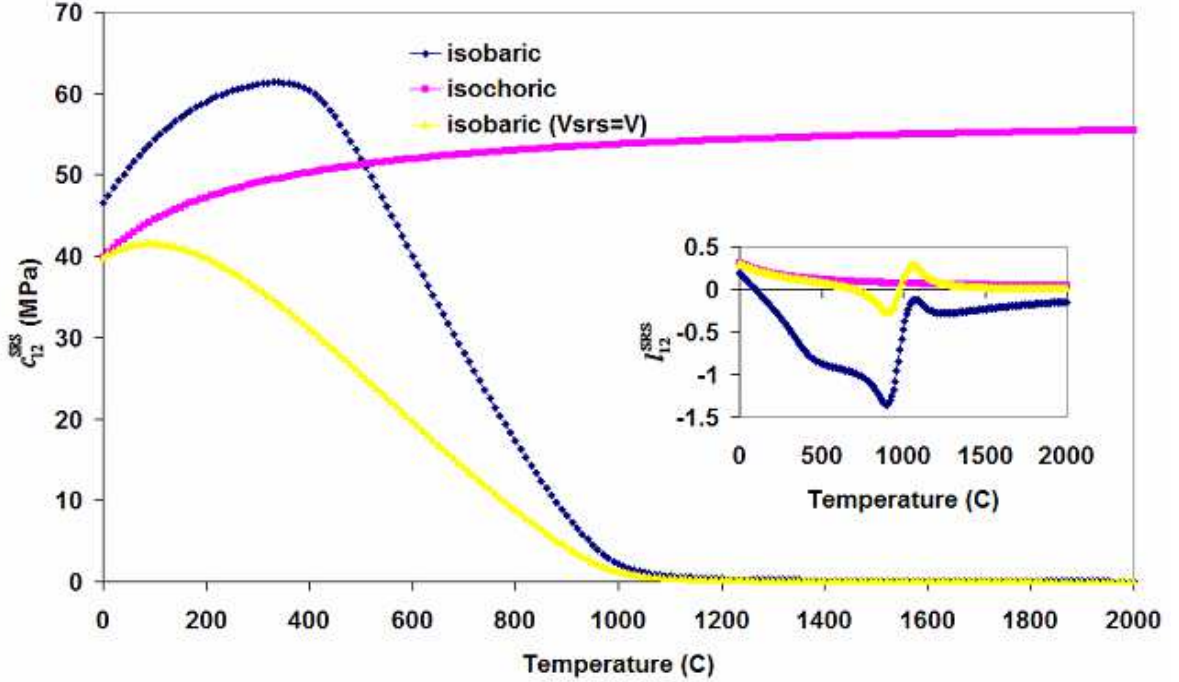}%
\caption{Mutual cohesive energy density, and correction $l_{12}$ using the IRS
as a function of \ temperature for a 50-50 blend. We take $M_{1}=M_{2}=100$,
$e_{11}=-2.2\times10^{-21}$ J, $e_{22}=-2.6\times10^{-21}$ J $v_{0}%
=1.6\times10^{-28}$ m$^{\text{3}}$ and $q=10.$ The pressure is fixed at $1.0$
atm. }%
\label{F20}%
\end{center}
\end{figure}

\subsubsection{RMA Limit of $E_{\text{int}}^{(\text{M})}$}

In the RMA limit, along with the monomer equality assumption
(\ref{monomer_DenEq}), which implies that $\phi_{\text{m}}=\phi_{\text{m,SRS}%
},$ it is easily seen that
\[
E_{\text{int}}^{(\text{M})}v_{0}{}^{\underrightarrow{\text{RMA}}}qe_{12}%
\phi_{\text{m}1}\phi_{\text{m}2}=qe_{12}\phi_{\text{m}1}^{\text{P}}%
\phi_{\text{m}2}^{\text{P}}x(1-x).
\]
As a consequence, $E_{\text{int}}^{(\text{M})}/2x(1-x)$ reduces to the RMA
value of $(-c_{12})$ given in (\ref{RMA_CohesiveDensities}). This means that
the quantity $c_{12}^{\text{SRS}}$ defined via%
\begin{equation}
c_{12}^{\text{SRS}}\equiv-E_{\text{int}}^{(\text{M})}/2\varphi_{1}\varphi_{2}
\label{cohesiveIRS}%
\end{equation}
has the required property that it not only reduces to the correct RMA value of
the van Laar-Hildebrand$\ c_{12},$ but it also vanishes with $e_{12}$. [As
usual, we assume the identification (\ref{VolumeFraction}).]

\subsubsection{New Mutual Cohesive Energy Density: $c_{12}^{\text{SRS}}$}

We now take (\ref{cohesiveIRS}) as the general definition of a more suitable
quantity to play the role of cohesive energy density. This quantity is a true
measure of the effect produced by mutual interaction energy, and which also
vanishes with $e_{12}.$ Away from the RMA limit, $c_{12}$ and $c_{12}%
^{\text{SRS}}$ are not going to be the same. It is evident that $c_{12}%
^{\text{SRS}}$ depends not only on $T,P,$ or $T,V,$ but also on the
composition $x\equiv\phi_{\text{m}1}/\phi_{\text{m}},$ $q$, and the energies
$e_{ij}.$ For isochoric calculations, we will ensure that the mixture and SRS
have the same monomer density $\phi_{\text{m}}.$ In this case, $V_{\text{SRS}%
}=V.$ For isobaric calculations, we will, as usual, ensure that they have the
same pressure, so that the two volumes need not be the same. However, we will
also consider $V_{\text{SRS}}=V$ for isobaric calculations to see the effect
of this on $c_{12}^{\text{SRS}}.$ This process is what we have labelled
isobaric ($V_{\text{SRS}}=$ $V$) in the Fig. \ref{F19}. We show $c_{12}%
^{\text{SRS}}$ for various processes in Fig. \ref{F20}. We see that it
continues to increase monotonically for the isochoric case, and reaches an
asymptotic value of about 50 MPa, while van Laar-Hildebrand $c_{12}$ in Fig.
\ref{F18} is almost constant and about 90 MPa. An increase in solubility with
temperature at constant volume is captured by $c_{12}^{\text{SRS}}$ but not by
$c_{12}.$ The behavior of the two quantities for the isobaric case is also
profoundly different. While $c_{12}$ monotonically decreases with temperature,
$c_{12}^{\text{SRS}}$ is most certainly not monotonic. It goes through a
maximum around $400%
{{}^o}%
$C before continuing to decrease. This suggests that the solubility increases
before decreasing. This should come as no surprise as we explain later in the
following section.

Note that while the definition of $c_{12}^{\text{SRS}}$ is independent of any
approximate theory, the definition of $c_{12}$ depends on a form
(\ref{mixEnergy1}) whose validity is of questionable origin as it depends on
RST. It should be noted that the value of $c_{12}^{\text{SRS}}$ truly
represents the energy change that occurs (in SRS) due to the presence of the
mutual interaction between the two components. The changes brought about due
to $e_{12}\neq0$ should not be confused with the changes that occur when the
two pure components are mixed ( $e_{12}\neq0),$ because the latter involves
not only the process of mixing when $e_{12}=0$ (this gives rise to SRS\ ) but
also involves the rearrangements of the two components due to the mutual
interaction. It is the latter rearrangement due to mutual interaction that
determines $c_{12}^{\text{SRS}}.$ This is easily appreciated by studying the
void density profiles for SRS and the mixture in Fig. \ref{F14}. As usual,
$e_{12}$ for the mixture is defined by (\ref{london_Conj}).~The hypothetical
process of mixing (under the condition $e_{12}=0$) also contributes to the
energy of mixing, since the energy of this state is not the same as the sum of
the two pure systems' energies, because SRS is affected by mixing the pure
components. This is easily seen by computing the energy of mixing $\Delta
E_{\text{M,SRS}}$ for SRS for which $e_{12}=0$ but not $e_{11},e_{22}$:%
\[
V_{\text{SRS}}\Delta E_{\text{M,SRS}}\equiv V_{\text{SRS}}E_{\text{SRS}}%
-V_{1}^{\text{P}}E_{1}^{\text{P}}-V_{2}^{\text{P}}E_{2}^{\text{P}},
\]
where $V_{i}^{\text{P}},E_{i}^{\text{P}}$ are the volume and energy density of
the pure $i$th component. This part determines the energy of mixing due to
pure mixing at $e_{12}=0,$\ and but it has nothing to do with the mutual
interaction $e_{12}.$ Thus, this contribution should not be allowed to
determine in part the mutual cohesive energy density. It is easy to see that%
\[
\Delta E_{\text{M}}\equiv E_{\text{int}}^{\text{M}}+(V_{\text{SRS}}/V)\Delta
E_{\text{M,SRS}},
\]
in which the second contribution is not a true measure of the mutual
interaction. Nevertheless, it is considered a part of the van Laar-Hildebrand
cohesive energy density. Using (\ref{mixEnergy1}) and (\ref{cohesiveIRS}) in
the above equation, we find that we can express $c_{12}^{\text{SRS}}$ in terms
of the van Laar-Hildebrand cohesive energy densities as follows:%
\begin{equation}
c_{12}^{\text{SRS}}(e_{12})=c_{12}(e_{12})-c_{12}(e_{12}=0)+(1-\phi_{\text{m}%
}/\phi_{\text{m,SRS}})(c_{11}^{\text{P}}+c_{22}^{\text{P}})/2.
\label{IRS-Hilde_Rel}%
\end{equation}
Here, $c_{12}^{\text{SRS}}(e_{12})$\ is the SRS-based $c_{12}^{\text{SRS}}$
introduced in (\ref{cohesiveIRS}), and $c_{12}(e_{12})$ is energy-of-mixing
based $c_{12}$ in (\ref{mixEnergy1}). In the special case $\phi_{\text{m}%
}=\phi_{\text{m,SRS}},$ the last term vanishes, and the SRS-$c_{12}$ is the
difference of the van Laar-Hildebrand-$c_{12}.$ In any case, because of the
above relation, studying van Laar-Hildebrand-$c_{12}$ also allows us to learn
about the SRS-$c_{12}.$ Therefore, we have mostly investigate $c_{12}$ in the
present work. However, it should be realized that we also need van
Laar-Hildebrand-$c_{12}$ for SRS, data for which ate not available.

The arguments similar to the presented above suggest that we can similarly use
the difference $P_{\text{int}}-$ $P_{\text{int,SRS}},$ where
$P_{\text{int,SRS}}$\ is the interaction pressure in SRS at the same volume
$V$, as a measure of the cohesive pressure for the system. Alternatively, we
can use SRS at the same pressure $P$ as the mixture, whose volume
$V_{\text{SRS}}$ is different from $V$. We can use the negative of the
difference $V-V_{\text{SRS}}$ as a measure for cohesion. However, we will
study these quantities elsewhere and not here.

\subsection{Cohesion and Volume}

We have argued above that the changes in the volume can also be taken as a
measure of cohesion. We now follow this line of thought. The volume of mixing
is governed by two factors, as discussed elsewhere
\cite{Gujrati1998,Gujrati2003}: (i) the size disparity between the two
components, and (ii) the interactions. To disentangle the two contributions,
we can consider the difference
\[
\Delta_{\text{ath}}V\equiv V-V_{\text{ath}},
\]
where $V_{\text{ath}}$ is the volume of the athermal system (no interaction)
at the same $T,P$. (In the athermal limit, $\Delta V_{\text{M,ath}%
}=V_{\text{ath}}$ $-V_{1}^{\text{P}}-V_{2}^{\text{P}}$ will be identically
zero if the two polymer components have identical sizes $\left(  M_{1}%
=M_{2}\right)  $, and will be negative if they are different.) The difference
$\Delta_{\text{ath}}V$ is governed only by the presence of interactions
($e_{11},e_{22},$ and $e_{12}$). A positive $\Delta_{\text{ath}}V$ will imply
an effective repulsion, and a negative $\Delta_{\text{ath}}V$ will imply an
effective attraction. Thus, $\Delta_{\text{ath}}V$ can also be used as a
measure of the cohesiveness of the mixture.

\begin{figure}
[ptb]
\begin{center}
\includegraphics[
trim=0.902732in 5.820900in 0.903547in 0.903464in,
height=3.9444in,
width=6.3693in
]%
{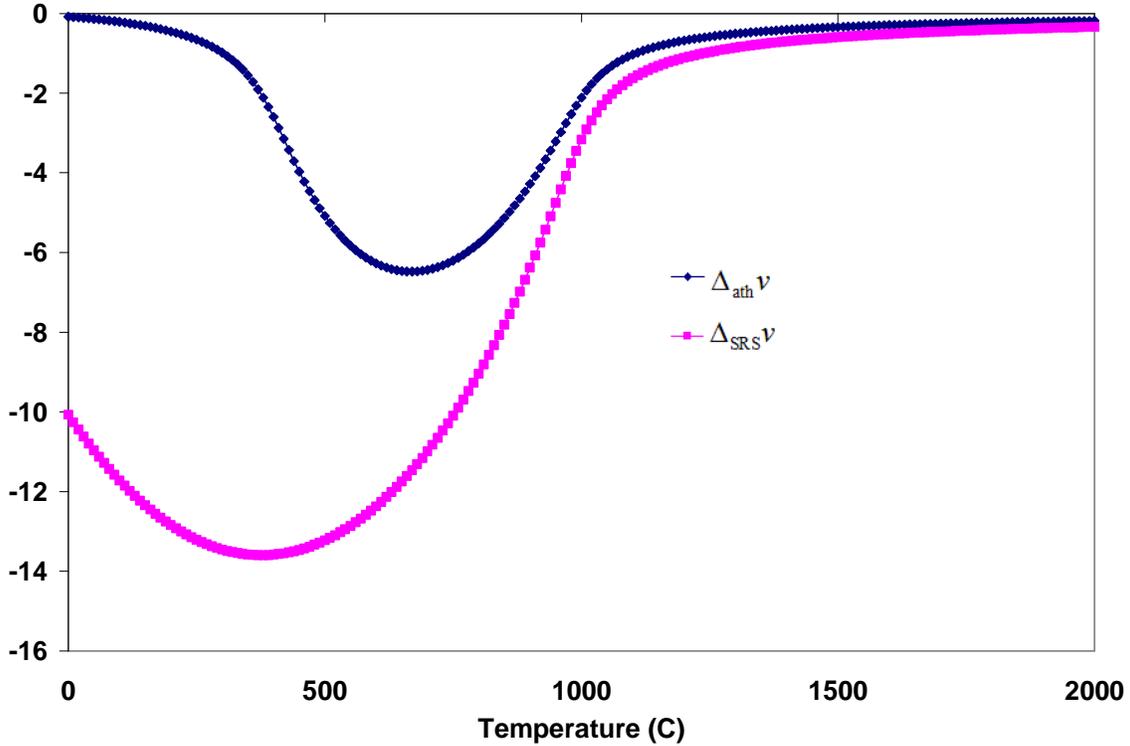}%
\caption{Volume difference $\Delta_{\text{ath}}v$ and $\Delta_{\text{SIRS}}v$
for a 50-50 blend as a function of temperature. We have $M_{1}=M_{2}=100$,
$e_{11}=-2.2\times10^{-21}$ J, $e_{22}=-2.6\times10^{-21}$ J $v_{0}%
=1.6\times10^{-28}$ m$^{\text{3}}$ and $q=10.$ The pressure is fixed at $1.0$
atm. }%
\label{F13}%
\end{center}
\end{figure}

It is easy to see that
\begin{equation}
V_{\text{ath}}/V=\phi_{\text{m}}/\phi_{\text{m,ath}}=(1-\phi_{\text{0}%
})/(1-\phi_{\text{0,ath}}), \label{Volume_ratio_ath}%
\end{equation}
in terms of total monomer density or void density. Using this, we can also
calculate the difference $\Delta_{\text{ath}}v\equiv\Delta_{\text{ath}}V/V$
per unit volume in our theory, and is shown in Fig. \ref{F13} in an isobaric
process at $1.0$ atm. We note that $\varepsilon_{01}=\varepsilon
_{02}=\varepsilon_{12}=0$ ($e_{11}=e_{12}=e_{22}=0$) for the athermal state,
while $\varepsilon_{01}=1.1\times10^{-21}$ J, $\varepsilon_{02}=1.3\times
10^{-21}$ J, and $\varepsilon_{12}\simeq0.01\times10^{-21}$ J ($e_{11}%
=-2.2\times10^{-21}$J$,$ $e_{12}=-2.6\times10^{-21}$J$,e_{12}\simeq
-2.39\times10^{-21}$ J) for the mixture. Recall that we have decided to define
$e_{12}$ for the mixture by (\ref{london_Conj}).~The difference $\Delta
_{\text{ath}}v$ is obviously controlled by all three energies $e_{ij}$, and
not just by $e_{12}.$ Since the mixture has attractive interactions
($e_{ij}<0$), its volume $V$ is smaller than $V_{\text{ath}}$ so that
$\Delta_{\text{ath}}v<0$ as seen in Fig. \ref{F13}.

To determine the effect of only $e_{12},$ we use SRS in place of the athermal
state and introduce the difference
\[
\Delta_{\text{SRS}}v=1-(V_{\text{SRS}}/V),
\]
which is also shown in Fig. \ref{F13}. Here, the mixture is compared with the
SRS\ state in which $\varepsilon_{01}=1.1\times10^{-21}$ J, $\varepsilon
_{02}=1.3\times10^{-21}$ J, but $\varepsilon_{12}=2.4\times10^{-21}$ J
($e_{11}=-2.2\times10^{-21}$J$,$ $e_{12}=-2.6\times10^{-21}$J$,$ $e_{12}=0$).
Since the mixture has extra attractive interaction ($e_{12}<0$) than the SRS
state ($e_{12}=0$), $V$ is smaller than $V_{\text{SRS}},$ so that the negative
value of $\Delta_{\text{SRS}}v$\ should not be a surprise. The minimum in
$\Delta_{\text{SRS}}v$ is due to the behavior of the void density, which is
shown in Fig. \ref{F14} for SRS and the mixture. In the two insets, we also
show the relative volume of mixing at constant pressure (1.0 atm) for SRS
(lower inset) and the mixture (upper inset). We notice that below around $500%
{{}^\circ}%
$C, the free volume in SRS changes faster than in the mixture, while above it,
the converse is the case. This changeover in the rate of change causes the dip
and the minimum in $\Delta_{\text{SRS}}v.$ Similarly, the dip in
$\Delta_{\text{ath}}v$ is caused by the changeover in the rate of change of
free volume in the mixture and the athermal state.

The minimum in $\Delta_{\text{SRS}}v$, see Fig. \ref{F13}, also explains the
intermediate maximum in the isobaric $c_{12}^{\text{SRS}}$ in Fig. \ref{F20}.
The behavior of $c_{12}^{\text{SRS}}$ shows that the mutual cohesiveness
increases and then decreases with the temperature, and is merely a reflection
of the way the void densities\ in Fig. \ref{F14} behave with the temperature.%

%

\begin{figure}
[ptb]
\begin{center}
\includegraphics[
trim=0.902732in 5.820900in 0.903547in 0.903464in,
height=3.9444in,
width=6.3693in
]%
{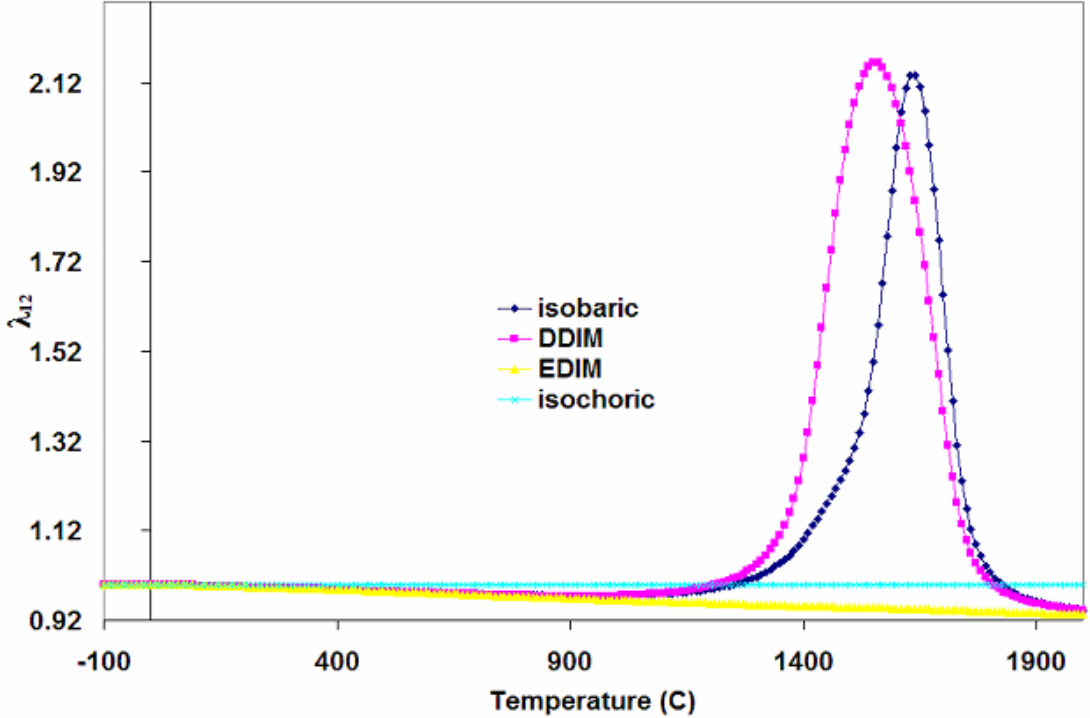}%
\caption{The scaled quantity $\gamma_{12}$ for various processes. This
quantity is very close to $1$ except at high temperatures where one of the
components seems to be near its boiling. For the system considered, we have
$M_{1}=M_{2}=100$, $e_{11}=-2.2\times10^{-21}$ J, $e_{22}=-2.6\times10^{-21}$
J $v_{0}=1.6\times10^{-28}$ m$^{\text{3}}$ and $q=14.$ The pressure is fixed
at $1.0$ atm.}%
\label{F21}%
\end{center}
\end{figure}
%

\begin{figure}
[ptb]
\begin{center}
\includegraphics[
trim=0.901917in 5.821965in 0.904361in 0.904527in,
height=3.9435in,
width=6.3693in
]%
{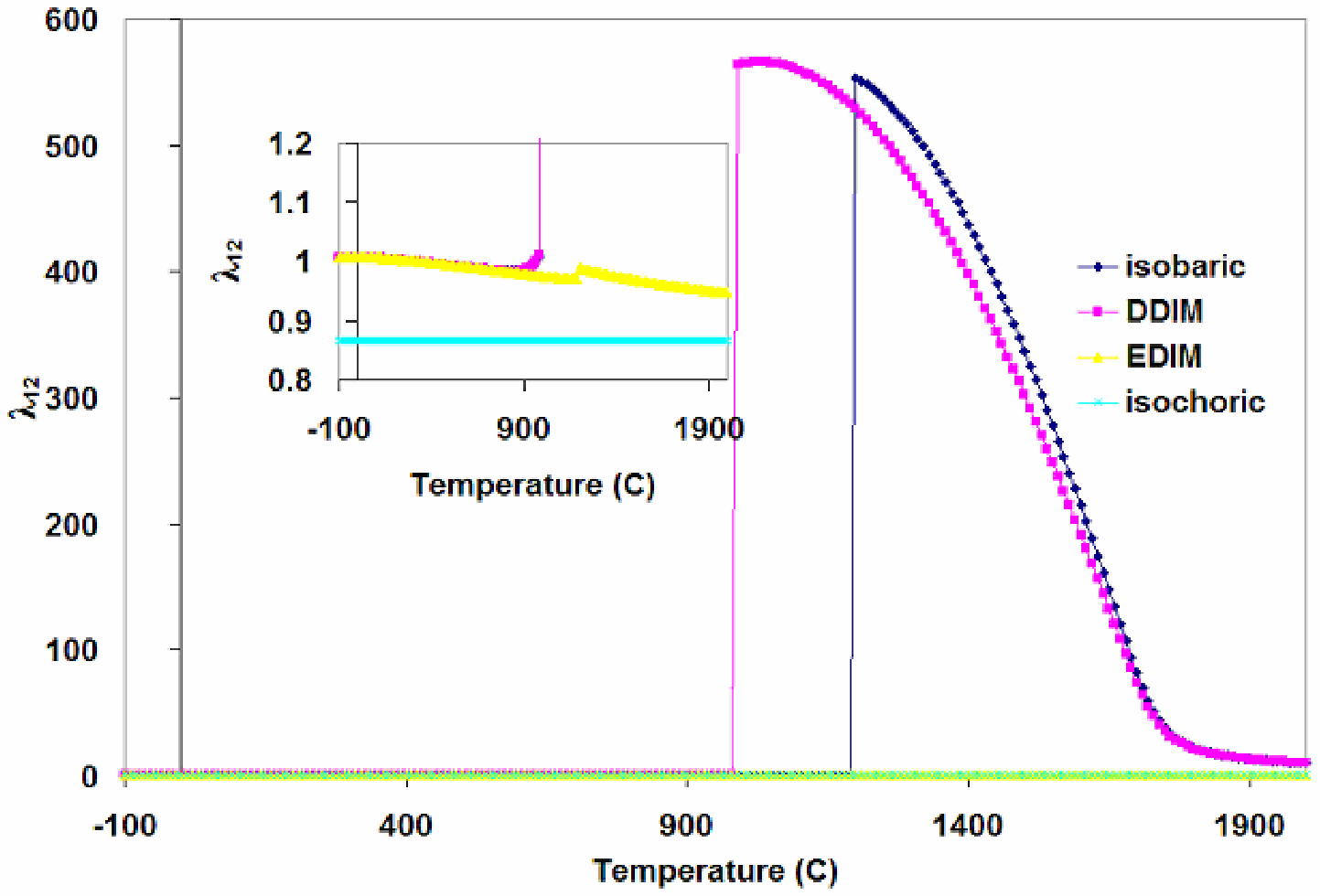}%
\caption{The scaled quantity $\gamma_{12}$ for various processes for a $50-50$
mixture. This quantity is very close to $1$ except at high temperatures where
the sharp discontinuity is due to one of its components undergoing boiling.
The system is $M_{1}=100,$ $M_{2}=10$, $e_{11}=-2.6\times10^{-21}$ J$=$
$e_{22},$ $v_{0}=1.6\times10^{-28}$ m$^{\text{3}}$ and $q=14.$ The pressure is
fixed at $1.0$ atm. }%
\label{F22}%
\end{center}
\end{figure}

\section{More Numerical Results: $\gamma_{12}$ and $l_{12}$}

We observe from (\ref{RMA_CohesiveDensities}) that the quantity
\[
\text{ }c_{12}v_{0}/(-qe_{12}\phi_{\text{m}}^{2}/2)
\]
approaches $1$ in the RMA limit. As we have seen earlier (compare Figs.
\ref{F9} and \ref{F10}), the quantity $q/2$ in the RMA limit should be
properly replaced by ($q/2-\nu)$ for finite $q.$ The new factor incorporates
the correction due to end groups. Therefore, we introduce a new dimensionless
quantity%
\[
\gamma_{12}\equiv-c_{12}v_{0}/(q/2-\nu)e_{12}\phi_{\text{m}}^{2},
\]
and investigate how close it is to $1.$ Any deviation is due to nonrandom
mixing and will provide a clue to its importance. In Fig. \ref{F21}, we plot
$\gamma_{12}$ as a function of temperature for various processes. The
coordination number is taken to be $q=14$, and the initial void density is
$\phi_{0}=0.00031333.$ This system is identical to the one studied in Fig.
\ref{F18}. For the isochoric process, $\gamma_{12}$ is almost a constant and
very close to (but below) $1.$ It decreases gradually for EDIM and becomes
almost 0.92 at about $t_{\text{C}}=2000^{\circ}$C. The situation is very
different for isobaric and DDIM\ processes where $\gamma_{12}$ shows
enhancement at a temperature near the boiling of one of the pure components.
To be convinced that the peak in $\gamma_{12}$ in Fig. \ref{F21} is indeed due
to boiling, we plot it for $q=14$ in Fig. \ref{F22}. We see a discontinuity in
$\gamma_{12}$ for all cases except for isochoric case, as is evident from the
inset (in which isobaric results are not shown). The discontinuity is due to
the boiling transition which occurs at different temperatures for the
different cases. These discontinuities become rounded as in Fig. \ref{F21},
when we are close to boiling transitions.%

\begin{figure}
[ptb]
\begin{center}
\includegraphics[
trim=0.902732in 5.820900in 0.903547in 0.903464in,
height=3.9444in,
width=6.3693in
]%
{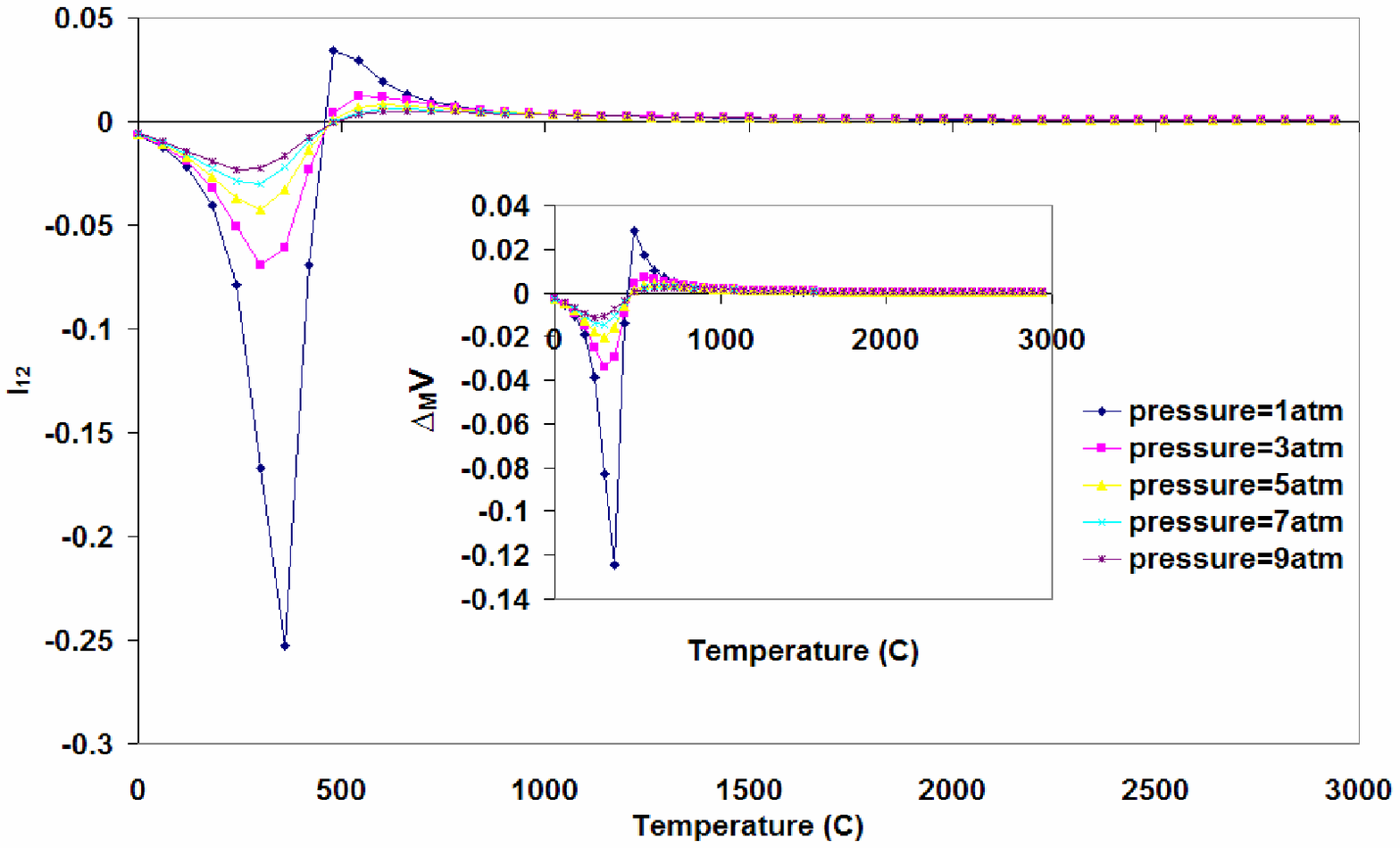}%
\caption{Binary deviation $l_{12\text{ }}$ and the relative volume of mixing
in the inset for a $50-50$ mixture. For the system considered, we have
$M_{1}=M_{2}=100$, $e_{11}=-2.2\times10^{-21}$ J, $e_{22}=-2.6\times10^{-21}$
J $v_{0}=1.6\times10^{-28}$ m$^{\text{3}}$ and $q=6.$ }%
\label{F23}%
\end{center}
\end{figure}
%

\begin{figure}
[ptb]
\begin{center}
\includegraphics[
trim=0.902732in 5.820900in 0.903547in 0.903464in,
height=3.9444in,
width=6.3693in
]%
{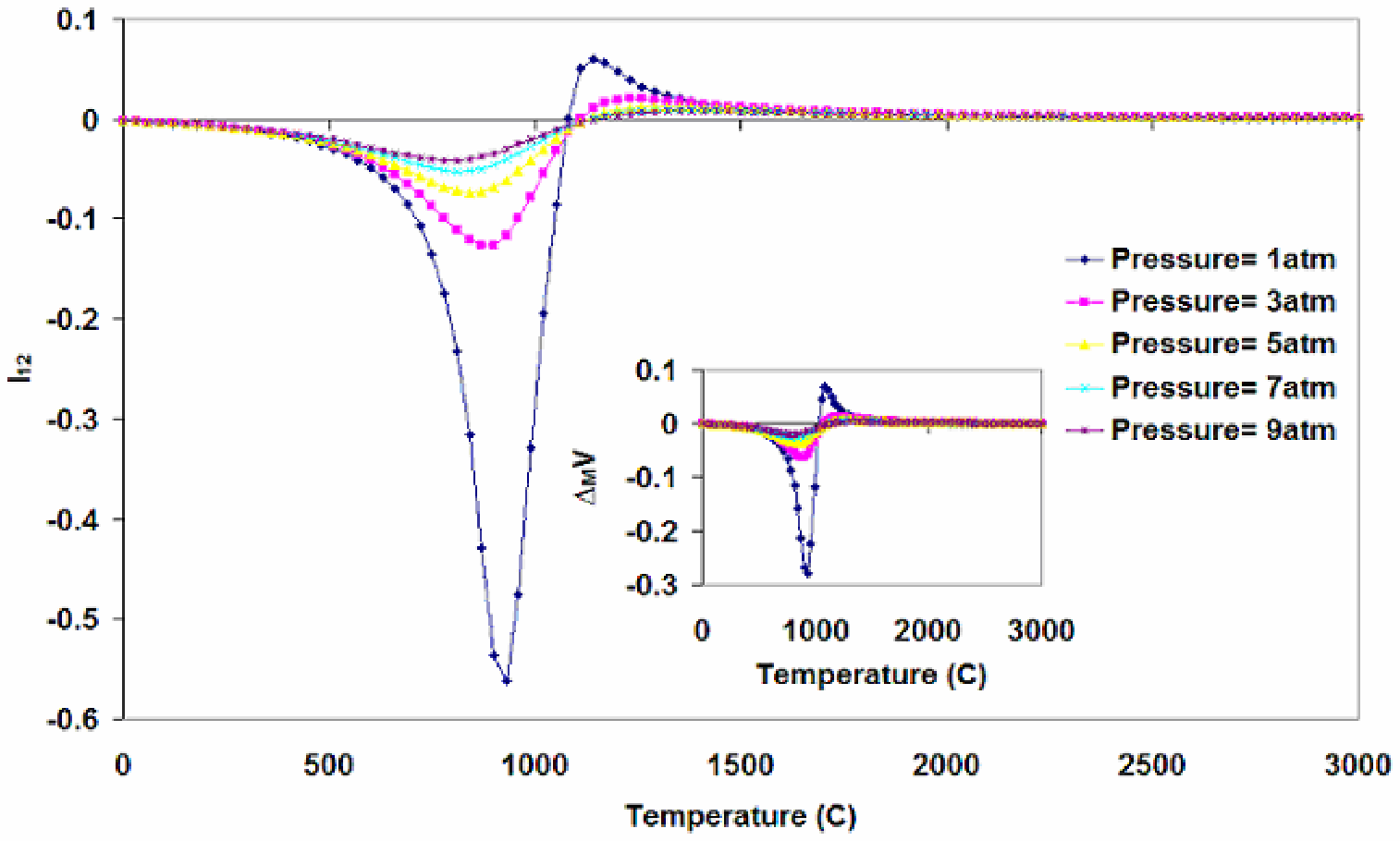}%
\caption{Binary deviation $l_{12\text{ }}$ and the relative volume of mixing
in the inset for a $50-50$ mixture. The system considered has $M_{1}%
=M_{2}=100$, $e_{11}=-2.2\times10^{-21}$ J, $e_{22}=-2.6\times10^{-21}$ J,
$v_{0}=1.6\times10^{-28}$ m$^{\text{3}}$ and $q=14.$ }%
\label{F24}%
\end{center}
\end{figure}

We now turn our attention to the binary correction $l_{12},$ which measures
the deviation from the London-Berthelot conjecture
(\ref{london_berthelot_Conj}). The existence of this deviation in the case
when we choose $e_{12\text{ }}$ to satisfy the London conjecture
(\ref{london_Conj}) will suggest that it is caused by thermodynamics which
makes cohesive energies different from their respective van der Walls energies
$e_{ij}$. Moreover, the behavior of $l_{12}$ also reflects partially the
behavior of the pure components, since the definition of $l_{12}$ utilizes the
pure component cohesive energies, see (\ref{Dev_l12}), even if $c_{12}$ is
defined without any association with them as is the case with $c_{12}%
^{\text{SRS}},$ whose definition is independent of pure component quantities
$c_{ii}^{\text{P}}$. Consider the inset in Fig. \ref{F16}, which shows
$l_{12}$ associated with $c_{12}.$ We observe that it is almost constant for
the two cases (DDIM\ and EDIM) for which mixing is isometric. However, it has
a strong variation for the isobaric case where mixing is not isometric. Thus,
we see the first hint of a dependence on volume of mixing in the behavior of
$l_{12}.$ Such a dependence on the volume of mixing is also seen in the inset
showing $l_{12}$ (corresponding to $c_{12})$\ in Fig. \ref{F18}. We are struck
by the presence of maxima and minima for the two isochoric cases. They appear
to be almost at same temperatures in both cases; notice the location of minima
and maxima in the isobaric $l_{12}$ and the volume of mixing.

To further illustrate the relationship between isobaric $l_{12}$ and the
volume of mixing $\Delta v_{\text{M}}$, we plot them as a function of the
temperature in Figs. \ref{F23} and \ref{F24} for $q=6$ and $q=14,$
respectively, for several values of the pressure $P.$ The strong correlation
between their maxima and minima are clearly evident. These figures illustrate
that the binary correction $l_{12}$ for the isobaric process closely follows
how the volume of mixing $\Delta v_{\text{M}}$ behaves. A negative $\Delta
v_{\text{M}}$ is a consequence of the effective attraction between the mixing
components, which corresponds to a larger value of their mutual cohesive
energy density $c_{12}$; this is reflected in a corresponding negative value
of $l_{12}.$ As the pressure increases, the correction decreases and the
system gets closer to satisfying the the London-Berthelot conjecture
(\ref{london_berthelot_Conj}). Thus, we can conclude that the binary
correction is negligible, but not zero, as long as the volume of mixing is
very small or even zero.

\section{Discussion and Conclusions}

We have carried out a comprehensive investigation of the classical concept of
cohesiveness as applied to a thermodynamic system composed of linear polymers.
In pure components, we are only dealing with one kind of interaction:
microscopic self-interactions between the monomers. Therefore, we are
interested in relating this microscopic self-interaction to a thermodynamic
quantity in order to estimate the strength of the microscopic
self-interaction. In binary mixtures (polymer solutions or blends), there are
two self interactions, and one mutual interaction. The microscopic self
interactions in the mixture do not depend on the state of the mixture.
Therefore, they are evidently the same as in pure components. Thus, it is
the\ microscopic mutual interaction that is new, and needs to be extracted by
physical measurements. However, what one measures experimentally is not the
microscopic interaction strengths, but macroscopic interaction strengths due
to thermodynamic modifications.

Traditionally, cohesive energies have been used as indicators of the
macroscopic interaction strengths. The definition of $c^{\text{P}}$ for pure
components, given in (\ref{cohesive_def}), does not depend on any particular
approximation or any theory. It is a general definition. Thus, it can be
measured directly if the interaction energy $\mathcal{E}_{\text{int}}$ can be
measured. It can also be calculated by the use of any theory. Only the
obtained value will depend on the nature of the theory.

On the other hand, the mutual cohesive energy density $c_{12}$ is defined by a
particular form of the energy of mixing. This gives rise to two serious
limitations of the definition. The first deficiency is caused by the use of
the form (\ref{mixEnergy1}), whose validity is questionable. Thus, the
extracted value of $c_{12}$ is based on this questionable form. The second
deficiency is that this value also depends on pure component properties
$c_{ii}^{\text{P}},$ since the mutual energy is measured with respect to the
pure components. However, we have argued that the energy of the mixture will
be different from the sum of the pure component energies even if $e_{12}$ is
absent. Thus, $c_{12}$ does not necessarily measure mutual interactions. One
of our aims was to see if some thermodynamic quantity can be identified that
could play the role of a true $c_{12}$ that would vanish with $e_{12}.$ This
is the quantity $c_{12}^{\text{SRS}}$ that we have identified in this work.
There is no such problem for $c^{\text{P}},$ since it not only vanishes with
$e$, as we have seen before, but it also does not suffer from any approximation.

We follow our recent work \cite{RaneGuj2005} in the approach that we take
here. We confine ourselves to a lattice model, since the classical approach
taken by van Laar \cite{vanLaar} and Hildebrand \cite{Hildebrand} is also
based on lattice models in that their theory can only be justified
consistently by a lattice model. We then introduce two different reference
states to be used for pure components and mixtures, respectively. For pure
components, we introduce a hypothetical reference state in which this
self-interaction is absent. Since there is only one interaction, which is
absent in the reference state, this state is nothing but the athermal state of
the pure component. With the use of this state, we introduce the interaction
energy $\mathcal{E}_{\text{int}}$ defined via (\ref{InteractionEnergy}).
Because of the subtraction in (\ref{InteractionEnergy}), $\mathcal{E}%
_{\text{int}}$\ depends directly on the strength of the self-interaction, the
only interaction present in a pure component that we wish to estimate. Thus,
it is not a surprise that $\mathcal{E}_{\text{int}}$\ is an appropriate
thermodynamic quantity that can be used to estimate the strength of the
self-interaction. Consequently, we use this quantity to properly define the
pure component cohesive energy density $c^{\text{P}}$ at any temperature. We
have discussed how this definition resembles the conventional definition of
$c^{\text{P}}$ in the literature. For the mixture, we introduce the
self-interacting reference state (SRS), which allows us to define
$c_{12}^{\text{SRS}}$ that vanishes with $e_{12}.$ In contrast, the van
Laar-Hildebrand $c_{12}$ does not share this property.

As noted earlier, our approach allows us to calculate $c^{\text{P}}$ and
$c_{12}$\ even for macromolecules like polymers, which is of central interest
in this work or for polar solvents, which we do not consider here but hope to
consider in a separate publication. The cohesive energy density $c^{\text{P}}$
is a macroscopic, i.e. a thermodynamic quantity characterizing\ microscopic
interparticle interactions in a pure component. In general, being a
macroscopic quantity, $c^{\text{P}}$ is a function of the lattice coordination
number $q$, the degree of polymerization $M$, and the interaction energy $e,$
in addition to the thermodynamic state variables $T,$ and $P$ or $V.$ We have
investigated all these dependences in our study here. The same is also true
for $c_{12}.$

Based on the same philosophy, we need to introduce another hypothetical
reference state called the self-interacting reference state (SRS) for a blend,
which differs from a real blend in that the mutual interactions are absent,
but self-interactions are the same. The difference $\mathcal{E}_{\text{int}%
}^{\text{M}}$ introduced in (\ref{Mutual_InteractionEnergy}) allows us to
estimate the strength of the microscopic mutual interaction between the two
components of a blend; we denote this estimate by $c_{12}^{\text{SRS}}$ to
distinguish it from the customary quantity $c_{12}$ originally due to van
Laar, and Hildebrand and coworkers. However, $\mathcal{E}_{\text{int}%
}^{\text{M}}$ is not what is usually used to define the mutual cohesive energy
density $c_{12}.$ Rather, one considers the energy of mixing $\Delta
E_{\text{M}}$. We have compared the two approaches in this work.

The conventional van Laar-Hildebrand approach to solubility is based on the
use of the regular solution theory. Thus, several of its consequences suffer
from the limitations of the regular solution theory (RST). One of the most
severe limitations, as discussed in the Introduction, is that the theory
treats a solution as if it is incompressible. This is far from the truth for a
real system. Hence, one of the aims of this work is to investigate the finite
compressibility effects, i.e., the so-called "equation-of-state" effects. As a
real solution most definitely does not obey the regular solution theory, the
experimental results usually will not conform to the predictions of the
theory. Thus, we have revisited the solubility ideas within the framework of a
new theory developed in our group. This theory goes beyond the regular
solution theory and also incorporates equation-of-state effects. We have
already found that this theory is able to explain various experimental
observations, which the regular solution theory (Flory-Huggins theory for
polymers) cannot explain. Therefore, we believe that the predictions of this
theory are closer to real observations than those of the regular solution theory.

\subsection{Pure Components}

We have first studied the cohesive energy density $c^{\text{P}}$of a pure
component. Since it is determined by the energy density of vaporization, it is
a quantity independent of the temperature. It is also clear from the
discussion of the van der Waals fluid that $c^{\text{P}},$ defined by
(\ref{cohesive_def}), is independent of the temperature, since the parameter
$a$ in (\ref{vdW_a}) is considered $T$-independent. This particular aspect of
$c^{\text{P}}$ ensures the equality of $P_{\text{IN}}\equiv-P_{\text{int}}%
\ $and $(\partial\mathcal{E}/\partial V)_{T};$ see (\ref{PINdEdV}). For a pure
component, Hildebrand \cite{Hildebrand1916} has argued that the solubility of
a given solute in different solvents is determined by relative magnitudes of
internal pressures, at least for nonpolar fluids. Thus, we have also
investigated the internal pressure. However, the equality (\ref{PINdEdV}) is
not valid in general as we have shown above. Thus, the internal pressure,
while a reliable and alternative measure of the cohesion of the system in its
own right, cannot be equated with $(\partial\mathcal{E}/\partial V)_{T}.$
Their equality is shown to hold only in RMA. If we allow $a$ in (\ref{vdW_a})
to have a temperature-dependence, the equality (\ref{PINdEdV}) will no longer
remain valid. Moreover, it can be easily checked that under this situation,
$P_{\text{IN}}~$and $c^{\text{P}}$ for the van der Waals fluid will no longer
be the same.

We have investigated the pure component $c^{\text{P}}$ using our recursive
theory. We find that it is very different depending on whether we consider an
isochoric process, in which case it is almost constant with $T$, or an
isobaric process, in which case it gradually decreases to zero at very high
$T$. In general, we find that
\[
c_{V}^{\text{P}}\ \geq c_{P}^{\text{P}}.
\]
\ We have found that isochoric $c^{\text{P}}$\ and $P_{\text{IN}}$ are almost
constants as a function of $T.$ It provides a strong argument in support of
their usefulness as a suitable candidate for the cohesive pressure. But
$c^{\text{P}}$\ and $P_{\text{IN}}$ do not remain almost constant in every
process, as the isobaric results in Fig. \ref{F2} clearly establish. The same
figure also shows in its inset that the isobaric $c^{\text{P}}$ exhibits a
discontinuity due to boiling, as expected. Unfortunately, most of the
experiments are done under isobaric conditions; hence the use of isochoric
cohesive quantities may not be useful, and even misleading and care has to be
exercised. We also see that $c^{\text{P}}$ changes a lot over the temperature
range and replacing it by its value at the boiling point may be not very
useful in all cases.

The cohesive energy also depends on the molecular weight, and usually
decreases with $M$ as expected; see Fig. \ref{F7}. This is the situation for
$q=14,$ and should be contrasted with the behavior shown in the inset in Fig.
\ref{F2}, where we show $c_{P}^{\text{P}}$ for $M=10,$ and $100,$ but for
$q=10.$ All other parameters are the same as in Fig. \ref{F7}. What we observe
is that at low temperatures, $c_{P}^{\text{P}}$ for $M=10$ lies above the
$c_{P}^{\text{P}}$ for $M=100,$ while the situation is reversed at high
temperatures. This crossover is due to the boiling transition that the smaller
$M$ pure component must undergo at about $600%
{{}^\circ}%
$C. Thus, $c_{P}^{\text{P}}$ decreases with increasing $M$ only far below the
boiling temperatures. The dependence on the lattice coordination number $q$ is
also not surprising: it increases with $q$ but requires end group correction
in that $c^{\text{P}}$ is proportional to $(q/2-\nu);$ see Fig. \ref{F9}$.$
This means that the solubility function $\delta$ also depends on the process,
and its value at the boiling temperature of the system will be dramatically
different in the two processes. We have also found, see Fig. \ref{F6}, that
$c^{\text{P}}$ is not truly a quadratic function of the monomer density, one
of the predictions of the regular solution theory.

\subsection{Mixtures (Solutions or Blends)}

We now turn to the mutual cohesive energy density $c_{12}.$ Here, the
situation is muddled since the definition of $c_{12}$ is based on a form
(\ref{mixEnergy1}) of the energy of mixing $\Delta E_{\text{M}},$\ whose
validity is questionable beyond RST. This should be contrasted with the
definition of $c^{\text{P}},$defined by (\ref{cohesive_def}), which is
independent of any particular theory. Therefore, $c^{\text{P}}$ can be
calculated in any theory without any further approximation except those
inherent in the theory. It can also be measured directly by measuring
$\mathcal{E}_{\text{int}},$\ and does not require any further approximation to
extract it. On the other hand, the calculation of $c_{12}$ in any theory
requires calculating $\Delta E_{\text{M}}$\ in that theory; thus, this
calculation is based on the approximations inherent in the theory. However,
one must still extract $c_{12}$ from the calculated $\Delta E_{\text{M}}$ by
expressing\ $\Delta E_{\text{M}}$\ in the form (\ref{mixEnergy1}). As
discussed above, this form is justified only in RST and $c_{12}$ in
(\ref{mixEnergy1}) at this level of the approximation is indeed a direct
measure of the mutual interaction energy $e_{12}.$ Whether $c_{12}$ defined by
the form (\ref{mixEnergy1}) still has a direct dependence on $e_{12}$ is one
of the questions we have investigated here by introducing SRS. For $c_{12}$ to
be a direct measure of the mutual interaction energy $e_{12},$ we have to
ensure that it vanish in SRS. What we have shown is that $c_{12}$ does not
vanish in SRS as seen in Fig. \ref{F19}. A new measure of the mutual cohesive
energy density $c_{12}^{\text{SRS}}$ is introduced in (\ref{cohesiveIRS}),
which has the desired property.

The reference state SRS behaves very different from the mixture or its
athermal analog, as clearly seen from the Figs. \ref{F19},\ref{F14}, and
\ref{F13}. In particular, SRS has strong repulsive interactions between the
two species. This results in the SRS volume to be much larger than that of the
mixture. Therefore, it is possible that the two components may undergo phase
separation in SRS. Its use to define $c_{12}^{\text{SRS}}$, therefore, should
only be limited to the case where SRS is a single phase state so as to compare
with the mixture. We do not report any result when SRS is not a single phase.

We have found that $c_{12}^{\text{SRS}}$ has the correct behavior that it
first rises and then decreases with $T$ for a compressible blend. As explained
above, this is a reflection of the way the void density behaves with the
temperature, as shown in Fig. \ref{F13}. The rise and fall of cohesion is not
apparent in the temperature-dependence of $c_{12}$. It has also been shown
that $c_{12}^{\text{SRS}}$\ can be expressed in terms of $c_{12}.$ Therefore,
we have basically explored the behavior of $c_{12}$ more than that of
$c_{12}^{\text{SRS}}.$\ As expected, and as shown in Fig. \ref{F12}, $c_{12}$
increases with $q.$ However, it is not just simply a linear function of $q$,
as seen in Fig. \ref{F12}. It is also clear from the behavior in this figure
that $c_{12}$ changes its curvature with $T.$ Thus, $c_{12}$ is not linear
with the inverse temperature $\beta$ over a wide range of temperatures.

We have been specifically interested in the contribution due to volume of
mixing. For this reason, we have considered three different kinds of isometric
mixing (zero volume of mixing), two of which we have called EDIM and DDIM.
These are introduced in Sect. V. The third isometric mixing is studied as part
of isochoric processes, which we have also considered. The fourth process is
an isobaric process in which mixing is at constant pressure. We have also
compared $c_{12}$ with a related internal pressure quantity $P_{\text{IN}%
}^{\text{M}},$ which is another measure of the mutual cohesive energy density
or pressure$.$ For a particular process, the two quantities behave similarly,
though their magnitudes are different in that $P_{\text{IN}}^{\text{M}}%
>c_{12};$ see Fig. \ref{F18-1}. We have found that isobaric and EDIM
quantities are somewhat similar, and both are different from the
DDIM\ quantities. All three quantities have a strong temperature dependence,
but the isochoric quantity is very different in that $c_{12}$ or
$P_{\text{IN}}^{\text{M}}$ for the latter remains almost a constant with $T.$

We have paid special attention to the violation of the Scatchard-Hildebrand
conjecture (\ref{mixEnergy0}) even when the microscopic interactions obey the
London conjecture (\ref{london_Conj}). We find that even under isometric
mixing (EDIM), the energy of mixing can become negative, as seen in Fig.
\ref{F18}. There is another possible source of violation of the
Scatchard-Hildebrand conjecture seen in Fig. \ref{F15}. Here, the pure
components in the isobaric and DDIM processes remain the same, so that the
pure component solubilities $\delta$ are the same. Therefore, from
(\ref{mixEnergy0}) we would conclude that the energy of mixing should be zero,
regardless of the process. What we find is that the energy of mixing is not
only not zero, it is also different in the two processes. This violation is
due to non-random mixing caused by size and/or interaction strength disparity.

We have also investigated the behavior of the binary correction $l_{12}.$ We
find that it need not be small. In particular, we find that it becomes very
large in isobaric and DDIM processes as seen in Fig. \ref{F18}. As the
pressure increases, $l_{12}$ decreases in magnitude and the deviation from the
London-Berthelot conjecture (\ref{london_berthelot_Conj}) becomes smaller.
This effect shows that the deviation from the London-Berthelot conjecture is
due to the presence of compressibility. Thus, the behavior of the free volume
determines that of $l_{12}.$ In the isobaric case, we have discovered a very
interesting fact. The behavior of $l_{12}$ mimics the behavior of the volume
of mixing as seen clearly in Figs. \ref{F23}, and \ref{F24}. This observation
requires further investigation to see if there is something unusual about
isobaric processes or it is jut due to non-isometric mixing in the presence of
free volume..

In summary, we have found that it is not only the non-isometric mixing by
itself that controls the behavior of the cohesive quantities, but also the
non-random contributions.

\begin{acknowledgments}
Acknowledgement is made to the National Science Foundation for support (A.A.
Smith) of this research through the University of Akron REU Site for Polymer
Science (DMR-0352746).
\end{acknowledgments}

\end{document}